\documentclass[%
 aip,
 amsmath,amssymb,
 reprint,%
]{revtex4-1}

\usepackage{graphicx}
\usepackage{dcolumn}
\usepackage{bm}

\usepackage[utf8]{inputenc}
\usepackage[T1]{fontenc}
\usepackage{mathptmx}
\usepackage{etoolbox}
\usepackage{appendix}

\usepackage{color}

\makeatletter
\def\@email#1#2{%
 \endgroup
 \patchcmd{\titleblock@produce}
  {\frontmatter@RRAPformat}
  {\frontmatter@RRAPformat{\produce@RRAP{*#1\href{mailto:#2}{#2}}}\frontmatter@RRAPformat}
  {}{}
}%
\makeatother
\begin{document}

\preprint{AIP/123-QED}

\title[Emulation for Tokamak Plasma Scenario and Control Design]{Emulation Techniques for Scenario and Classical Control Design of Tokamak Plasmas}
\author{A. Agnello}
 \email{adriano.agnello@stfc.ac.uk}
\affiliation{%
STFC Hartree Centre, Sci-Tech Daresbury, Keckwick Lane, Daresbury, Warrington (UK) WA4 4AD
}%
 
\author{N.~C. Amorisco}%

\author{A. Keats}

\author{G. K. Holt}
\affiliation{%
STFC Hartree Centre, Sci-Tech Daresbury, Keckwick Lane, Daresbury, Warrington (UK) WA4 4AD
}%

\author{J. Buchanan}
\affiliation{%
United Kingdom Atomic Energy Authority, Culham Science Centre, Abingdon (UK) OX14 3DB
}%

\author{S. Pamela}
\affiliation{%
United Kingdom Atomic Energy Authority, Culham Science Centre, Abingdon (UK) OX14 3DB
}%

\author{C. Vincent}
\affiliation{%
United Kingdom Atomic Energy Authority, Culham Science Centre, Abingdon (UK) OX14 3DB
}%

\author{G. McArdle}
\affiliation{%
United Kingdom Atomic Energy Authority, Culham Science Centre, Abingdon (UK) OX14 3DB
}%

\date{\today}

\begin{abstract}
The optimisation of scenarios and design of real-time-control in tokamaks, especially for machines still in design phase, requires a comprehensive exploration of \textcolor{black}{solutions to the Grad-Shafranov (GS) equation over a high-dimensional space of plasma and coil parameters}.
\textcolor{black}{Emulators can bypass the numerical issues in the GS equation, if a large enough library of equilibria is available.} We train an ensemble of neural networks to emulate the typical shape-control targets (separatrix at midplane, X-points, divertor strike point, flux expansion, poloidal beta) as a function of plasma parameters and active coil currents for the range of plasma configurations relevant to spherical tokamaks with a super-X divertor, with percent-level accuracy.
This allows a quick calculation of the classical-control shape matrices, potentially allowing real-time calculation at any point in a shot with sub-ms latency.
We devise a hyperparameter sampler to select the optimal network architectures and quantify uncertainties on the model predictions. To generate the relevant training set, we devise a Markov-Chain Monte Carlo algorithm to produce large libraries of forward Grad-Shafranov solutions without the need for user intervention. The algorithm promotes equilibria with desirable properties, while avoiding parameter combinations resulting in  problematic profiles or numerical issues in the integration of the GS equation.
\end{abstract}

\maketitle
%
%
\section{\label{sec:Intro}Introduction}


\textcolor{black}{The Grad-Shafranov equation is ubiquitous in the description of magneto-hydrodynamical (MHD) equilibria, from tokamaks to astrophysical plasmas \cite{Freidberg_2008,Schindler2010}. It is a nonlinear, partial-differential equation with nontrivial boundary conditions, and its accurate solution is necessary in order to describe the current-density and pressure profiles of a plasma in a magnetic field. In the rest of this paper, we concentrate on issues inherent the use of the Grad-Shafranov equation in tokamak plasmas, and how they can be alleviated by the use of suitably constructed equilibrium libraries and emulator architectures.} Several tokamak designs are being developed in order to bring magnetic-confinement fusion to net power generation. In preparation for energy production from tokamaks, such as STEP (\citet{10.1088/978-0-7503-2719-0ch8}) and SPARC (\citet{2020JPlPh..86e8602C, 2022NucFu..62d2003R}), precursor machines have been in operation to probe fundamental physics, different operational regimes, and control strategies (e.g. JET \cite{JET_1992}, ASDEX-U \cite{Gruber1997}, NSTX-U \cite{Kaye2005, 2017NucFu..57j2006M}, MAST-U \cite{MAST2005,MAST2015,Fishpool2013}, D\textsc{iii}-D \cite{DiiiD1995}, TCV \cite{2022NucFu..62d2018R}, Globus-M \cite{GlobusM_2009, GlobusM_2019}, ST40 \cite{ST40_2019, ST40_2023}). \textcolor{black}{In a tokamak, equilibria are established on $\approx\mu$s Alfv\'en timescales that are much faster than those of the typical current drives and dissipation, so the plasma can indeed be modelled as a sequence of MHD equilibrium solutions\cite{Ariola2008,Freidberg_2008}. }


Tokamak fusion experiments require real-time control of the burning plasma and of its exhaust. A range of physics-based algorithms have been designed to control the plasma's shape targets, its profile, and its exhausts, ultimately to operate the machines reliably and attain longer confinement times. On machines with a large number of campaigns, the experiment data themselves have been used to build emulators of the control targets and predict the state corresponding to a set of actuator requests \citep{Seo_2021,Wai2022,Wan2023,Wei2023}, bypassing the use of expensive and sometimes numerically unstable PDE solvers.
This route, however, requires large data volumes and is not viable for newer and upcoming machines. There, shot design typically proceeds by constructing sequences of desired equilibria and designing classical (linear) control for selected regimes. Here, we propose to complement this with physics-based emulation.
Based on a set of synthetic equilibria covering the whole operational space of a tokamak,
we present strategies to build emulators of the plasma and magnetic field configurations,
with the aim of facilitating and accelerating scenario design and control.
 The same approach is also applied to the exploration of different magnetic configurations that can be used in studies of divertor detachment. \textcolor{black}{For the reader's convenience, Table 1 summarises the symbols and notations chosen. Throughout this paper we use underlined symbols to denote multidimensional arrays of parameters.}

\begin{table}
\renewcommand\arraystretch{1.2}
\begin{tabular}{l | l }
\hline
\textbf{symbol} & \textbf{meaning} \\ 
\hline
$R$ & major radius (from tokamak axis) [m] \\
$Z$ & vertical coordinate (from tokamak midplane) [m] \\
$I_p$  & plasma current [MA] \\
$f_{vac}$ & $B_{\phi}R$ in vacuum \\
$p_{a}$ & plasma pressure at magnetic axis [kPa] \\
$I_{sol}$ & solenoid current [A] \\
$\underline{I}_{c}$ & array of poloidal-field coil currents [A] \\
$\underline{X}$ & array of plasma- and coil-parameters of a given equilibrium\\
$\underline{y}$ & array of target quantities derived from a GS equation solution \\
\hline
\hline
\end{tabular}

\caption{\label{tab:notation} Symbols and notations used in this paper. }
\end{table}

\subsection{\label{sec:ABCD}Classical Control}
Coil-current control is traditionally designed by examining how a set of  desired targets change under a small change in active-coil currents. Usually, this is done under a set of approximations, in particular that the passive-structure transients decay fast enough and that the plasma \textcolor{black}{Ohmic dissipation} is negligible over the millisecond timescales of real-time control. \textcolor{black}{These hypotheses are generally satisfied by the choice of solenoid and coil "sweeps", and by the closed-loop control to keep the coil currents close to the desired ones and hence minimise the shielding effects of passive structures \citep{Ariola2008}.} The control design equations assume a lumped-parameter circuit-equation for the coupling of plasma current $I_{p}$ and \textcolor{black}{desired shaping} coil currents $\underline{I}_{c},$ so that 
under a small change $\delta \underline{I}_{c}$ in coil currents, a given target $y$ changes as
\begin{equation}
 \label{eq:controldesign}   \delta y = \left(\frac{\partial y}{\partial \underline{I}_{c}}- \frac{\partial y}{\partial I_{p}}\frac{\underline{L}_{cp}^{T}}{L_{p}}\right)\delta \underline{I}_{c} 
\end{equation}
with suitably defined inductances $L.$ Different authors adopt different definitions for the inductances \cite{Lister_2002, Ariola2008, Freidberg_2008,albanese2015create}, which we discuss in the Appendix \ref{sect:circuits} for the sake of completeness.

The objective of real-time control is to find \textcolor{black}{changes $\delta \underline{I}_{c}$ in the array of \textit{requested} shaping-coil currents, on top of a pre-programmed coil sweep, that correspond} to a desired change $\delta \underline{y}$ in the \textcolor{black}{array of} targets of interest.
This is done by considering the SVD pseudo-inverse of the sensitivity matrix ${\mathbf{S}}=\partial\underline{y}/\partial\underline{I}_{c},$ yielding \textit{virtual circuits} with the desired linear combinations of \textcolor{black}{coil-}current changes \cite{Humphreys2007, ELDON2020111797, MCARDLE2020111764, ANAND2023113724}. \textcolor{black}{Once the desiderd virtual circuits are obtained, they are used as "requested" coil current inputs in classical closed-loop control \citep{Ariola2008}}. The details of the chosen targets are specific to each tokamak, but are all broadly related to geometric or poloidal magnetic flux constraints on the shape of the separatrix and its extension into the divertor region. The sensitivity matrix is an inherently local quantity, as it is computed around one given equilibrium. In practice, computing a sensitivity matrix around any foreseen equilibrium configuration in a shot is computationally prohibitive. Therefore, at experiment design, the coil "wave-forms" are constructed by prescribing the plasma shape and divertor strike points, and the virtual circuits are computed for a finite set of expected configurations along these trajectories, using sensitivity matrices calculated by finite differences. 
The desired shape and strike point may be computed at different values of the Ohmic circuit currents, if they have significant stray field (as e.g. on MAST-U), otherwise different sets of virtual circuits can be designed to control the main shape of the plasma and the divertor.


\subsection{\label{sec:divertor}Divertor Studies}
Modern research tokamaks are endowed with a divertor, to remove impurities and manage the high heat flux \cite{Pitcher1997, Gruber1997, Ambrosino2008, Goldston2010, Petrie2013, Fishpool2013}. The transport of heat along a flux tube can be generally captured e.g. by the so called "two-point model", which explicitly demonstrates the importance of flux expansion and connection length \cite{Pitcher1997,Kotov2009,Stangeby2018}. Refined analyses can be performed with more complete and multi-species descriptions of transport along field lines, or in the full meridional plane \cite{chankin2006,havlichkova2015,Dudson2023, Myatra2023}. The design and control of divertors is an active field of research, but one overall objective is the attainment of high flux expansion, i.e. the widening of a flux tube from the scrape-off layer towards the machine tiles \cite{Pitcher1997,Moulton2017}. Finding magnetic configurations that can provide long connections and high flux expansions, compatibly with the operational limits and plasma shape constraints, is an integral part of scenario design.


\subsection{\label{sec:thispaper}This paper}
From the above examples, two general issues emerge: (\textsc{i}) is it possible to quickly obtain magnetic configurations and targets of interest at any point over a wide range of control parameters? and (\textsc{ii}) is it possible to traverse the operational space to search for desirable configurations in the presence of non-trivial boundaries (as e.g. resulting from the need for a diverted plasma)?

In order to minimise manual intervention and also enable uncertainty quantification in follow-up work, we develop smooth emulations of targets of interest over the parameter range, that in principle can be deployed with sub-ms latency. This in turn requires a library of synthetic equilibria, for which we develop an efficient sampling algorithm. The synthetic equilibria, numerical challenges and the resulting sampling algorithm are discussed in Section~\ref{sect:equilibria}. The emulation is discussed in Section~\ref{sect:emulation}, including a sampling algorithm to search hyperparameter space. All results are presented in section~\ref{sect:results}, and implications are discussed in Section~\ref{sect:discussion}.

\begin{figure}
\includegraphics[width=0.475\textwidth]{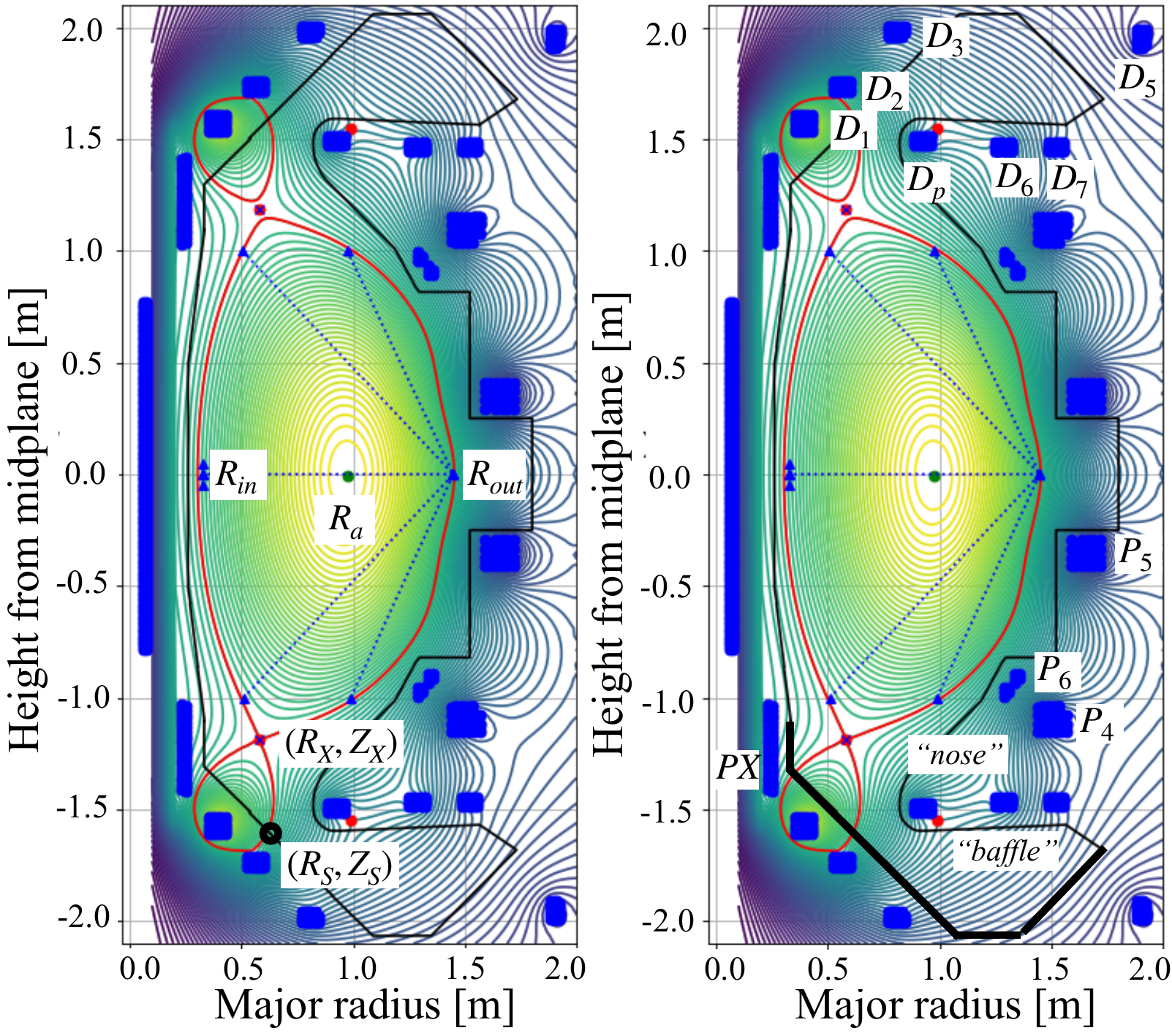}\\
\caption{\label{fig:MASTU} Tokamak configuration for this study, roughly based on MAST-U, and first synthetic equilibrium. The separatrix ($\psi=\psi_{b}=\psi_{X}$) is marked in red. \textit{Left:} names of main shape targets computed and considered for the subsequent emulation. \textit{Right:} divertor coils (D1-D7, Dp) and shape coils (P4-P6). All coils are up-down symmetric except P6, whose winding is up-down antisymmetric and is used for vertical position control. The divertor tiles are marked by thick black lines, and the divertor "nose" and "baffle" are indicated. The tokamak coil positions and windings are given by \textcolor{black}{the "\texttt{MASTU}" tokamak class in the public  \texttt{FreeGS} repository}. }
\end{figure}

We remark that all equilibria in this work are synthetic, i.e. they are computed assuming a perfectly known state of tokamak and plasma, and none are from a reconstruction of experiment shots. This also allows us to control the sources of uncertainty in the emulation, which only come from the numerics of the equilibrium construction and emulation procedures, without any systematics from the profile reconstruction. The equilibria used in this work are computed on a tokamak that {resembles} the MAST-U machine configuration, although with different coil positions and windings (displayed in Fig.~\ref{fig:MASTU}). \textcolor{black}{The full machine description is given in the public \texttt{FreeGS} code repository\footnote{{https://github.com/freegs-plasma/freegs}}. While we are interested in solutions that can be applied to the real MAST-U, here we deemed appropriate to use a synthetic machine as it is fully specified (no measurement uncertainties), while retaining the general complexity of a real-life tokamak. }
All of the profiles in this work are \textit{diverted}, i.e. their last-closed flux surface within the main chamber has at least one X-point. This is because, at least on MAST-U, the plasma can be brought from a limited to a diverted configuration shortly after the breakdown phase.

\section{ \label{sect:equilibria} Synthetic Dataset}

The plasma equilibrium configurations are described by the Grad-Shafranov (GS) equation, constrained by the choice of global parameters for the plasma and coil currents, as discussed in Section~\ref{sect:GS} below. The library of synthetic equilibria is built through a random walk in parameter space as discused in Section~\ref{sect:MCMC}, starting from one equilibrium with prescribed shape. To solve for the flux-function of the synthetic equilibria, we use the publicly available \texttt{FreeGS} codebase\footnote{ {https://github.com/freegs-plasma/freegs} } (Dudson et al. in prep.) and the static GS solver 
presented by \citet{Amorisco2023}. Wherever needed, we build additional functionality as described in the rest of the paper.

The procedure to build the equilibrium library is devised in order to address two specific issues. First, due to the nonlinearities of the GS equation, its integration converges more often if starting from a "reasonable" guess of the solution, which is available if a new trial equilibrium parameters are sampled close to those of a previously computed equilibrium. Second, a good emulation of the control targets is needed primarily in regimes with more desirable shape properties, and as the dimensionality of the parameter space increases, more na\"{i}ve Monte Carlo methods (e.g. acceptance-rejection, Latin hypercube, pseudo-random) will often sample equilibria far away from those with more desirable properties.

Several conventions are used in the literature for the flux-function $\psi$, here we define it such that the poloidal magnetic field satisfies $B_{R}=-\partial_{Z}\psi/R.$ In what follows, $f_{vac}$ denotes the amplitude $B_{\phi}R$ in vacuum, $\psi_{a}$ and $\psi_{b}$ are $\psi$ on the magnetic axis and at the boundary, and $\psi_{n}=(\psi-\psi_{a})/(\psi_{b}-\psi_{a}).$ We will work with free-boundary problems, where $\psi$ is to be determined together with a toroidal current density $j_{\phi}$ and the plasma boundary is defined as the surface where $\psi$ equals the one at the closest X-point to the magnetic axis. 

\subsection{\label{sect:GS} Grad-Shafranov solutions}
The flux-function $\psi$ and toroidal current density $j_{\phi}(R,Z)$ must statisfy the Grad-Shafranov equation
\begin{equation}
\Delta^{*}\psi=-\mu_{0}R j_{\phi}(R,Z) = -\mu_{0}R\left(j_{pl}+j_{coils}  \right)
\label{eq:GS}
\end{equation}
where $\Delta^{*}g=R\mathrm{div}(R^{-1}\nabla g),$ the toroidal plasma current is
\begin{equation}
j_{pl}(\psi,R)=Rp^{\prime}(\psi)+\frac{1}{\mu_{0}R}f^{\prime}f(\psi)
\end{equation}
and $p$ and $f$ are the pressure and $B_{\phi}R$ along a surface of constant $\psi$ respectively. By $j_{coils}$ we denote any current density in conductors that can be modelled through a set of azimuthally symmetric coils. Given a current density $j_\phi,$ the corresponding $\psi$ can be obtained by a (suitably defined) Green-function integration over the plasma and coil domain $\mathcal{D},$
\begin{equation}
\psi = \int_{\mathcal{D}}\mathcal{G}(R^{\prime},Z^{\prime},R,Z)j_{\phi}(R^{\prime},Z^{\prime})\mathrm{d}R^{\prime}\mathrm{d}Z^{\prime}\ \equiv \psi^{iGS}(j_{\phi})
\end{equation}
so if a dependence $\hat{j}_{\phi}(\psi,R)$ on $\psi$ is prescribed through $p(\psi)$ and $f(\psi)$, a mapping can be defined as
\begin{equation}
\psi(R,Z) \mapsto \psi^{iGS}(\hat{j}_{\phi}(\psi(R,Z),R))
\end{equation}
and solving the GS equation~(\ref{eq:GS}) amounts to finding
\begin{equation}
\psi(R,Z)\ \mathrm{s.t.}\ \psi^{iGS}(\hat{j}_{\phi}(\psi(R,Z),R))-\psi(R,Z) =0
\end{equation}
with additional constraints. 
In a \textit{forward} free-boundary GS problem, the currents in the coils are prescribed, as are the plasma current, $f_{vac}$ and plasma parameters.
In an \textit{inverse} (or \textit{interpretative}) GS problem, the coil currents or the plasma parameters are not provided and are to be found based on a suitable set of constraints on $\psi$ or on the magnetic field, e.g. X-point and isoflux constraints. During a shot, these in turn result from a real-time inference of the flux-function in vacuum from magnetic measurements (flux loops, pickup coils).

Approximate solutions to the inverse GS problem can be found by alternating a Picard iteration
\begin{equation}
\psi^{(n+1)} = \psi^{iGS}(\hat{j}_{\phi}(\psi^{(n)},R))
\end{equation}
and an adjustment of the coil currents towards the constraints. In our work, adopting the parameterization choices of \textcolor{black}{\citet{Lackner1976} and } \citet{Jeon2015}, the plasma parameters can be found explicitly once current and pressure on axis are prescribed. In the \textit{forward} GS problem, the Picard iterations are often unstable and a more robust root-finder is needed. Here, we use 
the static GS solver which is part of the \texttt{FreeGSNKE} package~\cite{Amorisco2023}, which uses the Newton-Krylov method and was developed on purpose to extend the \texttt{FreeGS} framework. 

We choose $p(\psi)\propto f^{\prime}f(\psi) \propto 1-\psi_{n}^{2}$ for the whole library of equilibria. In particular, with the notation of \citet{Jeon2015}:
\begin{eqnarray}
j_{pl}=\left( 1-\psi_{n}^{2} \right)\left( \lambda\beta_{0}R/R_{0}+(1-\beta_{0})\lambda R_{0}/R \right)
\end{eqnarray}
where $R_0$ is an arbitrary scale-length, and the scalings $\lambda,$ $\beta_{0}$ can be solved for explicitly at given $p_a$ and $I_{p}.$

Given the physics of the problem, the shape targets should only depend on the ratio of coil currents to plasma current, and on $\beta_{p}\propto p_{a}/I_{p}^{2},$ with the only caveat that the coil currents (rather than their ratio to $I_p$) must be kept within some operational limits. 
The first equilibrium is described by the constraints in Table~\ref{tab:first} and shown in Figure~\ref{fig:MASTU}, which also introduces some of the nomenclature. The procedure to build the library, by varying the coil currents, $f_{vac},$ plasma current and pressure on axis, uses the Newton-Krylov solver for the forward GS problem and is described below.

\begin{table}
\renewcommand\arraystretch{1.2}
\begin{tabular}{l | l | l}
\hline
\textbf{constraints} & \textbf{initial values} & \textbf{limits}\\  
\textbf{/ parameters} &  & \\  
\hline
X-point (R,Z) [m] & (0.58,1.189), (0.58,1.185) & 
\\ 
isoflux (R,Z) [m]  & (0.327,0), (1.4475,0), & \\
                   & (0.9725, 1.0017), (0.9892, -0.9993), & \\
                   & (0.5035, 0.9997), (0.5081, -1.0018)  & \\ 
\hline
$I_p$ [MA] & 0.748 & [0.1,1.125] \\
$f_{vac}$ & -0.475 & $\pm$0.75 \\
$p_{a}$ [kPa] & 9.59 & [1.0, 100.0] \\
$I_{sol}$ [A] & -1400 & $\pm$55$\times10^3$\\
\hline
\textbf{PF currents} & \textbf{initial values from inverse GS} & \textbf{limits} \\ 
\textcolor{black}{\textbf{(single-turn)}} &  &  \\  \hline
$I_{PX}$ [A] & 2102.4 & $\pm$3.5$\times10^3$  \\
$I_{D1}$ [A] & 7130.0 & $\pm$11$\times10^3$ \\
$I_{D2}$ [A] & 1816.1 & $\pm$6.4$\times10^3$ \\
$I_{D3}$ [A] & 371.3 & $\pm$6.4$\times10^3$ \\
$I_{Dp}$ [A] & -2029.3 & $\pm$6.4$\times10^3$ \\
$I_{D5}$ [A] & -1022.7 & $\pm$4.0$\times10^3$  \\
$I_{D6}$ [A] & -320.3 & $\pm$3.2$\times10^3$  \\
$I_{D7}$ [A] & -957.4 & $\pm$4.6$\times10^3$  \\
$I_{P4}$ [A] & -2391.3 & $\pm$11$\times10^3$ \\
$I_{P5}$ [A] & -4377.8 & $\pm$11$\times10^3$ \\
$I_{P6}$ [A] & 38.2 & $\pm$3.5$\times10^3$  \\
\hline
\end{tabular}

\caption{\label{tab:first} Parameters and constraints used to obtain the first synthetic equilibrium (\textit{top}, shown in Figure 1), and the resulting PF coil currents from the inverse GS solution (\textit{bottom}). All equilibria except the first one are obtained by integrating the forward GS problem under small changes of the coil+solenoid currents and plasma parameters. \textcolor{black}{All coil currents are meant as single-turn currents; the number of turns of each coil for this machine description is publicly available through the "MASTU" tokamak class in the \texttt{FreeGS} code repository.} }
\end{table}

\subsection{\label{sect:MCMC} Sequential Sampling of Synthetic Equilibria}

Any solution of the forward Grad-Shafranov equation can be identified with its input parameters, so in what follows we will use interchangeably
\begin{equation}
eq(n) \leftrightarrow (p_{a}^{(n)},I_{p}^{(n)},f_{vac}^{(n)},I_{sol}^{(n)},\underline{I}_{PF coils}^{(n)\ T}) \equiv \underline{X}_{n}^{\mathrm{T}}
\end{equation}

Each synthetic equilibrium is given a score, $\mathcal{L}(eq.),$ that privileges better-behaved configurations. In general, the score is given by
\begin{eqnarray}
\nonumber \mathcal{L}(eq.) & \propto &  \mathcal{L}(I_{coils}) \mathcal{L}(\psi_{n}(wall))  \mathcal{L}(q_{a})\\ 
\nonumber & & \times \mathcal{L}(R_{X},Z_{X}|\mathrm{X-point}) \mathcal{L}(R_{in}, R_{out}) \\ 
 & & \times \mathcal{L}(conn. length) \mathcal{L}(flux\  exp.)\ .
\end{eqnarray}
The details of $\mathcal{L}(eq.)$ are not important here, and the particular functional forms of its factors \textcolor{black}{can be} varied across different \textcolor{black}{sampling} runs to ensure that interesting regions of parameter space are adequately sampled. \textcolor{black}{An example is given in the Appendix.} 
In general, the score $\mathcal{L}(eq.)$ increases with the connection length, and in general it privileges equilibria with strike points on the divertor tiles, inboard mid-plane point away from the central column, X-point between the divertor "nose" and the column tile, small displacement of the magnetic axis from the midplane, and safety factor on axis $q_{a}>1.$  A penalty term in the coil currents is included in order to ensure that the current in PX is $\approx 0$~A when the current in the solenoid is $>2000$~A or $<-2000$~A, to demonstrate the inclusion of physical constraints necessary when running a real machine. The connection length is computed from one scrape-off-layer width of 3~mm from the midplane outboard radius. 

In order to sample $\mathcal{L}(eq.)$, a Monte Carlo Markov Chain (MCMC) is built, such that a new proposed equilibrium "prop.eq.(n+1)" is accepted with probability $\min (1, \mathcal{L}(eq.(n+1))/\mathcal{L}(eq.(n)))$. At each step, the Grad-Shafranov equation is integrated using the previous equilibrium as a first guess. In the formulae below, $X_{j,n}$ denotes the $j-$th coordinate in parameter space (i.e. plasma parameters or coil currents) of the $n$-th equilibrium. In a Metropolis-Hastings (MH) sampler, a walker draws a step in parameter space from a chosen distribution, while in the affine-invariant samplers of \citet{GW2010} each of $2W+2$ walkers draws the trial step based on the relative positions of $W$ of the other walkers.
Here, a combination of the two is adapted into a sequential sampler as:
\begin{eqnarray}
\nonumber step_{MH_j}(n) & \in & \mathcal{U}[-1.0,1.0](max_{j}-min_{j})/100 \\
 & & \forall\ j=1,...,\mathrm{dim_{in}}\\
\mu_{GW, j}(n) & = & \frac{1}{P}\sum_{p=1}^{P}X_{j,n-p}\\
\nonumber step_{GW , j}(n) & = & \sqrt{\frac{3}{P}} \sum_{p}u_{p}(X_{j,p}-\mu_{GW, j}(n) ),\\
 & & \ u_{p}\in \mathcal{U}[-1.0,1.0] \ \ \forall\ j,p\\
step(n) & = & step_{MH}(n) +step_{GW}(n)\\
 prop.eq.(n+1) & = & eq(n)+step(n)\ ,
\end{eqnarray}

with $P=2\rm{dim}_{in}+2=32$. The prefactor $\sqrt{3/P}$ is such that $\langle step_{GW,i}step_{GW,j} \rangle$ is the same as the sample covariance of the last $P$ equilibria.

This walker update rule, which makes the step size adaptive (i.e. dependent on the local steepness of the score $\mathcal{L}_{eq}$), is used only after 100 steps of pure MH sampling. 
A uniform distribution $\mathcal{U}$ is preferred to a normal distribution for the steps, because it samples more efficiently step sizes above the variance while also truncating steps that would be too large and therefore challenging for the convergence of the Grad-Shafranov integration. If the Grad-Shafranov integration fails to converge, the walker is restarted from one of the equilibria in the chain, chosen at random.

Strictly speaking, $step_{GW}$ is not the same as in the Goodman \& Weare sampling algorithm, but it has the advantage that the chain is parameterised by a single walker, which eases the task of indexing the output configurations $\psi_{n}(R,Z).$ Even though this MCMC may not necessarily satisfy the detailed balance conditions, it results in an efficient exploration.

\section{\label{sect:emulation} Emulation}
We examine two solutions to obtain fast fitting functions to the targets computed on synthetic equilibria, that can also enable extrapolation and uncertainty quantification: scaling relations and neural networks. 
All of the neural networks examined in this work are built with the \texttt{TensorFlow python} APIs. \textcolor{black}{We devote a separate subsection to the design of an annealed hyperparameter sampler, which is devised on purpose for this work to produce an ensemble of emulators.}

\subsection{Scaling Relations}
Some of the targets show a simple behaviour with the inputs, which may be described by simple power-laws. To investigate these trends in the targets, we perform power-law fits via a least-squares linear regression in their logarithms. The only regressor that is fit with an exponential, instead of a power-law, is the solenoid current rescaled by the plasma current. The overall scaling relation to be fit is then, for each target $\theta:$
\begin{equation}
\label{eq:scalings} \theta \propto {I_{p}}^{\alpha_{p}}\left(\frac{p_{a}}{I_{p}^{2}}\right)^{\alpha_{bp}} f_{vac}^{\alpha_{f}}\  \mathrm{exp}\left(\alpha_{sol}(I_{sol}/I_{p})\right)  \prod_{c}(I_{c}/I_{p})^{\alpha_{c}}
\end{equation}
While a power-law scaling can be an adequate hypothesis for most parameters, the solenoid current spans a wide range of positive and negative values and the behaviour of some of the parameters is clearly asymmetric in the solenoid current, so an exponential was the preferred choice. This combination still ensures that the fits of log-targets are all linear combinations of the regressors ($\alpha_{p},$ $\alpha_{bp},$ $\alpha_{f},$ $\alpha_{sol},$ $\underline{\alpha}_{c}$). We remark that, in this work, all quantities are from synthetic profiles, so the uncertainties are arbitrarily low (except for numerical round-off) and the best-fit regressors are the same whether the fit is performed in target space or log-target space.

\subsection{Neural Networks}

\begin{figure}
\includegraphics[width=0.4\textwidth]{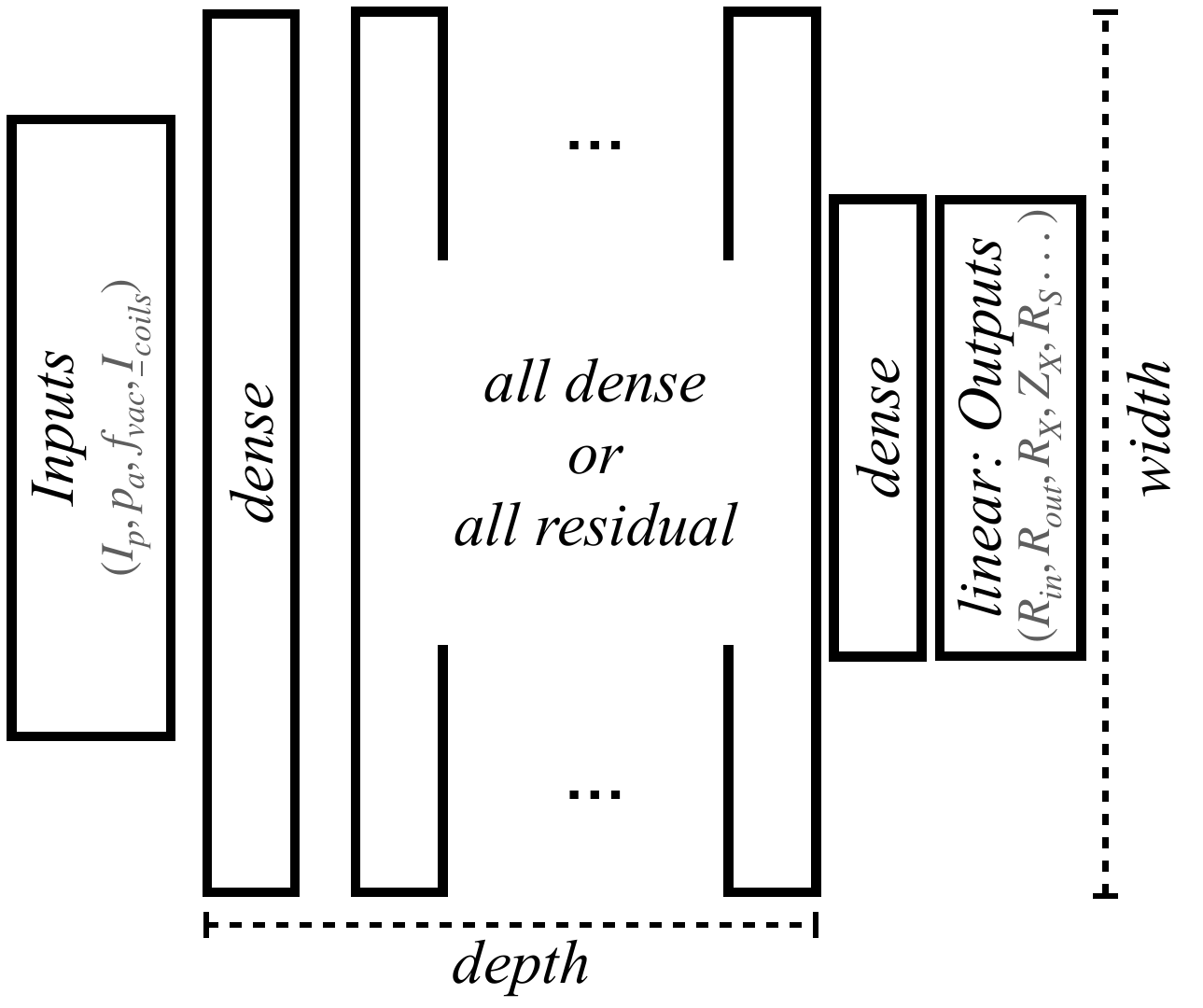}\\
\caption{\label{fig:ANNschema} \textcolor{black}{Nomenclature and g}eneral architecture of the NNs considered in this work. All have one input layer with identity activation $\sigma(\mathbf{x})=\mathbf{x},$ one dense layer, then $d-1$ layers (all dense or all redisual, with the same activation), one dense and one affine layer with the same shape as the outputs. The \textcolor{black}{user can choose the activation function among} a sigmoid, a ReLU, a leaky ReLU, an eLU, and a SReLU (see eq.~\ref{eq:srelu}). The depth, width, learning rate, $L^2$ regularisation and gradient batch size are all hyperparameters to be explored. }
\end{figure}

Neural networks \textcolor{black}{(hereafter NNs)} are amenable to the task of emulation because: (1) Universal Approximation theorems make them appealing candidates in general; (2) they can also be used to extrapolate in regions of parameter space that are less densely sampled; (3) off-the-shelf packages can compute all derivatives analytically, overcoming issues of numerical noise; (4) one can build simple enough architectures that can be deployed with low latency on a real-time controller; (5) uncertainties can be quantified whenever an ensemble of neural-network emulators has been trained, on different random subsamples of a dataset or with different hyperparameters.

While Universal Approximation theorems are known, three remarks will be useful for this work. First, a neural network with sigmoid activations is a universal approximator of continuous functions over a compact domain, so any target that may diverge close to the domain boundary or with a discontinuity would need a suitable reparameterisation of the inputs and outputs. This is true also for networks with a \textit{Rectified Linear Unit} activation $ReLU(x)=(x+|x|)/2,$ also as approximators to $L^{p}$ functions in the $L^{p}$ norm, which provides only a global goodness-of-fit measure whenever the functions have discontinuities. This is the case e.g. for the connection length, which can change abruptly as soon as the scrape-off layer grazes the divertor "nose" region. Another example is the strike-point location, which can be quite sensitive to small changes in $R_{out}$ or $\lambda_{SOL}$ and can exhibit a sudden discontinuity (primarily in $Z$) whenever the divertor coils create another X-point in the divertor region. Third, while there are known lower bounds to the network width that is required for universal approximation, there is no known lower bound on the depth (nor a known upper bound on depth and width) for finite datasets or prescribed fitting tolerance, so an extensive exploration of the network architecture is needed.

The neural networks used here are fully-connected and feed-forward, and are trained to fit all chosen targets simultaneously for each parameter input. All inputs and targets are standardised before training. They can be trained to fit the targets or their logarithm, using the inputs as-is or rescaled by powers of the plasma current as above; the inputs can be further passed as they are, or under logarithm or arcsinh. The simplest architecture (a combination of ReLU activations on log-inputs and log-outputs) is also the simplest extension of the scaling relations considered above. We examine different possible activation functions: the ReLU defined above; a \textit{leaky ReLU} (i.e. $ReLU(x)+0.3ReLU(-x)$) ; an \textit{exponential LU} ($eLU(x)=ReLU(x)+(e^{x}-1)\mathbf{1}_{x<0}$); a sigmoid $-\log(1+\mathrm{e}^{-x});$ and a smoothed ReLU
\begin{eqnarray}
\nonumber SReLU(x,\alpha) & := & \log(\exp(x)+\exp(\alpha x))\\
\label{eq:srelu} & = & \alpha x + \log(1+\exp((1-\alpha)x))
\end{eqnarray}
which behaves asymptotically like a leaky ReLU but with smooth gradients everywhere and is a simple combination of \texttt{TensorFlow} built-in primitives. Two kinds of architecture are available: the first has $d$ hidden layers all with the same activation function; the second has one first hidden layer as above, and the remaining $d-1$ as residual blocks
\begin{equation}
\mathbf{x} \mapsto \mathbf{x}+ \sigma(\mathbf{w}\cdot\mathbf{x}+w_{0})\ .
\end{equation}

With a (large but) finite data-set, wider or deeper networks are more prone to overfitting, so regularization terms must also be introduced. A quadratic regularization is added to the loss function, which in turn can be a mean-squared error or a mean-absolute error. The hyperparameter exploration procedure is described below, \textcolor{black}{together with the choices made to split the data into training, evaluation, and test sets}. 

\subsection{Hyperparameter Tuning}
The chosen NN architecture has several hyperparameters to be tuned: depth (how many hidden layers), width (how many nodes per hidden layer), batch size, learning rate $l_{r},$ regularization $l_{L2}.$ A grid search is expensive and prone to arbitrary choices in e.g. how one would set the grid mesh in $l_{r}$ and $l_{L2}.$ A `random search' with uniform draws is less expensive, but still rather noisy and hence inefficient. Ideally, we want a sampler that already knows about islands of hyperparameter space that perform better, but with e.g. a MCMC we would discard an arbitrary number of computed models. Off-the-shelf packages for a "Bayesian" hyperparameter search exist, however they are not necessarily suited to this particular problem. Population-Based Training \cite{PBT} replaces the worst performing models with mutations of the best-performing ones, and so it is mostly useful for the scheduling of $l_r$ and $l_{L2},$ while in principle it cannot resample models with varying width and depth. Hyperband \cite{hyperband} starts from a random draw of hyperparameters and progressively prunes the worst performing models in favour of training the best performers for more epochs. So-called "Bayesian" optimizers, like \texttt{Optuna} \cite{optuna}, use a density estimator to approximate the loss \textit{vs} hyperparameters and therefore inform the next hyperparameter draw, however this neglects the inherent scatter in losses (even on the same hyperparameter tuple) from different initial NN weights and different random splits in training and validation sets.

We build a different implementation for this work, where in principle all trained models can contribute to inform a random draw of the next hyperparameter trial. Below, $h_{n}$ denotes the $n$-th hyperparameter combination being examined. The first $P_{hp}$ choices $h_{1},...,h_{Php}$ are drawn from a prior distribution within some chosen ranges. Then, given the $n$ choices $\underline{h}$ and losses examined so far, with their validation losses $L^{(j)}$ ($j=1,...n$), the choice (n+1) for $n=P_{hp},...,N_{mc}$ uses an explore-exploit strategy. In particular, with probability $\mathcal{P}_{n}$ the hyperparameter tuple $h_{n+1}$ is drawn like the first $P_{hp},$ and with probability $1-\mathcal{P}_{n}$ it is drawn as follows:
\begin{eqnarray}
\alpha_{n} & = & 1.0 + 9.0\times 2n/(n+N_{mc})\\
p_{j} & = & \mathrm{exp}(-\alpha_{n}L_{j})\ \ \forall j=1,...,n\\
z_{l} & \in & \{\underline{h} | \underline{p}\}\ \ \forall l=1,...,P_{hp}\\
\mu & = & \frac{1}{P_{hp}}\sum_{l=1}^{P_{hp}}  z_{l} \\
\nonumber h_{n+1} & = & \mu+\ \sqrt{\frac{3}{ P_{hp}}} \sum_{l=1}^{P_{hp}} u_{l}(z_{l}-\mu)\ , \\
 & \  & u_{l}\in\mathcal{U}[-1.0,1.0]\ \ \forall l=1,...,P_{hp}\\
\underline{h} & \mapsto & \underline{h} \cup \{h_{n+1}\}
\end{eqnarray}
Here $x\in\{\underline{h}|\underline{p}\}$ means that a hyperparameter $h_{j}$ in $\underline{h}$ is drawn with probability $p_{j}$ and its value is assigned to $x.$ In other words, in the "exploit" case, $P_{hp}$ models are chosen among the $n$ available (privileging models with lower validation losses) and are used to build a random vector for the next trial model. The steepness $\alpha_{n}$ allows for a more demanding selection of the $P_{hp}$ vertices $z_{l}.$ 
The "explore" probability threshold is parameterised as
\begin{equation}
\mathcal{P}_{n} = \mathrm{max}(1,P_{hp}/n)\varepsilon_{0}
\end{equation}
For this work, $P_{hp}=10$ and $\varepsilon_{0}=0.1$ have been chosen. {The loss is the mean-absolute deviation evaluated on the validation set, described below.} \textcolor{black}{Similarly to the sequential MCMC above, the prefactors $\sqrt{3/P_{hp}}$ are such that the hyperparameter steps are combinations with standardised random coefficients. We choose this approach over the MCMC just because it retains all of the sampled hyperparameters, and every explored hyperparameter can (in principle) contribute to the choice of a new hyperparameter combination.}

\textcolor{black}{From the synthetic dataset, a random 20\% has been set aside as a {test} set, over which the performance of the sampled models is evaluated at the end of the hyperparameter exploration. Of the remaining 80\%, whenever a new model is instantiated, a random 20\% (i.e. 16\% of the original data) is taken as a {validation} set for early-stopping.} For each hyperparameter trial, the training is run until $it_{m}$ such that $loss_{val}(it_{m})\geq loss_{val}(it_{m}-20)$ and $it_{m}\leq 500.$ The final loss $L_{j}$ for $h_{j}$ is then $loss_{val}(mod_{j},it_{m}).$ \textcolor{black}{During training, the neural-network gradients are evaluated on mini-batches, whose size is also one of the hyperparameters. The choice of a random validation subset for every new model instantiation is made to render the hyperparameter search robust against specialising on a particular training-validation split.}



\section{\label{sect:results} Results}

\begin{figure}
\includegraphics[width=0.475\textwidth]{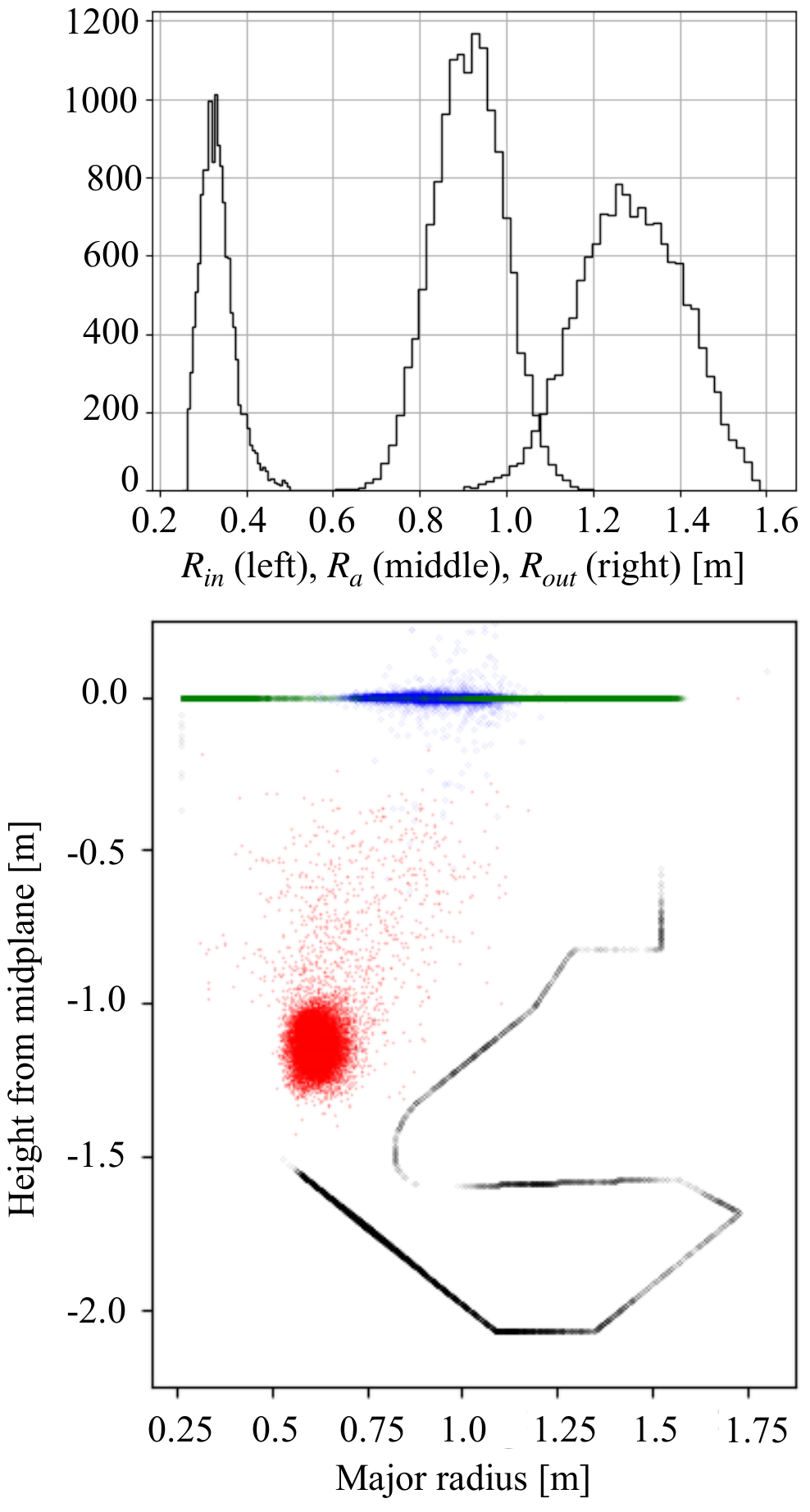}
\caption{\label{fig:shape} Distribution of the shape targets computed on the library of synthetic profiles. In the lower panel, the black symbols indicate the strike-point positions (a majority is on the lower tiles), which result in the distribution of allowed X-point positions (red swarm). The green and points around $Z=0$ mark the range of inner/outer separatrix radii and magnetic axis, respectively. Their distribution is shown in the histograms in the top panel. }
\end{figure}
Approximately thirty thousand synthetic profiles are computed, together with a number of targets and other integrated properties for each profile. The NN emulation and hyperparameter sampling is done on a subset of the computed targets, while others are provided with the
synthetic library as they can be useful for diagnostic purposes, e.g. to select particular equilibria for divertor detachment simulations. The distribution of parameter values explored by the MCMC sampler is shown in Figure~\ref{fig:mcmchist} in the Appendix.

\subsection{Synthetic libraries}

Figure~\ref{fig:shape} summarizes the distribution of the shape targets computed on the synthetic equilibria. Solutions with a strike point on the divertor "nose" or "baffle" are highly penalised in the library generation but are not rejected, in order for the subsequent emulators to learn which parameter combinations can lead to unfavourable divertor conditions. Histograms of the separatrix midplane inboard and outboard radii, as well as the magnetic axis radius, are also shown. The sharp cutoff at $R\approx0.26$~m is where the limiter of the inner column is touched. The distribution of (lower) X-point locations is merely reflective of the score function $\mathcal{L}$ privileging configurations with the strike point hitting the bottom divertor tiles. The connection length 
can exhibit a sharp drop whenever the SOL grazes the divertor "nose", 
mostly due to equilibrium configurations with large outboard separatrix radius, and long tails due to configurations where an additional X-point is created in the divertor region.

\subsection{Scaling Relations}
Some of the targets behaviour is captured by the scaling relations of equation~(\ref{eq:scalings}), based on the $R^2$ metric of the least-squares fit of the logarithm of the parameters, defined as
\begin{equation}
R^{2}_{log}=1-\mathrm{var}(\log(\theta/\theta_{fit}))/\mathrm{var}(\log\theta)
\end{equation}
In particular:

\begin{equation}
R_{a}\propto \left(\frac{p_{a}}{I_{p}^{2}}\right)^{0.1} \mathrm{e}^{-0.7I_{sol}/I_{p}} \left(\frac{I_{P5}}{I_{p}}\right)^{-0.33} 
\end{equation}

\begin{equation}
\frac{I_{p}Z_{a}}{I_{P6}}\propto \left(\frac{p_{a}}{I_{p}^{2}}\right)^{0.17}\mathrm{e}^{-4.5I_{sol}/I_{p}}\left(\frac{I_{D1}}{I_{p}}\right)^{-0.1}\left(\frac{I_{P5}}{I_{p}}\right)^{-1.64}\left(\frac{I_{Dp}}{I_{P4}}\right)^{0.04} 
\end{equation}

\begin{equation}
R_{out}\propto \left(\frac{p_{a}}{I_{p}^{2}}\right)^{0.07}\mathrm{e}^{-0.65I_{sol}/I_{p}}\left(\frac{I_{P5}}{I_{p}}\right)^{-0.4} 
\end{equation}

\begin{equation}
Z_{X}\propto \mathrm{e}^{-1.0I_{sol}/I_{p}}\left(\frac{I_{D1}}{I_{p}}\right)^{-0.1} 
\end{equation}

\begin{equation}
\psi_{a}\propto \left(\frac{p_{a}}{I_{p}^{2}}\right)^{0.13}I_{p}  \left(\frac{I_{D1}}{I_{p}}\right)^{0.1}\left(\frac{I_{P5}}{I_{p}}\right)^{-0.44} 
\end{equation}

\begin{equation}
l_{i}(2)\propto \left(\frac{p_{a}}{I_{p}^{2}}\right)^{+0.11} \mathrm{e}^{0.4I_{sol}/I_{p}} \left(\frac{I_{P5}}{I_{p}}\right)^{-0.23}  
\end{equation}

\begin{equation}
q_{a} \propto \left(\frac{p_{a}}{I_{p}^{2}}\right)^{-0.35} \left(\frac{f_{vac}}{I_{p}}\right)^{0.82} \mathrm{e}^{-1.5I_{sol}/I_{p}}\left(\frac{I_{D1}}{I_{p}}\right)^{-0.1} \left(\frac{I_{P5}}{I_{p}}\right)^{0.54} 
\end{equation}

with $R^{2}_{log}$ scores of 0.7, 0.96, 0.8, 0.7, 0.9, 0.74, 0.88 respectively. Some of the exponents are trivially expected (e.g. $Z_{a}\propto I_{P6}/I_{p}$ or $\psi_{a}\propto I_{p}$), while others reflect the chosen tokamak configuration (e.g. the impact of the poloidal-field coils P4, P5 and divertor coil D1). The scaling of $q_a$ is roughly consistent with what would be expected from the small-radius approximation $q_{a} \approx f_{vac}(\partial_{Z}\psi_{n})^{-1}/( 2\pi R_{a}|\psi_{a}-\psi_{b}|)$ $\propto f_{vac}Z_{x}^{2}/( R_{a}|\psi_{a}-\psi_{b}|),$ although with slightly different exponents, including the small diamagnetic effects in the plasma ($\partial \log q_{a}/\partial\log(f_{Vac}/I_{p})=0.82<1$). Some of the exponents are trivially expected (e.g. $Z_{a}\propto I_{P6}/I_{p}$ or $\psi_{a}\propto I_{p}$), while others reflect the chosen tokamak configuration (e.g. the impact of the poloidal-field coils P4, P5 and divertor coil D1). The scaling of $q_a$ is roughly consistent with what would be expected from the small-radius approximation $q_{a} \approx f_{vac}(\partial_{Z}\psi_{n})^{-1}/( 2\pi R_{a}|\psi_{a}-\psi_{b}|)$ $\propto f_{vac}Z_{x}^{2}/( R_{a}|\psi_{a}-\psi_{b}|),$ although with slightly different exponents, including the small diamagnetic effects in the plasma ($\partial \log q_{a}/\partial\log(f_{Vac}/I_{p})=0.82<1$). 

Other targets have very uncertain fits, in particular $R_{in}$ and $R_{X}.$ The behaviours with $I_{sol}/I_{p}$ are the most uncertain and they change at high solenoid currents, reflecting the role of different coil combinations in keeping the separatrix within the vessel for different strengths of the solenoid stray field. These scaling relations can be used to aid the real-time reconstruction of $p_a$ (or equivalently $\beta_p$ and $q_a$) with a good first guess from the magnetic reconstruction of $R_{out}$ and $R_{a}$. However, they are not necessarily accurate enough to be used directly in control design, especially because the dependence on solenoid current and in particular because the inboard midplane radius $R_{in}$ is not well fit.

\subsection{Neural Networks}

\begin{figure*}
\includegraphics[width=\textwidth]{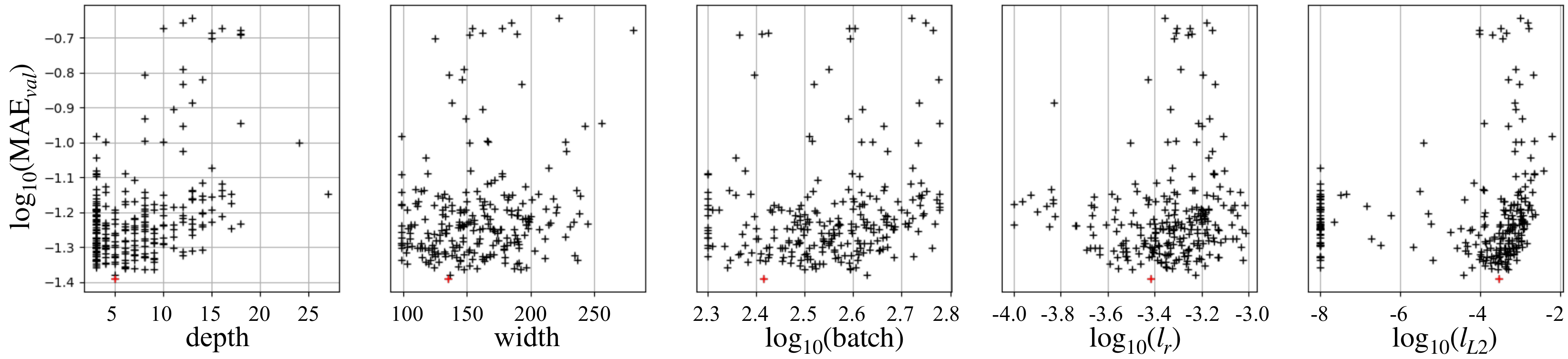}\\
\caption{\label{fig:ANNhyperpar} Hyperparameter exploration. The models tend to privilege shallow networks with $\mathcal{O}(100)$ nodes per hidden layer. }
\end{figure*}

\begin{figure*}
\includegraphics[width=\textwidth]{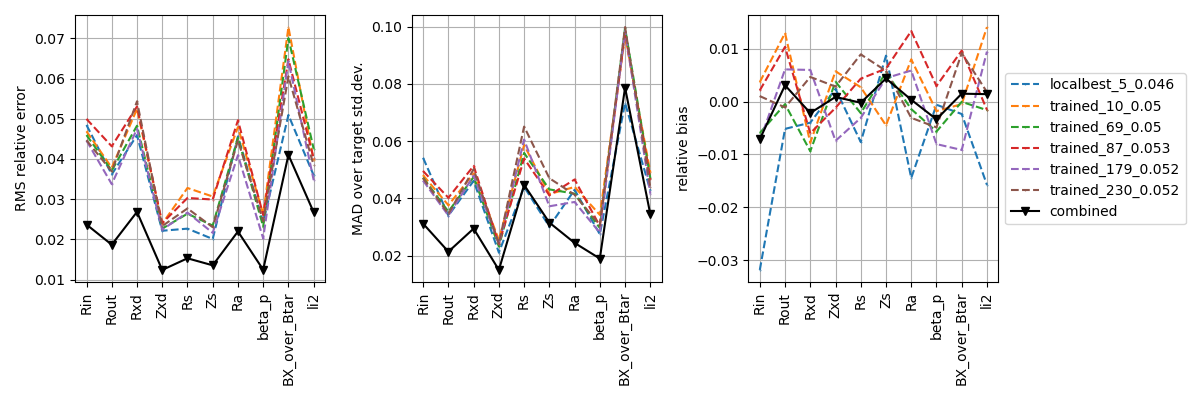}\\
\includegraphics[width=\textwidth]{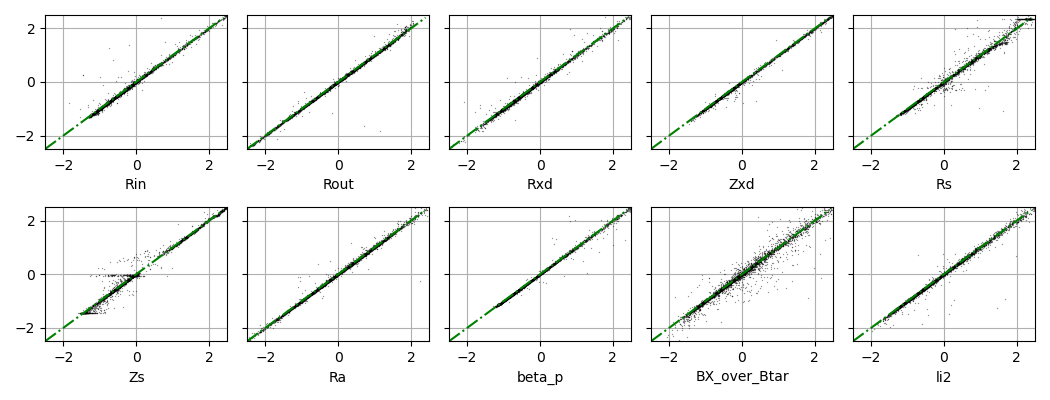}\\
\caption{\label{fig:ANNgeom} Results of training and hyperparameter exploration when fitting on geometric targets and integrated quantities. \textit{Top:} Test metrics on a sample of best-fitting models and on their aggregate prediction. \textit{Bottom:} standardised outputs \textit{vs} those predicted by the NN emulation, averaged over the best models. The label \texttt{BX\_over\_Btar} denotes the magnetic flux expansion between the lower X-point and the strike point. \textcolor{black}{The selected models in the legend are labeled by their index in the hyperparameter exploration followed by the total $loss_{val}.$ Many of the sampled models have a small $loss_{val}$ (Fig.~4), but a very strict cut on the six very-best models (indicated in the top-right legend) is already enough to yield sub-percent bias while keeping the overall ensemble model small enough for real-time applications.} }
\end{figure*}

Figure~\ref{fig:ANNhyperpar} illustrates the hyperparameter exploration, quantified by the validation loss. Shallow but wide network architectures are preferred. A {leaky ReLU} provides the best results, as it does not not have vanishing gradients. In principle, a standard ReLU activation would perform equally well, but that requires wider hidden layers, in order to reproduce enough ReLU combinations to resemble a leaky ReLU and prune nodes with vanishing gradients. 

Figure~\ref{fig:ANNgeom} shows test loss metrics and calibration plots for each target, for a subset of the best-performing models. 
The calibration plots are displayed in terms of standardised targets.
The metrics are computed on a separate subset, which was not used during training and early-stopping. The test metrics are defined as
\begin{itemize}
    \item rms relative error:\\ $\sqrt{\mathrm{var}(\theta-\theta_{fit})/\mathrm{var}(\theta)} \equiv \sqrt{(1-R^{2})}$
    \item mean-absolute-deviation over std.dev:\\ $\langle |\theta_{fit}-\theta|\rangle/\sqrt{\mathrm{var}(\theta)}$
    \item relative bias: $\langle \theta-\theta_{fit} \rangle/\sqrt{\mathrm{var}(\theta)}$
\end{itemize}
 These values are preferred over the $R^2$ metric (which has been used in other emulation studies \cite{Wai2022}), as it would be $>0.995$ in all cases. The combined prediction is obtained for each target as a simple average of its predicted values from the best-fitting neural networks.
 
The general percent-level performance on the geometric targets and integrated quantities is expected, given the dataset. Given the intrinsic standard deviations of the shape coordinates over the synthetic library (midplane radii and lower X-point position), these metrics translate into $\approx2-3$mm accuracy, i.e. comparable to the accuracy of the direct calculation on the numeric Grad-Shafranov solutions with a 2-4cm resolution grid. The flux expansion (\texttt{BX\_over\_Btar}) is generally the most uncertain as it is a ratio of magnetic fields (computed as derivatives of the flux-function), and because the strike-point location can depend appreciably on the upstream location and on whether a secondary X-point is created in the divertor region. Whenever this happens, a small difference in upstream location or in divertor coil currents can result in a large difference in the strike-point Z-location, hitting the divertor baffle instead of the bottom tiles. In general, the radial strike-point location is fit better than the Z-location.

\section{\label{sect:discussion} Discussion}

We have presented a stable procedure to sample synthetic equilibria from a high-dimensional parameter space with operational constraints, where the "score" function $\mathcal{L}$ can also be chosen to privilege configurations with more desirable properties. This sequential MCMC approach is also convenient because it curbs some of the numerical instability inherent in solving the non-linear GS equation. Passive-structure currents may be considered as additional parameters, if they can be reduced to a small enough number of modes. While we have shown a case study with a machine resembling MAST-U in this work, this framework is agnostic to the chosen machine configuration. In principle, it is also agnostic to the equilibrium solver, provided it can perform a numerically stable (and accurate) forward GS solution. The advantage of \texttt{FreeGS} over other codes is that, being fully in \texttt{python}, it can be seamlessly integrated with the MCMC sampling. The same MCMC sampler can also be used for design optimisation purposes, e.g. tokamak configurations that minimise inter-coil forces while maintaining the plasma shape within desirable bounds. It can also be used for the Bayesian fitting of models to data (to enable uncertainty quantification) whenever the models involve nonlinear differential equations with numerical instabilities.

\subsection{Profile libraries and divertor detachment studies}
Flux-function profiles are also available, for divertor studies, as well as the flux expansion along the connection length from the midplane. This synthetic library can aid in refining some working hypotheses, such as a flux-expansion that is constant or proportional to radial position, that are commonly adopted in divertor detachment simulations \cite{Moulton2017}. The choice of an appropriate score function $\mathcal{L}$ is particularly relevant if super-X configurations are to be adequately sampled. For example, while we have considered the flux expansion from X-point to strike point, the exploration of scenarios for divertor detachment may consider the flux-expansion at some prescribed distance from the divertor tiles along the connection.

\subsection{Emulating the physics}
Scaling relations can be used to understand the general behaviour of some targets, and may also be used to aid the real-time inference of pressure on axis from a magnetic reconstruction of the separatrix. However, they are not as useful for targets that are more sensitive to the wall, such as strike point position and inboard midplane radius, and in general they are too uncertain to be used directly in control design. 
\textcolor{black}{Power-law relations on other quantities have been determined experimentally, on historic ASDEX-U discharges, with comparable $R^2$ values by \citet{McCarthy2018}, using least-squares linear regression in log-log space, with a focus on inferring and controlling transport through $I_{p},$ $q_a$ and $B_{\phi}.$ In the presence of noisy regressors, \citet{Verdoolaege2021} argue for more robust Bayesian fitting techniques, which can affect the quality of control from power-law scalings on experimental data -- rather than purely synthetic data as in our case.}

Neural-network emulation works generally to within percent-level accuracy, but some caution is necessary for targets that can exhibit abrupt changes (strike point Z-location) or that are particularly sensitive to high solenoid currents ($R_{in}$). The annealed hyperparameter sampler developed for this work provides an ensemble of emulators, which can also be used to quantify uncertainties in the emulation.

\textcolor{black}{During experiments, a noisy hint of the strike point position is given by real-time magnetic reconstruction. The emulators presented here may be re-trained on subsets corresponding to different wall segments, or adding a "tile segment" feature in the training set. Splitting the wall into segments would not affect the calculation of virtual circuits, since their derivative with respect to other inputs is going to be almost-always zero.}

\subsection{Choice of internal profiles}
Ultimately, the quality of the emulation can only be as good as the quality of the synthetic inputs, at best. For this reason, in this work we chose to decouple the accuracy of the physics from the aspects more strictly related to the dataset construction and emulation.  
The parameterisation of current density yields internal inductance values $l_{i}(2)$ in the range $0.55\pm0.15,$ more typical of L-mode plasma in purely Ohmic discharges. The sheared profiles of non-inductive scenarios may be better described by \citep{Freidberg_2008}
\begin{equation}
p^{\prime},\ f^{\prime}f\ \propto \left( 1-\psi_{n}^{2} \right)(1+\zeta\psi_{n}^{2})
\end{equation}
with $\zeta\approx 3.$
This functional form only adds one parameter, so it is not inherently difficult to implement within the same MCMC sampling described in this work.

\subsection{From emulation to control}

Networks with ReLU activation functions perform best at approximating the control targets. However, a perfect ReLU NN approximant would have discontinuous derivatives on a dense subset of parameter space, which can be a problem e.g. for virtual circuit emulation. The SReLU activation introduced in this paper alleviates this issue, but in principle there is no upper bound to the norm of the gradients from this choice either. Two solutions can be: (\textsc{i}) estimating as averages of finite-difference derivatives over an ensemble of approximators; (\textsc{ii}) introducing penalties that privilege smoother solutions, either through additional terms in the loss or through a variational intermediate layer \cite{Graves2011, Kameoka2019}. While the latter is left for further development, the former is already possible with the NNs ensembles from this work.

In the interest of keeping the analysis of physics emulation from the issues of real-time control algorithms, we chose not to present emulations of the virtual circuits themselves, as they involve derivatives of quantities that include effective inductances, for which different control-design choices have been made in the literature \cite{Lister_2002, deTommasi2007, Freidberg_2008, Ariola2008, Felici_2010,Carpanese_2020}, as summarised in the Appendix.

{We remark that this issue is already present in state-of-the-art calculations of virtual circuits, which are done via finite-element differences that can be affected by numerical noise and departures from linearity. Fully-differentiable GS solvers would deliver exact derivatives at each equilibrium configuration, which can then be emulated directly, however they would already need to be emulated for fast deployment in real-time control, also with safety limits from estimated model uncertainties. The very same techniques presented in this paper can be used directly for these two purposes.} 


\begin{acknowledgments}
This work was carried out within a Joint Endeavour between the Hartree Centre (Science and Technology Facilities Council, UKRI) and the UK Atomic Energy Authority.
\end{acknowledgments}

\section*{Author Statement}
The Authors have no conflcit of interests to disclose.\\
We adopt the CRediT author statement outlined at:\\
\url{https://www.elsevier.com/authors/policies-and-guidelines/credit-author-statement}


\textbf{\textit{Conceptualisation.}} the problem was formulated and formalised by A.Agnello, N.C.Amorisco, G.McArdle and C.Vincent. G.Holt contributed to the selection of targets relevant to divertor studies. 

\textbf{\textit{Methodology, Software.}} The MCMC routines and emulators have been developed by A.Agnello with contributions by A.Keats. The FreeGSNKE Newton-Krylov solver over the FreeGS infrastructure has been developed by N.C.Amorsico, with fast core-routines for FreeGS developed by A.Agnello. 






\textbf{\textit{Data Curation.}} The data-sets and model-import scripts have been assembled by A.Agnello ad A.Keats. Examples of equilibrium configurations and oeprational limits relevant to MAST-U have been provided by S.Pamela, J.Buchanan, C.Vincent.

\textbf{\textit{Writing and Reviewing.}} The manuscript has been written by A.Agnello, with reviews by N.C.Amorisco, J.Buchanan and S.Pamela.






\section*{Data Availability Statement}
The synthetic libraries, code to generate them, and the code for the neural-network training and hyperparameter exploration is available upon paper publication at\\
\url{https://github.com/farscape-project/freegs-emu}

\bibliography{Agnello_ICDDPS4_v2_plain}

\begin{thebibliography}{57}%
\makeatletter
\providecommand \@ifxundefined [1]{%
 \@ifx{#1\undefined}
}%
\providecommand \@ifnum [1]{%
 \ifnum #1\expandafter \@firstoftwo
 \else \expandafter \@secondoftwo
 \fi
}%
\providecommand \@ifx [1]{%
 \ifx #1\expandafter \@firstoftwo
 \else \expandafter \@secondoftwo
 \fi
}%
\providecommand \natexlab [1]{#1}%
\providecommand \enquote  [1]{``#1''}%
\providecommand \bibnamefont  [1]{#1}%
\providecommand \bibfnamefont [1]{#1}%
\providecommand \citenamefont [1]{#1}%
\providecommand \href@noop [0]{\@secondoftwo}%
\providecommand \href [0]{\begingroup \@sanitize@url \@href}%
\providecommand \@href[1]{\@@startlink{#1}\@@href}%
\providecommand \@@href[1]{\endgroup#1\@@endlink}%
\providecommand \@sanitize@url [0]{\catcode `\\12\catcode `\$12\catcode
  `\&12\catcode `\#12\catcode `\^12\catcode `\_12\catcode `\%12\relax}%
\providecommand \@@startlink[1]{}%
\providecommand \@@endlink[0]{}%
\providecommand \url  [0]{\begingroup\@sanitize@url \@url }%
\providecommand \@url [1]{\endgroup\@href {#1}{\urlprefix }}%
\providecommand \urlprefix  [0]{URL }%
\providecommand \Eprint [0]{\href }%
\providecommand \doibase [0]{http://dx.doi.org/}%
\providecommand \selectlanguage [0]{\@gobble}%
\providecommand \bibinfo  [0]{\@secondoftwo}%
\providecommand \bibfield  [0]{\@secondoftwo}%
\providecommand \translation [1]{[#1]}%
\providecommand \BibitemOpen [0]{}%
\providecommand \bibitemStop [0]{}%
\providecommand \bibitemNoStop [0]{.\EOS\space}%
\providecommand \EOS [0]{\spacefactor3000\relax}%
\providecommand \BibitemShut  [1]{\csname bibitem#1\endcsname}%
\let\auto@bib@innerbib\@empty
\bibitem [{\citenamefont {{Freidberg}}(2008)}]{Freidberg_2008}%
  \BibitemOpen
  \bibfield  {author} {\bibinfo {author} {\bibfnamefont {J.~P.}\ \bibnamefont
  {{Freidberg}}},\ }\href@noop {} {\emph {\bibinfo {title} {{Plasma Physics and
  Fusion Energy}}}}\ (\bibinfo {year} {2008})\BibitemShut {NoStop}%
\bibitem [{\citenamefont {{Schindler}}(2010)}]{Schindler2010}%
  \BibitemOpen
  \bibfield  {author} {\bibinfo {author} {\bibfnamefont {K.}~\bibnamefont
  {{Schindler}}},\ }\href@noop {} {\emph {\bibinfo {title} {{Physics of Space
  Plasma Activity}}}}\ (\bibinfo {year} {2010})\BibitemShut {NoStop}%
\bibitem [{\citenamefont {Wilson}\ \emph {et~al.}(2020)\citenamefont {Wilson},
  \citenamefont {Chapman}, \citenamefont {Denton}, \citenamefont {Morris},
  \citenamefont {Patel}, \citenamefont {Voss}, \citenamefont {Waldon},\ and\
  \citenamefont {the STEP~Team}}]{10.1088/978-0-7503-2719-0ch8}%
  \BibitemOpen
  \bibfield  {author} {\bibinfo {author} {\bibfnamefont {H.}~\bibnamefont
  {Wilson}}, \bibinfo {author} {\bibfnamefont {I.}~\bibnamefont {Chapman}},
  \bibinfo {author} {\bibfnamefont {T.}~\bibnamefont {Denton}}, \bibinfo
  {author} {\bibfnamefont {W.}~\bibnamefont {Morris}}, \bibinfo {author}
  {\bibfnamefont {B.}~\bibnamefont {Patel}}, \bibinfo {author} {\bibfnamefont
  {G.}~\bibnamefont {Voss}}, \bibinfo {author} {\bibfnamefont {C.}~\bibnamefont
  {Waldon}}, \ and\ \bibinfo {author} {\bibnamefont {the STEP~Team}},\
  }\bibfield  {title} {\enquote {\bibinfo {title} {Step—on the pathway to
  fusion commercialization},}\ }in\ \href {\doibase
  10.1088/978-0-7503-2719-0ch8} {\emph {\bibinfo {booktitle} {Commercialising
  Fusion Energy}}},\ \bibinfo {series and number} {2053-2563}\ (\bibinfo
  {publisher} {IOP Publishing},\ \bibinfo {year} {2020})\ pp.\ \bibinfo {pages}
  {8--1 to 8--18}\BibitemShut {NoStop}%
\bibitem [{\citenamefont {{Creely}}\ \emph {et~al.}(2020)\citenamefont
  {{Creely}}, \citenamefont {{Greenwald}}, \citenamefont {{Ballinger}},
  \citenamefont {{Brunner}}, \citenamefont {{Canik}}, \citenamefont {{Doody}},
  \citenamefont {{F{\"u}l{\"o}p}}, \citenamefont {{Garnier}}, \citenamefont
  {{Granetz}}, \citenamefont {{Gray}}, \citenamefont {{Holland}}, \citenamefont
  {{Howard}}, \citenamefont {{Hughes}}, \citenamefont {{Irby}}, \citenamefont
  {{Izzo}}, \citenamefont {{Kramer}}, \citenamefont {{Kuang}}, \citenamefont
  {{Labombard}}, \citenamefont {{Lin}}, \citenamefont {{Lipschultz}},
  \citenamefont {{Logan}}, \citenamefont {{Lore}}, \citenamefont {{Marmar}},
  \citenamefont {{Montes}}, \citenamefont {{Mumgaard}}, \citenamefont
  {{Paz-Soldan}}, \citenamefont {{Rea}}, \citenamefont {{Reinke}},
  \citenamefont {{Rodriguez-Fernandez}}, \citenamefont {{S{\"a}rkim{\"a}ki}},
  \citenamefont {{Sciortino}}, \citenamefont {{Scott}}, \citenamefont
  {{Snicker}}, \citenamefont {{Snyder}}, \citenamefont {{Sorbom}},
  \citenamefont {{Sweeney}}, \citenamefont {{Tinguely}}, \citenamefont
  {{Tolman}}, \citenamefont {{Umansky}}, \citenamefont {{Vallhagen}},
  \citenamefont {{Varje}}, \citenamefont {{Whyte}}, \citenamefont {{Wright}},
  \citenamefont {{Wukitch}}, \citenamefont {{Zhu}},\ and\ \citenamefont {{Sparc
  Team}}}]{2020JPlPh..86e8602C}%
  \BibitemOpen
  \bibfield  {author} {\bibinfo {author} {\bibfnamefont {A.~J.}\ \bibnamefont
  {{Creely}}}, \bibinfo {author} {\bibfnamefont {M.~J.}\ \bibnamefont
  {{Greenwald}}}, \bibinfo {author} {\bibfnamefont {S.~B.}\ \bibnamefont
  {{Ballinger}}}, \bibinfo {author} {\bibfnamefont {D.}~\bibnamefont
  {{Brunner}}}, \bibinfo {author} {\bibfnamefont {J.}~\bibnamefont {{Canik}}},
  \bibinfo {author} {\bibfnamefont {J.}~\bibnamefont {{Doody}}}, \bibinfo
  {author} {\bibfnamefont {T.}~\bibnamefont {{F{\"u}l{\"o}p}}}, \bibinfo
  {author} {\bibfnamefont {D.~T.}\ \bibnamefont {{Garnier}}}, \bibinfo {author}
  {\bibfnamefont {R.}~\bibnamefont {{Granetz}}}, \bibinfo {author}
  {\bibfnamefont {T.~K.}\ \bibnamefont {{Gray}}}, \bibinfo {author}
  {\bibfnamefont {C.}~\bibnamefont {{Holland}}}, \bibinfo {author}
  {\bibfnamefont {N.~T.}\ \bibnamefont {{Howard}}}, \bibinfo {author}
  {\bibfnamefont {J.~W.}\ \bibnamefont {{Hughes}}}, \bibinfo {author}
  {\bibfnamefont {J.~H.}\ \bibnamefont {{Irby}}}, \bibinfo {author}
  {\bibfnamefont {V.~A.}\ \bibnamefont {{Izzo}}}, \bibinfo {author}
  {\bibfnamefont {G.~J.}\ \bibnamefont {{Kramer}}}, \bibinfo {author}
  {\bibfnamefont {A.~Q.}\ \bibnamefont {{Kuang}}}, \bibinfo {author}
  {\bibfnamefont {B.}~\bibnamefont {{Labombard}}}, \bibinfo {author}
  {\bibfnamefont {Y.}~\bibnamefont {{Lin}}}, \bibinfo {author} {\bibfnamefont
  {B.}~\bibnamefont {{Lipschultz}}}, \bibinfo {author} {\bibfnamefont {N.~C.}\
  \bibnamefont {{Logan}}}, \bibinfo {author} {\bibfnamefont {J.~D.}\
  \bibnamefont {{Lore}}}, \bibinfo {author} {\bibfnamefont {E.~S.}\
  \bibnamefont {{Marmar}}}, \bibinfo {author} {\bibfnamefont {K.}~\bibnamefont
  {{Montes}}}, \bibinfo {author} {\bibfnamefont {R.~T.}\ \bibnamefont
  {{Mumgaard}}}, \bibinfo {author} {\bibfnamefont {C.}~\bibnamefont
  {{Paz-Soldan}}}, \bibinfo {author} {\bibfnamefont {C.}~\bibnamefont {{Rea}}},
  \bibinfo {author} {\bibfnamefont {M.~L.}\ \bibnamefont {{Reinke}}}, \bibinfo
  {author} {\bibfnamefont {P.}~\bibnamefont {{Rodriguez-Fernandez}}}, \bibinfo
  {author} {\bibfnamefont {K.}~\bibnamefont {{S{\"a}rkim{\"a}ki}}}, \bibinfo
  {author} {\bibfnamefont {F.}~\bibnamefont {{Sciortino}}}, \bibinfo {author}
  {\bibfnamefont {S.~D.}\ \bibnamefont {{Scott}}}, \bibinfo {author}
  {\bibfnamefont {A.}~\bibnamefont {{Snicker}}}, \bibinfo {author}
  {\bibfnamefont {P.~B.}\ \bibnamefont {{Snyder}}}, \bibinfo {author}
  {\bibfnamefont {B.~N.}\ \bibnamefont {{Sorbom}}}, \bibinfo {author}
  {\bibfnamefont {R.}~\bibnamefont {{Sweeney}}}, \bibinfo {author}
  {\bibfnamefont {R.~A.}\ \bibnamefont {{Tinguely}}}, \bibinfo {author}
  {\bibfnamefont {E.~A.}\ \bibnamefont {{Tolman}}}, \bibinfo {author}
  {\bibfnamefont {M.}~\bibnamefont {{Umansky}}}, \bibinfo {author}
  {\bibfnamefont {O.}~\bibnamefont {{Vallhagen}}}, \bibinfo {author}
  {\bibfnamefont {J.}~\bibnamefont {{Varje}}}, \bibinfo {author} {\bibfnamefont
  {D.~G.}\ \bibnamefont {{Whyte}}}, \bibinfo {author} {\bibfnamefont {J.~C.}\
  \bibnamefont {{Wright}}}, \bibinfo {author} {\bibfnamefont {S.~J.}\
  \bibnamefont {{Wukitch}}}, \bibinfo {author} {\bibfnamefont {J.}~\bibnamefont
  {{Zhu}}}, \ and\ \bibinfo {author} {\bibnamefont {{Sparc Team}}},\ }\bibfield
   {title} {\enquote {\bibinfo {title} {{Overview of the SPARC tokamak}},}\
  }\href {\doibase 10.1017/S0022377820001257} {\bibfield  {journal} {\bibinfo
  {journal} {Journal of Plasma Physics}\ }\textbf {\bibinfo {volume} {86}},\
  \bibinfo {eid} {865860502} (\bibinfo {year} {2020})}\BibitemShut {NoStop}%
\bibitem [{\citenamefont {{Rodriguez-Fernandez}}\ \emph
  {et~al.}(2022)\citenamefont {{Rodriguez-Fernandez}}, \citenamefont
  {{Creely}}, \citenamefont {{Greenwald}}, \citenamefont {{Brunner}},
  \citenamefont {{Ballinger}}, \citenamefont {{Chrobak}}, \citenamefont
  {{Garnier}}, \citenamefont {{Granetz}}, \citenamefont {{Hartwig}},
  \citenamefont {{Howard}}, \citenamefont {{Hughes}}, \citenamefont {{Irby}},
  \citenamefont {{Izzo}}, \citenamefont {{Kuang}}, \citenamefont {{Lin}},
  \citenamefont {{Marmar}}, \citenamefont {{Mumgaard}}, \citenamefont {{Rea}},
  \citenamefont {{Reinke}}, \citenamefont {{Riccardo}}, \citenamefont {{Rice}},
  \citenamefont {{Scott}}, \citenamefont {{Sorbom}}, \citenamefont
  {{Stillerman}}, \citenamefont {{Sweeney}}, \citenamefont {{Tinguely}},
  \citenamefont {{Whyte}}, \citenamefont {{Wright}},\ and\ \citenamefont
  {{Yuryev}}}]{2022NucFu..62d2003R}%
  \BibitemOpen
  \bibfield  {author} {\bibinfo {author} {\bibfnamefont {P.}~\bibnamefont
  {{Rodriguez-Fernandez}}}, \bibinfo {author} {\bibfnamefont {A.~J.}\
  \bibnamefont {{Creely}}}, \bibinfo {author} {\bibfnamefont {M.~J.}\
  \bibnamefont {{Greenwald}}}, \bibinfo {author} {\bibfnamefont
  {D.}~\bibnamefont {{Brunner}}}, \bibinfo {author} {\bibfnamefont {S.~B.}\
  \bibnamefont {{Ballinger}}}, \bibinfo {author} {\bibfnamefont {C.~P.}\
  \bibnamefont {{Chrobak}}}, \bibinfo {author} {\bibfnamefont {D.~T.}\
  \bibnamefont {{Garnier}}}, \bibinfo {author} {\bibfnamefont {R.}~\bibnamefont
  {{Granetz}}}, \bibinfo {author} {\bibfnamefont {Z.~S.}\ \bibnamefont
  {{Hartwig}}}, \bibinfo {author} {\bibfnamefont {N.~T.}\ \bibnamefont
  {{Howard}}}, \bibinfo {author} {\bibfnamefont {J.~W.}\ \bibnamefont
  {{Hughes}}}, \bibinfo {author} {\bibfnamefont {J.~H.}\ \bibnamefont
  {{Irby}}}, \bibinfo {author} {\bibfnamefont {V.~A.}\ \bibnamefont {{Izzo}}},
  \bibinfo {author} {\bibfnamefont {A.~Q.}\ \bibnamefont {{Kuang}}}, \bibinfo
  {author} {\bibfnamefont {Y.}~\bibnamefont {{Lin}}}, \bibinfo {author}
  {\bibfnamefont {E.~S.}\ \bibnamefont {{Marmar}}}, \bibinfo {author}
  {\bibfnamefont {R.~T.}\ \bibnamefont {{Mumgaard}}}, \bibinfo {author}
  {\bibfnamefont {C.}~\bibnamefont {{Rea}}}, \bibinfo {author} {\bibfnamefont
  {M.~L.}\ \bibnamefont {{Reinke}}}, \bibinfo {author} {\bibfnamefont
  {V.}~\bibnamefont {{Riccardo}}}, \bibinfo {author} {\bibfnamefont {J.~E.}\
  \bibnamefont {{Rice}}}, \bibinfo {author} {\bibfnamefont {S.~D.}\
  \bibnamefont {{Scott}}}, \bibinfo {author} {\bibfnamefont {B.~N.}\
  \bibnamefont {{Sorbom}}}, \bibinfo {author} {\bibfnamefont {J.~A.}\
  \bibnamefont {{Stillerman}}}, \bibinfo {author} {\bibfnamefont
  {R.}~\bibnamefont {{Sweeney}}}, \bibinfo {author} {\bibfnamefont {R.~A.}\
  \bibnamefont {{Tinguely}}}, \bibinfo {author} {\bibfnamefont {D.~G.}\
  \bibnamefont {{Whyte}}}, \bibinfo {author} {\bibfnamefont {J.~C.}\
  \bibnamefont {{Wright}}}, \ and\ \bibinfo {author} {\bibfnamefont {D.~V.}\
  \bibnamefont {{Yuryev}}},\ }\bibfield  {title} {\enquote {\bibinfo {title}
  {{Overview of the SPARC physics basis towards the exploration of
  burning-plasma regimes in high-field, compact tokamaks}},}\ }\href {\doibase
  10.1088/1741-4326/ac1654} {\bibfield  {journal} {\bibinfo  {journal} {Nuclear
  Fusion}\ }\textbf {\bibinfo {volume} {62}},\ \bibinfo {eid} {042003}
  (\bibinfo {year} {2022})}\BibitemShut {NoStop}%
\bibitem [{\citenamefont {{\textsc{JET} Team}}(1992)}]{JET_1992}%
  \BibitemOpen
  \bibfield  {author} {\bibinfo {author} {\bibnamefont {{\textsc{JET} Team}}},\
  }\bibfield  {title} {\enquote {\bibinfo {title} {{Fusion energy production
  from a deuterium-tritium plasma in the JET tokamak}},}\ }\href {\doibase
  10.1088/0029-5515/32/2/I01} {\bibfield  {journal} {\bibinfo  {journal}
  {Nuclear Fusion}\ }\textbf {\bibinfo {volume} {32}},\ \bibinfo {pages}
  {187--203} (\bibinfo {year} {1992})}\BibitemShut {NoStop}%
\bibitem [{\citenamefont {{Gruber}}\ \emph {et~al.}(1997)\citenamefont
  {{Gruber}}, \citenamefont {{Mertens}}, \citenamefont {{Neuhauser}},
  \citenamefont {{Ryter}}, \citenamefont {{Suttrop}}, \citenamefont
  {{Albrecht}}, \citenamefont {{Alexander}}, \citenamefont {{Asmussen}},
  \citenamefont {{Becker}}, \citenamefont {{Behler}}, \citenamefont
  {{Behringer}}, \citenamefont {{Bergmann}}, \citenamefont
  {{Bessenrodt-Weberpals}}, \citenamefont {{Borras}}, \citenamefont {{Bosch}},
  \citenamefont {{Braams}}, \citenamefont {{Brambilla}}, \citenamefont
  {{Braun}}, \citenamefont {{Brinkschulte}}, \citenamefont {{B{\"u}chl}},
  \citenamefont {{Buhler}}, \citenamefont {{Carlson}}, \citenamefont
  {{Chodura}}, \citenamefont {{Coster}}, \citenamefont {{Cupido}},
  \citenamefont {{De Blank}}, \citenamefont {{De Pena Hempel}}, \citenamefont
  {{Deschka}}, \citenamefont {{Dorn}}, \citenamefont {{Drube}}, \citenamefont
  {{Dux}}, \citenamefont {{Engelhardt}}, \citenamefont {{Engstler}},
  \citenamefont {{Fahrbach}}, \citenamefont {{Feist}}, \citenamefont
  {{Fiedler}}, \citenamefont {{Franzen}}, \citenamefont {{Fuchs}},
  \citenamefont {{Fussmann}}, \citenamefont {{Gafert}}, \citenamefont
  {{Gehre}}, \citenamefont {{Gernhardt}}, \citenamefont {{G{\"u}nter}},
  \citenamefont {{Haas}}, \citenamefont {{Hallatschek}}, \citenamefont
  {{Hartmann}}, \citenamefont {{Heinemann}}, \citenamefont {{Herppich}},
  \citenamefont {{Herrmann}}, \citenamefont {{Herrmann}}, \citenamefont
  {{Hirsch}}, \citenamefont {{Hoek}}, \citenamefont {{Hoenen}}, \citenamefont
  {{Hofmeister}}, \citenamefont {{Hohenoecker}}, \citenamefont {{Holzhauer}},
  \citenamefont {{Ignacz}}, \citenamefont {{Jacobi}}, \citenamefont {{Junker}},
  \citenamefont {{Kakoulidis}}, \citenamefont {{Kallenbach}}, \citenamefont
  {{Karakatsanis}}, \citenamefont {{Kardaun}}, \citenamefont {{Kass}},
  \citenamefont {{Kaufmann}}, \citenamefont {{Khutoretski}}, \citenamefont
  {{Kollotzek}}, \citenamefont {{Koeppendoerfer}}, \citenamefont {{Kraus}},
  \citenamefont {{Krieger}}, \citenamefont {{Kurzan}}, \citenamefont
  {{Kyriakakis}}, \citenamefont {{Lackner}}, \citenamefont {{Lang}},
  \citenamefont {{Lang}}, \citenamefont {{Laux}}, \citenamefont {{Lengyel}},
  \citenamefont {{Leuterer}}, \citenamefont {{Maraschek}}, \citenamefont
  {{Markoulaki}}, \citenamefont {{Mast}}, \citenamefont {{McCarthy}},
  \citenamefont {{Meisel}}, \citenamefont {{Meister}}, \citenamefont
  {{Merkel}}, \citenamefont {{Mueller}}, \citenamefont {{Muenich}},
  \citenamefont {{Murmann}}, \citenamefont {{Napiontek}}, \citenamefont
  {{Neu}}, \citenamefont {{Neu}}, \citenamefont {{Niethammer}}, \citenamefont
  {{Noterdaeme}}, \citenamefont {{Pasch}}, \citenamefont {{Pautasso}},
  \citenamefont {{Peeters}}, \citenamefont {{Pereverzev}}, \citenamefont
  {{Pitcher}}, \citenamefont {{Poschenrieder}}, \citenamefont {{Raupp}},
  \citenamefont {{Reinmueller}}, \citenamefont {{Riedl}}, \citenamefont
  {{Rohde}}, \citenamefont {{Roehr}}, \citenamefont {{Roth}}, \citenamefont
  {{Salzmann}}, \citenamefont {{Sandmann}}, \citenamefont {{Schilling}},
  \citenamefont {{Schittenhelm}}, \citenamefont {{Schloegl}}, \citenamefont
  {{Schneider}}, \citenamefont {{Schneider}}, \citenamefont {{Schneider}},
  \citenamefont {{Schramm}}, \citenamefont {{Schweinzer}}, \citenamefont
  {{Schweizer}}, \citenamefont {{Schwoerer}}, \citenamefont {{Scott}},
  \citenamefont {{Seidel}}, \citenamefont {{Serra}}, \citenamefont {{Sesnic}},
  \citenamefont {{Silva}}, \citenamefont {{Sokoll}}, \citenamefont {{Speth}},
  \citenamefont {{Staebler}}, \citenamefont {{Steuer}}, \citenamefont
  {{Stober}}, \citenamefont {{Streibl}}, \citenamefont {{Thoma}}, \citenamefont
  {{Treutterer}}, \citenamefont {{Troppmann}}, \citenamefont {{Tsois}},
  \citenamefont {{Ulrich}}, \citenamefont {{Varela}}, \citenamefont
  {{Verbeek}}, \citenamefont {{Vollmer}}, \citenamefont {{Wedler}},
  \citenamefont {{Weinlich}}, \citenamefont {{Wenzel}}, \citenamefont
  {{Wesner}}, \citenamefont {{Wolf}}, \citenamefont {{Wunderlich}},
  \citenamefont {{Xantopoulus}}, \citenamefont {{Yu}}, \citenamefont
  {{Zasche}}, \citenamefont {{Zehetbauer}}, \citenamefont {{Zehrfeld}},
  \citenamefont {{Zohm}},\ and\ \citenamefont {{Zouhar}}}]{Gruber1997}%
  \BibitemOpen
  \bibfield  {author} {\bibinfo {author} {\bibfnamefont {O.}~\bibnamefont
  {{Gruber}}}, \bibinfo {author} {\bibfnamefont {V.}~\bibnamefont {{Mertens}}},
  \bibinfo {author} {\bibfnamefont {J.}~\bibnamefont {{Neuhauser}}}, \bibinfo
  {author} {\bibfnamefont {F.}~\bibnamefont {{Ryter}}}, \bibinfo {author}
  {\bibfnamefont {W.}~\bibnamefont {{Suttrop}}}, \bibinfo {author}
  {\bibfnamefont {M.}~\bibnamefont {{Albrecht}}}, \bibinfo {author}
  {\bibfnamefont {M.}~\bibnamefont {{Alexander}}}, \bibinfo {author}
  {\bibfnamefont {K.}~\bibnamefont {{Asmussen}}}, \bibinfo {author}
  {\bibfnamefont {G.}~\bibnamefont {{Becker}}}, \bibinfo {author}
  {\bibfnamefont {K.}~\bibnamefont {{Behler}}}, \bibinfo {author}
  {\bibfnamefont {K.}~\bibnamefont {{Behringer}}}, \bibinfo {author}
  {\bibfnamefont {A.}~\bibnamefont {{Bergmann}}}, \bibinfo {author}
  {\bibfnamefont {M.}~\bibnamefont {{Bessenrodt-Weberpals}}}, \bibinfo {author}
  {\bibfnamefont {K.}~\bibnamefont {{Borras}}}, \bibinfo {author}
  {\bibfnamefont {H.~S.}\ \bibnamefont {{Bosch}}}, \bibinfo {author}
  {\bibfnamefont {B.}~\bibnamefont {{Braams}}}, \bibinfo {author}
  {\bibfnamefont {M.}~\bibnamefont {{Brambilla}}}, \bibinfo {author}
  {\bibfnamefont {F.}~\bibnamefont {{Braun}}}, \bibinfo {author} {\bibfnamefont
  {H.}~\bibnamefont {{Brinkschulte}}}, \bibinfo {author} {\bibfnamefont
  {K.}~\bibnamefont {{B{\"u}chl}}}, \bibinfo {author} {\bibfnamefont
  {A.}~\bibnamefont {{Buhler}}}, \bibinfo {author} {\bibfnamefont
  {A.}~\bibnamefont {{Carlson}}}, \bibinfo {author} {\bibfnamefont
  {R.}~\bibnamefont {{Chodura}}}, \bibinfo {author} {\bibfnamefont
  {D.}~\bibnamefont {{Coster}}}, \bibinfo {author} {\bibfnamefont
  {L.}~\bibnamefont {{Cupido}}}, \bibinfo {author} {\bibfnamefont {H.~J.}\
  \bibnamefont {{De Blank}}}, \bibinfo {author} {\bibfnamefont
  {S.}~\bibnamefont {{De Pena Hempel}}}, \bibinfo {author} {\bibfnamefont
  {S.}~\bibnamefont {{Deschka}}}, \bibinfo {author} {\bibfnamefont
  {C.}~\bibnamefont {{Dorn}}}, \bibinfo {author} {\bibfnamefont
  {R.}~\bibnamefont {{Drube}}}, \bibinfo {author} {\bibfnamefont
  {R.}~\bibnamefont {{Dux}}}, \bibinfo {author} {\bibfnamefont
  {W.}~\bibnamefont {{Engelhardt}}}, \bibinfo {author} {\bibfnamefont
  {J.}~\bibnamefont {{Engstler}}}, \bibinfo {author} {\bibfnamefont {H.~U.}\
  \bibnamefont {{Fahrbach}}}, \bibinfo {author} {\bibfnamefont {H.~U.}\
  \bibnamefont {{Feist}}}, \bibinfo {author} {\bibfnamefont {S.}~\bibnamefont
  {{Fiedler}}}, \bibinfo {author} {\bibfnamefont {P.}~\bibnamefont
  {{Franzen}}}, \bibinfo {author} {\bibfnamefont {J.~C.}\ \bibnamefont
  {{Fuchs}}}, \bibinfo {author} {\bibfnamefont {G.}~\bibnamefont {{Fussmann}}},
  \bibinfo {author} {\bibfnamefont {J.}~\bibnamefont {{Gafert}}}, \bibinfo
  {author} {\bibfnamefont {O.}~\bibnamefont {{Gehre}}}, \bibinfo {author}
  {\bibfnamefont {J.}~\bibnamefont {{Gernhardt}}}, \bibinfo {author}
  {\bibfnamefont {S.}~\bibnamefont {{G{\"u}nter}}}, \bibinfo {author}
  {\bibfnamefont {G.}~\bibnamefont {{Haas}}}, \bibinfo {author} {\bibfnamefont
  {K.}~\bibnamefont {{Hallatschek}}}, \bibinfo {author} {\bibfnamefont
  {J.}~\bibnamefont {{Hartmann}}}, \bibinfo {author} {\bibfnamefont
  {B.}~\bibnamefont {{Heinemann}}}, \bibinfo {author} {\bibfnamefont
  {G.}~\bibnamefont {{Herppich}}}, \bibinfo {author} {\bibfnamefont
  {A.}~\bibnamefont {{Herrmann}}}, \bibinfo {author} {\bibfnamefont
  {W.}~\bibnamefont {{Herrmann}}}, \bibinfo {author} {\bibfnamefont
  {S.}~\bibnamefont {{Hirsch}}}, \bibinfo {author} {\bibfnamefont
  {M.}~\bibnamefont {{Hoek}}}, \bibinfo {author} {\bibfnamefont
  {F.}~\bibnamefont {{Hoenen}}}, \bibinfo {author} {\bibfnamefont
  {F.}~\bibnamefont {{Hofmeister}}}, \bibinfo {author} {\bibfnamefont
  {H.}~\bibnamefont {{Hohenoecker}}}, \bibinfo {author} {\bibfnamefont
  {E.}~\bibnamefont {{Holzhauer}}}, \bibinfo {author} {\bibfnamefont
  {P.}~\bibnamefont {{Ignacz}}}, \bibinfo {author} {\bibfnamefont
  {D.}~\bibnamefont {{Jacobi}}}, \bibinfo {author} {\bibfnamefont
  {W.}~\bibnamefont {{Junker}}}, \bibinfo {author} {\bibfnamefont
  {M.}~\bibnamefont {{Kakoulidis}}}, \bibinfo {author} {\bibfnamefont
  {A.}~\bibnamefont {{Kallenbach}}}, \bibinfo {author} {\bibfnamefont
  {N.}~\bibnamefont {{Karakatsanis}}}, \bibinfo {author} {\bibfnamefont
  {O.}~\bibnamefont {{Kardaun}}}, \bibinfo {author} {\bibfnamefont
  {T.}~\bibnamefont {{Kass}}}, \bibinfo {author} {\bibfnamefont
  {M.}~\bibnamefont {{Kaufmann}}}, \bibinfo {author} {\bibfnamefont
  {A.}~\bibnamefont {{Khutoretski}}}, \bibinfo {author} {\bibfnamefont
  {H.}~\bibnamefont {{Kollotzek}}}, \bibinfo {author} {\bibfnamefont
  {W.}~\bibnamefont {{Koeppendoerfer}}}, \bibinfo {author} {\bibfnamefont
  {W.}~\bibnamefont {{Kraus}}}, \bibinfo {author} {\bibfnamefont
  {K.}~\bibnamefont {{Krieger}}}, \bibinfo {author} {\bibfnamefont
  {B.}~\bibnamefont {{Kurzan}}}, \bibinfo {author} {\bibfnamefont
  {G.}~\bibnamefont {{Kyriakakis}}}, \bibinfo {author} {\bibfnamefont
  {K.}~\bibnamefont {{Lackner}}}, \bibinfo {author} {\bibfnamefont {P.~T.}\
  \bibnamefont {{Lang}}}, \bibinfo {author} {\bibfnamefont {R.~S.}\
  \bibnamefont {{Lang}}}, \bibinfo {author} {\bibfnamefont {M.}~\bibnamefont
  {{Laux}}}, \bibinfo {author} {\bibfnamefont {L.}~\bibnamefont {{Lengyel}}},
  \bibinfo {author} {\bibfnamefont {F.}~\bibnamefont {{Leuterer}}}, \bibinfo
  {author} {\bibfnamefont {M.}~\bibnamefont {{Maraschek}}}, \bibinfo {author}
  {\bibfnamefont {M.}~\bibnamefont {{Markoulaki}}}, \bibinfo {author}
  {\bibfnamefont {K.~F.}\ \bibnamefont {{Mast}}}, \bibinfo {author}
  {\bibfnamefont {P.}~\bibnamefont {{McCarthy}}}, \bibinfo {author}
  {\bibfnamefont {D.}~\bibnamefont {{Meisel}}}, \bibinfo {author}
  {\bibfnamefont {H.}~\bibnamefont {{Meister}}}, \bibinfo {author}
  {\bibfnamefont {R.}~\bibnamefont {{Merkel}}}, \bibinfo {author}
  {\bibfnamefont {H.~W.}\ \bibnamefont {{Mueller}}}, \bibinfo {author}
  {\bibfnamefont {M.}~\bibnamefont {{Muenich}}}, \bibinfo {author}
  {\bibfnamefont {H.}~\bibnamefont {{Murmann}}}, \bibinfo {author}
  {\bibfnamefont {B.}~\bibnamefont {{Napiontek}}}, \bibinfo {author}
  {\bibfnamefont {G.}~\bibnamefont {{Neu}}}, \bibinfo {author} {\bibfnamefont
  {R.}~\bibnamefont {{Neu}}}, \bibinfo {author} {\bibfnamefont
  {M.}~\bibnamefont {{Niethammer}}}, \bibinfo {author} {\bibfnamefont {J.~M.}\
  \bibnamefont {{Noterdaeme}}}, \bibinfo {author} {\bibfnamefont
  {E.}~\bibnamefont {{Pasch}}}, \bibinfo {author} {\bibfnamefont
  {G.}~\bibnamefont {{Pautasso}}}, \bibinfo {author} {\bibfnamefont {A.~G.}\
  \bibnamefont {{Peeters}}}, \bibinfo {author} {\bibfnamefont {G.}~\bibnamefont
  {{Pereverzev}}}, \bibinfo {author} {\bibfnamefont {C.~S.}\ \bibnamefont
  {{Pitcher}}}, \bibinfo {author} {\bibfnamefont {W.}~\bibnamefont
  {{Poschenrieder}}}, \bibinfo {author} {\bibfnamefont {G.}~\bibnamefont
  {{Raupp}}}, \bibinfo {author} {\bibfnamefont {K.}~\bibnamefont
  {{Reinmueller}}}, \bibinfo {author} {\bibfnamefont {R.}~\bibnamefont
  {{Riedl}}}, \bibinfo {author} {\bibfnamefont {V.}~\bibnamefont {{Rohde}}},
  \bibinfo {author} {\bibfnamefont {H.}~\bibnamefont {{Roehr}}}, \bibinfo
  {author} {\bibfnamefont {J.}~\bibnamefont {{Roth}}}, \bibinfo {author}
  {\bibfnamefont {H.}~\bibnamefont {{Salzmann}}}, \bibinfo {author}
  {\bibfnamefont {W.}~\bibnamefont {{Sandmann}}}, \bibinfo {author}
  {\bibfnamefont {H.~B.}\ \bibnamefont {{Schilling}}}, \bibinfo {author}
  {\bibfnamefont {M.}~\bibnamefont {{Schittenhelm}}}, \bibinfo {author}
  {\bibfnamefont {D.}~\bibnamefont {{Schloegl}}}, \bibinfo {author}
  {\bibfnamefont {H.}~\bibnamefont {{Schneider}}}, \bibinfo {author}
  {\bibfnamefont {R.}~\bibnamefont {{Schneider}}}, \bibinfo {author}
  {\bibfnamefont {W.}~\bibnamefont {{Schneider}}}, \bibinfo {author}
  {\bibfnamefont {G.}~\bibnamefont {{Schramm}}}, \bibinfo {author}
  {\bibfnamefont {J.}~\bibnamefont {{Schweinzer}}}, \bibinfo {author}
  {\bibfnamefont {S.}~\bibnamefont {{Schweizer}}}, \bibinfo {author}
  {\bibfnamefont {R.}~\bibnamefont {{Schwoerer}}}, \bibinfo {author}
  {\bibfnamefont {B.~D.}\ \bibnamefont {{Scott}}}, \bibinfo {author}
  {\bibfnamefont {U.}~\bibnamefont {{Seidel}}}, \bibinfo {author}
  {\bibfnamefont {F.}~\bibnamefont {{Serra}}}, \bibinfo {author} {\bibfnamefont
  {S.}~\bibnamefont {{Sesnic}}}, \bibinfo {author} {\bibfnamefont
  {A.}~\bibnamefont {{Silva}}}, \bibinfo {author} {\bibfnamefont
  {M.}~\bibnamefont {{Sokoll}}}, \bibinfo {author} {\bibfnamefont
  {E.}~\bibnamefont {{Speth}}}, \bibinfo {author} {\bibfnamefont
  {A.}~\bibnamefont {{Staebler}}}, \bibinfo {author} {\bibfnamefont {K.~H.}\
  \bibnamefont {{Steuer}}}, \bibinfo {author} {\bibfnamefont {J.}~\bibnamefont
  {{Stober}}}, \bibinfo {author} {\bibfnamefont {B.}~\bibnamefont {{Streibl}}},
  \bibinfo {author} {\bibfnamefont {A.}~\bibnamefont {{Thoma}}}, \bibinfo
  {author} {\bibfnamefont {W.}~\bibnamefont {{Treutterer}}}, \bibinfo {author}
  {\bibfnamefont {M.}~\bibnamefont {{Troppmann}}}, \bibinfo {author}
  {\bibfnamefont {N.}~\bibnamefont {{Tsois}}}, \bibinfo {author} {\bibfnamefont
  {M.}~\bibnamefont {{Ulrich}}}, \bibinfo {author} {\bibfnamefont
  {P.}~\bibnamefont {{Varela}}}, \bibinfo {author} {\bibfnamefont
  {H.}~\bibnamefont {{Verbeek}}}, \bibinfo {author} {\bibfnamefont
  {O.}~\bibnamefont {{Vollmer}}}, \bibinfo {author} {\bibfnamefont
  {H.}~\bibnamefont {{Wedler}}}, \bibinfo {author} {\bibfnamefont
  {M.}~\bibnamefont {{Weinlich}}}, \bibinfo {author} {\bibfnamefont
  {U.}~\bibnamefont {{Wenzel}}}, \bibinfo {author} {\bibfnamefont
  {F.}~\bibnamefont {{Wesner}}}, \bibinfo {author} {\bibfnamefont
  {R.}~\bibnamefont {{Wolf}}}, \bibinfo {author} {\bibfnamefont
  {R.}~\bibnamefont {{Wunderlich}}}, \bibinfo {author} {\bibfnamefont
  {N.}~\bibnamefont {{Xantopoulus}}}, \bibinfo {author} {\bibfnamefont
  {Q.}~\bibnamefont {{Yu}}}, \bibinfo {author} {\bibfnamefont {D.}~\bibnamefont
  {{Zasche}}}, \bibinfo {author} {\bibfnamefont {T.}~\bibnamefont
  {{Zehetbauer}}}, \bibinfo {author} {\bibfnamefont {H.~P.}\ \bibnamefont
  {{Zehrfeld}}}, \bibinfo {author} {\bibfnamefont {H.}~\bibnamefont {{Zohm}}},
  \ and\ \bibinfo {author} {\bibfnamefont {M.}~\bibnamefont {{Zouhar}}},\
  }\bibfield  {title} {\enquote {\bibinfo {title} {{Divertor tokamak operation
  at high densities on ASDEX Upgrade}},}\ }\href {\doibase
  10.1088/0741-3335/39/12B/003} {\bibfield  {journal} {\bibinfo  {journal}
  {Plasma Physics and Controlled Fusion}\ }\textbf {\bibinfo {volume} {39}},\
  \bibinfo {eid} {B19} (\bibinfo {year} {1997})}\BibitemShut {NoStop}%
\bibitem [{\citenamefont {{Kaye}}\ \emph {et~al.}(2005)\citenamefont {{Kaye}},
  \citenamefont {{Bell}}, \citenamefont {{Bell}}, \citenamefont {{Bernabei}},
  \citenamefont {{Bialek}}, \citenamefont {{Biewer}}, \citenamefont
  {{Blanchard}}, \citenamefont {{Boedo}}, \citenamefont {{Bush}}, \citenamefont
  {{Carter}}, \citenamefont {{Choe}}, \citenamefont {{Crocker}}, \citenamefont
  {{Darrow}}, \citenamefont {{Davis}}, \citenamefont {{Delgado-Aparicio}},
  \citenamefont {{Diem}}, \citenamefont {{Ferron}}, \citenamefont {{Field}},
  \citenamefont {{Foley}}, \citenamefont {{Fredrickson}}, \citenamefont
  {{Gates}}, \citenamefont {{Gibney}}, \citenamefont {{Harvey}}, \citenamefont
  {{Hatcher}}, \citenamefont {{Heidbrink}}, \citenamefont {{Hill}},
  \citenamefont {{Hosea}}, \citenamefont {{Jarboe}}, \citenamefont {{Johnson}},
  \citenamefont {{Kaita}}, \citenamefont {{Kessel}}, \citenamefont {{Kubota}},
  \citenamefont {{Kugel}}, \citenamefont {{Lawson}}, \citenamefont {{LeBlanc}},
  \citenamefont {{Lee}}, \citenamefont {{Levinton}}, \citenamefont {{Maingi}},
  \citenamefont {{Manickam}}, \citenamefont {{Maqueda}}, \citenamefont
  {{Marsala}}, \citenamefont {{Mastrovito}}, \citenamefont {{Mau}},
  \citenamefont {{Medley}}, \citenamefont {{Menard}}, \citenamefont {{Meyer}},
  \citenamefont {{Mikkelsen}}, \citenamefont {{Mueller}}, \citenamefont
  {{Munsat}}, \citenamefont {{Nelson}}, \citenamefont {{Neumeyer}},
  \citenamefont {{Nishino}}, \citenamefont {{Ono}}, \citenamefont {{Park}},
  \citenamefont {{Park}}, \citenamefont {{Paul}}, \citenamefont {{Peebles}},
  \citenamefont {{Peng}}, \citenamefont {{Phillips}}, \citenamefont
  {{Pigarov}}, \citenamefont {{Pinsker}}, \citenamefont {{Ram}}, \citenamefont
  {{Ramakrishnan}}, \citenamefont {{Raman}}, \citenamefont {{Rasmussen}},
  \citenamefont {{Redi}}, \citenamefont {{Rensink}}, \citenamefont {{Rewoldt}},
  \citenamefont {{Robinson}}, \citenamefont {{Roney}}, \citenamefont
  {{Roquemore}}, \citenamefont {{Ruskov}}, \citenamefont {{Ryan}},
  \citenamefont {{Sabbagh}}, \citenamefont {{Schneider}}, \citenamefont
  {{Skinner}}, \citenamefont {{Smith}}, \citenamefont {{Sontag}}, \citenamefont
  {{Soukhanovskii}}, \citenamefont {{Stevenson}}, \citenamefont {{Stotler}},
  \citenamefont {{Stratton}}, \citenamefont {{Stutman}}, \citenamefont
  {{Swain}}, \citenamefont {{Synakowski}}, \citenamefont {{Takase}},
  \citenamefont {{Taylor}}, \citenamefont {{Tritz}}, \citenamefont {{von
  Halle}}, \citenamefont {{Wade}}, \citenamefont {{White}}, \citenamefont
  {{Wilgen}}, \citenamefont {{Williams}}, \citenamefont {{Wilson}},
  \citenamefont {{Zhu}}, \citenamefont {{Zweben}}, \citenamefont {{Akers}},
  \citenamefont {{Beiersdorfer}}, \citenamefont {{Betti}}, \citenamefont
  {{Bigelow}}, \citenamefont {{Bitter}}, \citenamefont {{Bonoli}},
  \citenamefont {{Bourdelle}}, \citenamefont {{Chang}}, \citenamefont
  {{Chrzanowski}}, \citenamefont {{Domier}}, \citenamefont {{Dudek}},
  \citenamefont {{Efthimion}}, \citenamefont {{Finkenthal}}, \citenamefont
  {{Fredd}}, \citenamefont {{Fu}}, \citenamefont {{Glasser}}, \citenamefont
  {{Goldston}}, \citenamefont {{Greenough}}, \citenamefont {{Grisham}},
  \citenamefont {{Gorelenkov}}, \citenamefont {{Guazzotto}}, \citenamefont
  {{Hawryluk}}, \citenamefont {{Hogan}}, \citenamefont {{Houlberg}},
  \citenamefont {{Humphreys}}, \citenamefont {{Jaeger}}, \citenamefont
  {{Kalish}}, \citenamefont {{Krasheninnikov}}, \citenamefont {{Lao}},
  \citenamefont {{Lawrence}}, \citenamefont {{Leuer}}, \citenamefont {{Liu}},
  \citenamefont {{Luhmann}}, \citenamefont {{Mazzucato}}, \citenamefont
  {{Oliaro}}, \citenamefont {{Pacella}}, \citenamefont {{Parsells}},
  \citenamefont {{Schaffer}}, \citenamefont {{Semenov}}, \citenamefont
  {{Shaing}}, \citenamefont {{Shapiro}}, \citenamefont {{Shinohara}},
  \citenamefont {{Sichta}}, \citenamefont {{Tang}}, \citenamefont {{Vero}},
  \citenamefont {{Walker}},\ and\ \citenamefont {{Wampler}}}]{Kaye2005}%
  \BibitemOpen
  \bibfield  {author} {\bibinfo {author} {\bibfnamefont {S.~M.}\ \bibnamefont
  {{Kaye}}}, \bibinfo {author} {\bibfnamefont {M.~G.}\ \bibnamefont {{Bell}}},
  \bibinfo {author} {\bibfnamefont {R.~E.}\ \bibnamefont {{Bell}}}, \bibinfo
  {author} {\bibfnamefont {S.}~\bibnamefont {{Bernabei}}}, \bibinfo {author}
  {\bibfnamefont {J.}~\bibnamefont {{Bialek}}}, \bibinfo {author}
  {\bibfnamefont {T.}~\bibnamefont {{Biewer}}}, \bibinfo {author}
  {\bibfnamefont {W.}~\bibnamefont {{Blanchard}}}, \bibinfo {author}
  {\bibfnamefont {J.}~\bibnamefont {{Boedo}}}, \bibinfo {author} {\bibfnamefont
  {C.}~\bibnamefont {{Bush}}}, \bibinfo {author} {\bibfnamefont {M.~D.}\
  \bibnamefont {{Carter}}}, \bibinfo {author} {\bibfnamefont {W.}~\bibnamefont
  {{Choe}}}, \bibinfo {author} {\bibfnamefont {N.}~\bibnamefont {{Crocker}}},
  \bibinfo {author} {\bibfnamefont {D.~S.}\ \bibnamefont {{Darrow}}}, \bibinfo
  {author} {\bibfnamefont {W.}~\bibnamefont {{Davis}}}, \bibinfo {author}
  {\bibfnamefont {L.}~\bibnamefont {{Delgado-Aparicio}}}, \bibinfo {author}
  {\bibfnamefont {S.}~\bibnamefont {{Diem}}}, \bibinfo {author} {\bibfnamefont
  {J.}~\bibnamefont {{Ferron}}}, \bibinfo {author} {\bibfnamefont
  {A.}~\bibnamefont {{Field}}}, \bibinfo {author} {\bibfnamefont
  {J.}~\bibnamefont {{Foley}}}, \bibinfo {author} {\bibfnamefont {E.~D.}\
  \bibnamefont {{Fredrickson}}}, \bibinfo {author} {\bibfnamefont {D.~A.}\
  \bibnamefont {{Gates}}}, \bibinfo {author} {\bibfnamefont {T.}~\bibnamefont
  {{Gibney}}}, \bibinfo {author} {\bibfnamefont {R.}~\bibnamefont {{Harvey}}},
  \bibinfo {author} {\bibfnamefont {R.~E.}\ \bibnamefont {{Hatcher}}}, \bibinfo
  {author} {\bibfnamefont {W.}~\bibnamefont {{Heidbrink}}}, \bibinfo {author}
  {\bibfnamefont {K.}~\bibnamefont {{Hill}}}, \bibinfo {author} {\bibfnamefont
  {J.~C.}\ \bibnamefont {{Hosea}}}, \bibinfo {author} {\bibfnamefont {T.~R.}\
  \bibnamefont {{Jarboe}}}, \bibinfo {author} {\bibfnamefont {D.~W.}\
  \bibnamefont {{Johnson}}}, \bibinfo {author} {\bibfnamefont {R.}~\bibnamefont
  {{Kaita}}}, \bibinfo {author} {\bibfnamefont {C.}~\bibnamefont {{Kessel}}},
  \bibinfo {author} {\bibfnamefont {S.}~\bibnamefont {{Kubota}}}, \bibinfo
  {author} {\bibfnamefont {H.~W.}\ \bibnamefont {{Kugel}}}, \bibinfo {author}
  {\bibfnamefont {J.}~\bibnamefont {{Lawson}}}, \bibinfo {author}
  {\bibfnamefont {B.~P.}\ \bibnamefont {{LeBlanc}}}, \bibinfo {author}
  {\bibfnamefont {K.~C.}\ \bibnamefont {{Lee}}}, \bibinfo {author}
  {\bibfnamefont {F.}~\bibnamefont {{Levinton}}}, \bibinfo {author}
  {\bibfnamefont {R.}~\bibnamefont {{Maingi}}}, \bibinfo {author}
  {\bibfnamefont {J.}~\bibnamefont {{Manickam}}}, \bibinfo {author}
  {\bibfnamefont {R.}~\bibnamefont {{Maqueda}}}, \bibinfo {author}
  {\bibfnamefont {R.}~\bibnamefont {{Marsala}}}, \bibinfo {author}
  {\bibfnamefont {D.}~\bibnamefont {{Mastrovito}}}, \bibinfo {author}
  {\bibfnamefont {T.~K.}\ \bibnamefont {{Mau}}}, \bibinfo {author}
  {\bibfnamefont {S.~S.}\ \bibnamefont {{Medley}}}, \bibinfo {author}
  {\bibfnamefont {J.}~\bibnamefont {{Menard}}}, \bibinfo {author}
  {\bibfnamefont {H.}~\bibnamefont {{Meyer}}}, \bibinfo {author} {\bibfnamefont
  {D.~R.}\ \bibnamefont {{Mikkelsen}}}, \bibinfo {author} {\bibfnamefont
  {D.}~\bibnamefont {{Mueller}}}, \bibinfo {author} {\bibfnamefont
  {T.}~\bibnamefont {{Munsat}}}, \bibinfo {author} {\bibfnamefont {B.~A.}\
  \bibnamefont {{Nelson}}}, \bibinfo {author} {\bibfnamefont {C.}~\bibnamefont
  {{Neumeyer}}}, \bibinfo {author} {\bibfnamefont {N.}~\bibnamefont
  {{Nishino}}}, \bibinfo {author} {\bibfnamefont {M.}~\bibnamefont {{Ono}}},
  \bibinfo {author} {\bibfnamefont {H.}~\bibnamefont {{Park}}}, \bibinfo
  {author} {\bibfnamefont {W.}~\bibnamefont {{Park}}}, \bibinfo {author}
  {\bibfnamefont {S.}~\bibnamefont {{Paul}}}, \bibinfo {author} {\bibfnamefont
  {T.}~\bibnamefont {{Peebles}}}, \bibinfo {author} {\bibfnamefont
  {M.}~\bibnamefont {{Peng}}}, \bibinfo {author} {\bibfnamefont
  {C.}~\bibnamefont {{Phillips}}}, \bibinfo {author} {\bibfnamefont
  {A.}~\bibnamefont {{Pigarov}}}, \bibinfo {author} {\bibfnamefont
  {R.}~\bibnamefont {{Pinsker}}}, \bibinfo {author} {\bibfnamefont
  {A.}~\bibnamefont {{Ram}}}, \bibinfo {author} {\bibfnamefont
  {S.}~\bibnamefont {{Ramakrishnan}}}, \bibinfo {author} {\bibfnamefont
  {R.}~\bibnamefont {{Raman}}}, \bibinfo {author} {\bibfnamefont
  {D.}~\bibnamefont {{Rasmussen}}}, \bibinfo {author} {\bibfnamefont
  {M.}~\bibnamefont {{Redi}}}, \bibinfo {author} {\bibfnamefont
  {M.}~\bibnamefont {{Rensink}}}, \bibinfo {author} {\bibfnamefont
  {G.}~\bibnamefont {{Rewoldt}}}, \bibinfo {author} {\bibfnamefont
  {J.}~\bibnamefont {{Robinson}}}, \bibinfo {author} {\bibfnamefont
  {P.}~\bibnamefont {{Roney}}}, \bibinfo {author} {\bibfnamefont {A.~L.}\
  \bibnamefont {{Roquemore}}}, \bibinfo {author} {\bibfnamefont
  {E.}~\bibnamefont {{Ruskov}}}, \bibinfo {author} {\bibfnamefont
  {P.}~\bibnamefont {{Ryan}}}, \bibinfo {author} {\bibfnamefont {S.~A.}\
  \bibnamefont {{Sabbagh}}}, \bibinfo {author} {\bibfnamefont {H.}~\bibnamefont
  {{Schneider}}}, \bibinfo {author} {\bibfnamefont {C.~H.}\ \bibnamefont
  {{Skinner}}}, \bibinfo {author} {\bibfnamefont {D.~R.}\ \bibnamefont
  {{Smith}}}, \bibinfo {author} {\bibfnamefont {A.}~\bibnamefont {{Sontag}}},
  \bibinfo {author} {\bibfnamefont {V.}~\bibnamefont {{Soukhanovskii}}},
  \bibinfo {author} {\bibfnamefont {T.}~\bibnamefont {{Stevenson}}}, \bibinfo
  {author} {\bibfnamefont {D.}~\bibnamefont {{Stotler}}}, \bibinfo {author}
  {\bibfnamefont {B.}~\bibnamefont {{Stratton}}}, \bibinfo {author}
  {\bibfnamefont {D.}~\bibnamefont {{Stutman}}}, \bibinfo {author}
  {\bibfnamefont {D.}~\bibnamefont {{Swain}}}, \bibinfo {author} {\bibfnamefont
  {E.}~\bibnamefont {{Synakowski}}}, \bibinfo {author} {\bibfnamefont
  {Y.}~\bibnamefont {{Takase}}}, \bibinfo {author} {\bibfnamefont
  {G.}~\bibnamefont {{Taylor}}}, \bibinfo {author} {\bibfnamefont
  {K.}~\bibnamefont {{Tritz}}}, \bibinfo {author} {\bibfnamefont
  {A.}~\bibnamefont {{von Halle}}}, \bibinfo {author} {\bibfnamefont
  {M.}~\bibnamefont {{Wade}}}, \bibinfo {author} {\bibfnamefont
  {R.}~\bibnamefont {{White}}}, \bibinfo {author} {\bibfnamefont
  {J.}~\bibnamefont {{Wilgen}}}, \bibinfo {author} {\bibfnamefont
  {M.}~\bibnamefont {{Williams}}}, \bibinfo {author} {\bibfnamefont {J.~R.}\
  \bibnamefont {{Wilson}}}, \bibinfo {author} {\bibfnamefont {W.}~\bibnamefont
  {{Zhu}}}, \bibinfo {author} {\bibfnamefont {S.~J.}\ \bibnamefont {{Zweben}}},
  \bibinfo {author} {\bibfnamefont {R.}~\bibnamefont {{Akers}}}, \bibinfo
  {author} {\bibfnamefont {P.}~\bibnamefont {{Beiersdorfer}}}, \bibinfo
  {author} {\bibfnamefont {R.}~\bibnamefont {{Betti}}}, \bibinfo {author}
  {\bibfnamefont {T.}~\bibnamefont {{Bigelow}}}, \bibinfo {author}
  {\bibfnamefont {M.}~\bibnamefont {{Bitter}}}, \bibinfo {author}
  {\bibfnamefont {P.}~\bibnamefont {{Bonoli}}}, \bibinfo {author}
  {\bibfnamefont {C.}~\bibnamefont {{Bourdelle}}}, \bibinfo {author}
  {\bibfnamefont {C.~S.}\ \bibnamefont {{Chang}}}, \bibinfo {author}
  {\bibfnamefont {J.}~\bibnamefont {{Chrzanowski}}}, \bibinfo {author}
  {\bibfnamefont {C.}~\bibnamefont {{Domier}}}, \bibinfo {author}
  {\bibfnamefont {L.}~\bibnamefont {{Dudek}}}, \bibinfo {author} {\bibfnamefont
  {P.~C.}\ \bibnamefont {{Efthimion}}}, \bibinfo {author} {\bibfnamefont
  {M.}~\bibnamefont {{Finkenthal}}}, \bibinfo {author} {\bibfnamefont
  {E.}~\bibnamefont {{Fredd}}}, \bibinfo {author} {\bibfnamefont {G.~Y.}\
  \bibnamefont {{Fu}}}, \bibinfo {author} {\bibfnamefont {A.}~\bibnamefont
  {{Glasser}}}, \bibinfo {author} {\bibfnamefont {R.~J.}\ \bibnamefont
  {{Goldston}}}, \bibinfo {author} {\bibfnamefont {N.~L.}\ \bibnamefont
  {{Greenough}}}, \bibinfo {author} {\bibfnamefont {L.~R.}\ \bibnamefont
  {{Grisham}}}, \bibinfo {author} {\bibfnamefont {N.}~\bibnamefont
  {{Gorelenkov}}}, \bibinfo {author} {\bibfnamefont {L.}~\bibnamefont
  {{Guazzotto}}}, \bibinfo {author} {\bibfnamefont {R.~J.}\ \bibnamefont
  {{Hawryluk}}}, \bibinfo {author} {\bibfnamefont {J.}~\bibnamefont {{Hogan}}},
  \bibinfo {author} {\bibfnamefont {W.}~\bibnamefont {{Houlberg}}}, \bibinfo
  {author} {\bibfnamefont {D.}~\bibnamefont {{Humphreys}}}, \bibinfo {author}
  {\bibfnamefont {F.}~\bibnamefont {{Jaeger}}}, \bibinfo {author}
  {\bibfnamefont {M.}~\bibnamefont {{Kalish}}}, \bibinfo {author}
  {\bibfnamefont {S.}~\bibnamefont {{Krasheninnikov}}}, \bibinfo {author}
  {\bibfnamefont {L.~L.}\ \bibnamefont {{Lao}}}, \bibinfo {author}
  {\bibfnamefont {J.}~\bibnamefont {{Lawrence}}}, \bibinfo {author}
  {\bibfnamefont {J.}~\bibnamefont {{Leuer}}}, \bibinfo {author} {\bibfnamefont
  {D.}~\bibnamefont {{Liu}}}, \bibinfo {author} {\bibfnamefont {N.~C.}\
  \bibnamefont {{Luhmann}}}, \bibinfo {author} {\bibfnamefont {E.}~\bibnamefont
  {{Mazzucato}}}, \bibinfo {author} {\bibfnamefont {G.}~\bibnamefont
  {{Oliaro}}}, \bibinfo {author} {\bibfnamefont {D.}~\bibnamefont {{Pacella}}},
  \bibinfo {author} {\bibfnamefont {R.}~\bibnamefont {{Parsells}}}, \bibinfo
  {author} {\bibfnamefont {M.}~\bibnamefont {{Schaffer}}}, \bibinfo {author}
  {\bibfnamefont {I.}~\bibnamefont {{Semenov}}}, \bibinfo {author}
  {\bibfnamefont {K.~C.}\ \bibnamefont {{Shaing}}}, \bibinfo {author}
  {\bibfnamefont {M.~A.}\ \bibnamefont {{Shapiro}}}, \bibinfo {author}
  {\bibfnamefont {K.}~\bibnamefont {{Shinohara}}}, \bibinfo {author}
  {\bibfnamefont {P.}~\bibnamefont {{Sichta}}}, \bibinfo {author}
  {\bibfnamefont {X.}~\bibnamefont {{Tang}}}, \bibinfo {author} {\bibfnamefont
  {R.}~\bibnamefont {{Vero}}}, \bibinfo {author} {\bibfnamefont
  {D.}~\bibnamefont {{Walker}}}, \ and\ \bibinfo {author} {\bibfnamefont
  {W.}~\bibnamefont {{Wampler}}},\ }\bibfield  {title} {\enquote {\bibinfo
  {title} {{Progress towards high performance plasmas in the National Spherical
  Torus Experiment (NSTX)}},}\ }\href {\doibase 10.1088/0029-5515/45/10/S14}
  {\bibfield  {journal} {\bibinfo  {journal} {Nuclear Fusion}\ }\textbf
  {\bibinfo {volume} {45}},\ \bibinfo {pages} {S168--S180} (\bibinfo {year}
  {2005})}\BibitemShut {NoStop}%
\bibitem [{\citenamefont {{Menard}}\ \emph {et~al.}(2017)\citenamefont
  {{Menard}}, \citenamefont {{Allain}}, \citenamefont {{Battaglia}},
  \citenamefont {{Bedoya}}, \citenamefont {{Bell}}, \citenamefont {{Belova}},
  \citenamefont {{Berkery}}, \citenamefont {{Boyer}}, \citenamefont
  {{Crocker}}, \citenamefont {{Diallo}}, \citenamefont {{Ebrahimi}},
  \citenamefont {{Ferraro}}, \citenamefont {{Fredrickson}}, \citenamefont
  {{Frerichs}}, \citenamefont {{Gerhardt}}, \citenamefont {{Gorelenkov}},
  \citenamefont {{Guttenfelder}}, \citenamefont {{Heidbrink}}, \citenamefont
  {{Kaita}}, \citenamefont {{Kaye}}, \citenamefont {{Kriete}}, \citenamefont
  {{Kubota}}, \citenamefont {{LeBlanc}}, \citenamefont {{Liu}}, \citenamefont
  {{Lunsford}}, \citenamefont {{Mueller}}, \citenamefont {{Myers}},
  \citenamefont {{Ono}}, \citenamefont {{Park}}, \citenamefont {{Podesta}},
  \citenamefont {{Raman}}, \citenamefont {{Reinke}}, \citenamefont {{Ren}},
  \citenamefont {{Sabbagh}}, \citenamefont {{Schmitz}}, \citenamefont
  {{Scotti}}, \citenamefont {{Sechrest}}, \citenamefont {{Skinner}},
  \citenamefont {{Smith}}, \citenamefont {{Soukhanovskii}}, \citenamefont
  {{Stoltzfus-Dueck}}, \citenamefont {{Yuh}}, \citenamefont {{Wang}},
  \citenamefont {{Waters}}, \citenamefont {{Ahn}}, \citenamefont {{Andre}},
  \citenamefont {{Barchfeld}}, \citenamefont {{Beiersdorfer}}, \citenamefont
  {{Bertelli}}, \citenamefont {{Bhattacharjee}}, \citenamefont {{Brennan}},
  \citenamefont {{Buttery}}, \citenamefont {{Capece}}, \citenamefont {{Canal}},
  \citenamefont {{Canik}}, \citenamefont {{Chang}}, \citenamefont {{Darrow}},
  \citenamefont {{Delgado-Aparicio}}, \citenamefont {{Domier}}, \citenamefont
  {{Ethier}}, \citenamefont {{Evans}}, \citenamefont {{Ferron}}, \citenamefont
  {{Finkenthal}}, \citenamefont {{Fonck}}, \citenamefont {{Gan}}, \citenamefont
  {{Gates}}, \citenamefont {{Goumiri}}, \citenamefont {{Gray}}, \citenamefont
  {{Hosea}}, \citenamefont {{Humphreys}}, \citenamefont {{Jarboe}},
  \citenamefont {{Jardin}}, \citenamefont {{Jaworski}}, \citenamefont {{Koel}},
  \citenamefont {{Kolemen}}, \citenamefont {{Ku}}, \citenamefont {{La Haye}},
  \citenamefont {{Levinton}}, \citenamefont {{Luhmann}}, \citenamefont
  {{Maingi}}, \citenamefont {{Maqueda}}, \citenamefont {{McKee}}, \citenamefont
  {{Meier}}, \citenamefont {{Myra}}, \citenamefont {{Perkins}}, \citenamefont
  {{Poli}}, \citenamefont {{Rhodes}}, \citenamefont {{Riquezes}}, \citenamefont
  {{Rowley}}, \citenamefont {{Russell}}, \citenamefont {{Schuster}},
  \citenamefont {{Stratton}}, \citenamefont {{Stutman}}, \citenamefont
  {{Taylor}}, \citenamefont {{Tritz}}, \citenamefont {{Wang}}, \citenamefont
  {{Wirth}},\ and\ \citenamefont {{Zweben}}}]{2017NucFu..57j2006M}%
  \BibitemOpen
  \bibfield  {author} {\bibinfo {author} {\bibfnamefont {J.~E.}\ \bibnamefont
  {{Menard}}}, \bibinfo {author} {\bibfnamefont {J.~P.}\ \bibnamefont
  {{Allain}}}, \bibinfo {author} {\bibfnamefont {D.~J.}\ \bibnamefont
  {{Battaglia}}}, \bibinfo {author} {\bibfnamefont {F.}~\bibnamefont
  {{Bedoya}}}, \bibinfo {author} {\bibfnamefont {R.~E.}\ \bibnamefont
  {{Bell}}}, \bibinfo {author} {\bibfnamefont {E.}~\bibnamefont {{Belova}}},
  \bibinfo {author} {\bibfnamefont {J.~W.}\ \bibnamefont {{Berkery}}}, \bibinfo
  {author} {\bibfnamefont {M.~D.}\ \bibnamefont {{Boyer}}}, \bibinfo {author}
  {\bibfnamefont {N.}~\bibnamefont {{Crocker}}}, \bibinfo {author}
  {\bibfnamefont {A.}~\bibnamefont {{Diallo}}}, \bibinfo {author}
  {\bibfnamefont {F.}~\bibnamefont {{Ebrahimi}}}, \bibinfo {author}
  {\bibfnamefont {N.}~\bibnamefont {{Ferraro}}}, \bibinfo {author}
  {\bibfnamefont {E.}~\bibnamefont {{Fredrickson}}}, \bibinfo {author}
  {\bibfnamefont {H.}~\bibnamefont {{Frerichs}}}, \bibinfo {author}
  {\bibfnamefont {S.}~\bibnamefont {{Gerhardt}}}, \bibinfo {author}
  {\bibfnamefont {N.}~\bibnamefont {{Gorelenkov}}}, \bibinfo {author}
  {\bibfnamefont {W.}~\bibnamefont {{Guttenfelder}}}, \bibinfo {author}
  {\bibfnamefont {W.}~\bibnamefont {{Heidbrink}}}, \bibinfo {author}
  {\bibfnamefont {R.}~\bibnamefont {{Kaita}}}, \bibinfo {author} {\bibfnamefont
  {S.~M.}\ \bibnamefont {{Kaye}}}, \bibinfo {author} {\bibfnamefont {D.~M.}\
  \bibnamefont {{Kriete}}}, \bibinfo {author} {\bibfnamefont {S.}~\bibnamefont
  {{Kubota}}}, \bibinfo {author} {\bibfnamefont {B.~P.}\ \bibnamefont
  {{LeBlanc}}}, \bibinfo {author} {\bibfnamefont {D.}~\bibnamefont {{Liu}}},
  \bibinfo {author} {\bibfnamefont {R.}~\bibnamefont {{Lunsford}}}, \bibinfo
  {author} {\bibfnamefont {D.}~\bibnamefont {{Mueller}}}, \bibinfo {author}
  {\bibfnamefont {C.~E.}\ \bibnamefont {{Myers}}}, \bibinfo {author}
  {\bibfnamefont {M.}~\bibnamefont {{Ono}}}, \bibinfo {author} {\bibfnamefont
  {J.~K.}\ \bibnamefont {{Park}}}, \bibinfo {author} {\bibfnamefont
  {M.}~\bibnamefont {{Podesta}}}, \bibinfo {author} {\bibfnamefont
  {R.}~\bibnamefont {{Raman}}}, \bibinfo {author} {\bibfnamefont
  {M.}~\bibnamefont {{Reinke}}}, \bibinfo {author} {\bibfnamefont
  {Y.}~\bibnamefont {{Ren}}}, \bibinfo {author} {\bibfnamefont {S.~A.}\
  \bibnamefont {{Sabbagh}}}, \bibinfo {author} {\bibfnamefont {O.}~\bibnamefont
  {{Schmitz}}}, \bibinfo {author} {\bibfnamefont {F.}~\bibnamefont {{Scotti}}},
  \bibinfo {author} {\bibfnamefont {Y.}~\bibnamefont {{Sechrest}}}, \bibinfo
  {author} {\bibfnamefont {C.~H.}\ \bibnamefont {{Skinner}}}, \bibinfo {author}
  {\bibfnamefont {D.~R.}\ \bibnamefont {{Smith}}}, \bibinfo {author}
  {\bibfnamefont {V.}~\bibnamefont {{Soukhanovskii}}}, \bibinfo {author}
  {\bibfnamefont {T.}~\bibnamefont {{Stoltzfus-Dueck}}}, \bibinfo {author}
  {\bibfnamefont {H.}~\bibnamefont {{Yuh}}}, \bibinfo {author} {\bibfnamefont
  {Z.}~\bibnamefont {{Wang}}}, \bibinfo {author} {\bibfnamefont
  {I.}~\bibnamefont {{Waters}}}, \bibinfo {author} {\bibfnamefont {J.~W.}\
  \bibnamefont {{Ahn}}}, \bibinfo {author} {\bibfnamefont {R.}~\bibnamefont
  {{Andre}}}, \bibinfo {author} {\bibfnamefont {R.}~\bibnamefont
  {{Barchfeld}}}, \bibinfo {author} {\bibfnamefont {P.}~\bibnamefont
  {{Beiersdorfer}}}, \bibinfo {author} {\bibfnamefont {N.}~\bibnamefont
  {{Bertelli}}}, \bibinfo {author} {\bibfnamefont {A.}~\bibnamefont
  {{Bhattacharjee}}}, \bibinfo {author} {\bibfnamefont {D.}~\bibnamefont
  {{Brennan}}}, \bibinfo {author} {\bibfnamefont {R.}~\bibnamefont
  {{Buttery}}}, \bibinfo {author} {\bibfnamefont {A.}~\bibnamefont {{Capece}}},
  \bibinfo {author} {\bibfnamefont {G.}~\bibnamefont {{Canal}}}, \bibinfo
  {author} {\bibfnamefont {J.}~\bibnamefont {{Canik}}}, \bibinfo {author}
  {\bibfnamefont {C.~S.}\ \bibnamefont {{Chang}}}, \bibinfo {author}
  {\bibfnamefont {D.}~\bibnamefont {{Darrow}}}, \bibinfo {author}
  {\bibfnamefont {L.}~\bibnamefont {{Delgado-Aparicio}}}, \bibinfo {author}
  {\bibfnamefont {C.}~\bibnamefont {{Domier}}}, \bibinfo {author}
  {\bibfnamefont {S.}~\bibnamefont {{Ethier}}}, \bibinfo {author}
  {\bibfnamefont {T.}~\bibnamefont {{Evans}}}, \bibinfo {author} {\bibfnamefont
  {J.}~\bibnamefont {{Ferron}}}, \bibinfo {author} {\bibfnamefont
  {M.}~\bibnamefont {{Finkenthal}}}, \bibinfo {author} {\bibfnamefont
  {R.}~\bibnamefont {{Fonck}}}, \bibinfo {author} {\bibfnamefont
  {K.}~\bibnamefont {{Gan}}}, \bibinfo {author} {\bibfnamefont
  {D.}~\bibnamefont {{Gates}}}, \bibinfo {author} {\bibfnamefont
  {I.}~\bibnamefont {{Goumiri}}}, \bibinfo {author} {\bibfnamefont
  {T.}~\bibnamefont {{Gray}}}, \bibinfo {author} {\bibfnamefont
  {J.}~\bibnamefont {{Hosea}}}, \bibinfo {author} {\bibfnamefont
  {D.}~\bibnamefont {{Humphreys}}}, \bibinfo {author} {\bibfnamefont
  {T.}~\bibnamefont {{Jarboe}}}, \bibinfo {author} {\bibfnamefont
  {S.}~\bibnamefont {{Jardin}}}, \bibinfo {author} {\bibfnamefont {M.~A.}\
  \bibnamefont {{Jaworski}}}, \bibinfo {author} {\bibfnamefont
  {B.}~\bibnamefont {{Koel}}}, \bibinfo {author} {\bibfnamefont
  {E.}~\bibnamefont {{Kolemen}}}, \bibinfo {author} {\bibfnamefont
  {S.}~\bibnamefont {{Ku}}}, \bibinfo {author} {\bibfnamefont {R.~J.}\
  \bibnamefont {{La Haye}}}, \bibinfo {author} {\bibfnamefont {F.}~\bibnamefont
  {{Levinton}}}, \bibinfo {author} {\bibfnamefont {N.}~\bibnamefont
  {{Luhmann}}}, \bibinfo {author} {\bibfnamefont {R.}~\bibnamefont {{Maingi}}},
  \bibinfo {author} {\bibfnamefont {R.}~\bibnamefont {{Maqueda}}}, \bibinfo
  {author} {\bibfnamefont {G.}~\bibnamefont {{McKee}}}, \bibinfo {author}
  {\bibfnamefont {E.}~\bibnamefont {{Meier}}}, \bibinfo {author} {\bibfnamefont
  {J.}~\bibnamefont {{Myra}}}, \bibinfo {author} {\bibfnamefont
  {R.}~\bibnamefont {{Perkins}}}, \bibinfo {author} {\bibfnamefont
  {F.}~\bibnamefont {{Poli}}}, \bibinfo {author} {\bibfnamefont
  {T.}~\bibnamefont {{Rhodes}}}, \bibinfo {author} {\bibfnamefont
  {J.}~\bibnamefont {{Riquezes}}}, \bibinfo {author} {\bibfnamefont
  {C.}~\bibnamefont {{Rowley}}}, \bibinfo {author} {\bibfnamefont
  {D.}~\bibnamefont {{Russell}}}, \bibinfo {author} {\bibfnamefont
  {E.}~\bibnamefont {{Schuster}}}, \bibinfo {author} {\bibfnamefont
  {B.}~\bibnamefont {{Stratton}}}, \bibinfo {author} {\bibfnamefont
  {D.}~\bibnamefont {{Stutman}}}, \bibinfo {author} {\bibfnamefont
  {G.}~\bibnamefont {{Taylor}}}, \bibinfo {author} {\bibfnamefont
  {K.}~\bibnamefont {{Tritz}}}, \bibinfo {author} {\bibfnamefont
  {W.}~\bibnamefont {{Wang}}}, \bibinfo {author} {\bibfnamefont
  {B.}~\bibnamefont {{Wirth}}}, \ and\ \bibinfo {author} {\bibfnamefont
  {S.~J.}\ \bibnamefont {{Zweben}}},\ }\bibfield  {title} {\enquote {\bibinfo
  {title} {{Overview of NSTX Upgrade initial results and modelling
  highlights}},}\ }\href {\doibase 10.1088/1741-4326/aa600a} {\bibfield
  {journal} {\bibinfo  {journal} {Nuclear Fusion}\ }\textbf {\bibinfo {volume}
  {57}},\ \bibinfo {eid} {102006} (\bibinfo {year} {2017})}\BibitemShut
  {NoStop}%
\bibitem [{\citenamefont {{Counsell}}\ \emph {et~al.}(2005)\citenamefont
  {{Counsell}}, \citenamefont {{Akers}}, \citenamefont {{Appel}}, \citenamefont
  {{Applegate}}, \citenamefont {{Axon}}, \citenamefont {{Baranov}},
  \citenamefont {{Brickley}}, \citenamefont {{Bunting}}, \citenamefont
  {{Buttery}}, \citenamefont {{Carolan}}, \citenamefont {{Challis}},
  \citenamefont {{Ciric}}, \citenamefont {{Conway}}, \citenamefont {{Cox}},
  \citenamefont {{Cunningham}}, \citenamefont {{Darke}}, \citenamefont
  {{Dnestrovskij}}, \citenamefont {{Dowling}}, \citenamefont {{Dudson}},
  \citenamefont {{Dunstan}}, \citenamefont {{Delchambre}}, \citenamefont
  {{Field}}, \citenamefont {{Foster}}, \citenamefont {{Gee}}, \citenamefont
  {{Gryaznevich}}, \citenamefont {{Helander}}, \citenamefont {{Hender}},
  \citenamefont {{Hole}}, \citenamefont {{Howell}}, \citenamefont {{Joiner}},
  \citenamefont {{Keeling}}, \citenamefont {{Kirk}}, \citenamefont {{Lehane}},
  \citenamefont {{Lisgo}}, \citenamefont {{Lloyd}}, \citenamefont {{Lott}},
  \citenamefont {{Maddison}}, \citenamefont {{Manhood}}, \citenamefont
  {{Martin}}, \citenamefont {{McArdle}}, \citenamefont {{McClements}},
  \citenamefont {{Meyer}}, \citenamefont {{Morris}}, \citenamefont {{Nelson}},
  \citenamefont {{O'Brien}}, \citenamefont {{Patel}}, \citenamefont
  {{Pinfold}}, \citenamefont {{Preinhaelter}}, \citenamefont {{Price}},
  \citenamefont {{Roach}}, \citenamefont {{Rozhansky}}, \citenamefont
  {{Saarelma}}, \citenamefont {{Saveliev}}, \citenamefont {{Scannell}},
  \citenamefont {{Sharapov}}, \citenamefont {{Shevchenko}}, \citenamefont
  {{Shibaev}}, \citenamefont {{Stammers}}, \citenamefont {{Storrs}},
  \citenamefont {{Sykes}}, \citenamefont {{Tabasso}}, \citenamefont
  {{Tallents}}, \citenamefont {{Taylor}}, \citenamefont {{Tournianski}},
  \citenamefont {{Turner}}, \citenamefont {{Turri}}, \citenamefont {{Valovic}},
  \citenamefont {{Volpe}}, \citenamefont {{Voss}}, \citenamefont {{Walsh}},
  \citenamefont {{Watkins}}, \citenamefont {{Wilson}}, \citenamefont {{Wisse}},
  \citenamefont {{MAST}}, \citenamefont {{NBI}},\ and\ \citenamefont {{ECRH
  Teams}}}]{MAST2005}%
  \BibitemOpen
  \bibfield  {author} {\bibinfo {author} {\bibfnamefont {G.~F.}\ \bibnamefont
  {{Counsell}}}, \bibinfo {author} {\bibfnamefont {R.~J.}\ \bibnamefont
  {{Akers}}}, \bibinfo {author} {\bibfnamefont {L.~C.}\ \bibnamefont
  {{Appel}}}, \bibinfo {author} {\bibfnamefont {D.}~\bibnamefont
  {{Applegate}}}, \bibinfo {author} {\bibfnamefont {K.~B.}\ \bibnamefont
  {{Axon}}}, \bibinfo {author} {\bibfnamefont {Y.}~\bibnamefont {{Baranov}}},
  \bibinfo {author} {\bibfnamefont {C.}~\bibnamefont {{Brickley}}}, \bibinfo
  {author} {\bibfnamefont {C.}~\bibnamefont {{Bunting}}}, \bibinfo {author}
  {\bibfnamefont {R.~J.}\ \bibnamefont {{Buttery}}}, \bibinfo {author}
  {\bibfnamefont {P.~G.}\ \bibnamefont {{Carolan}}}, \bibinfo {author}
  {\bibfnamefont {C.}~\bibnamefont {{Challis}}}, \bibinfo {author}
  {\bibfnamefont {D.}~\bibnamefont {{Ciric}}}, \bibinfo {author} {\bibfnamefont
  {N.~J.}\ \bibnamefont {{Conway}}}, \bibinfo {author} {\bibfnamefont
  {M.}~\bibnamefont {{Cox}}}, \bibinfo {author} {\bibfnamefont
  {G.}~\bibnamefont {{Cunningham}}}, \bibinfo {author} {\bibfnamefont
  {A.}~\bibnamefont {{Darke}}}, \bibinfo {author} {\bibfnamefont
  {A.}~\bibnamefont {{Dnestrovskij}}}, \bibinfo {author} {\bibfnamefont
  {J.}~\bibnamefont {{Dowling}}}, \bibinfo {author} {\bibfnamefont
  {B.}~\bibnamefont {{Dudson}}}, \bibinfo {author} {\bibfnamefont {M.~R.}\
  \bibnamefont {{Dunstan}}}, \bibinfo {author} {\bibfnamefont {E.}~\bibnamefont
  {{Delchambre}}}, \bibinfo {author} {\bibfnamefont {A.~R.}\ \bibnamefont
  {{Field}}}, \bibinfo {author} {\bibfnamefont {A.}~\bibnamefont {{Foster}}},
  \bibinfo {author} {\bibfnamefont {S.}~\bibnamefont {{Gee}}}, \bibinfo
  {author} {\bibfnamefont {M.~P.}\ \bibnamefont {{Gryaznevich}}}, \bibinfo
  {author} {\bibfnamefont {P.}~\bibnamefont {{Helander}}}, \bibinfo {author}
  {\bibfnamefont {T.~C.}\ \bibnamefont {{Hender}}}, \bibinfo {author}
  {\bibfnamefont {M.}~\bibnamefont {{Hole}}}, \bibinfo {author} {\bibfnamefont
  {D.~H.}\ \bibnamefont {{Howell}}}, \bibinfo {author} {\bibfnamefont
  {N.}~\bibnamefont {{Joiner}}}, \bibinfo {author} {\bibfnamefont
  {D.}~\bibnamefont {{Keeling}}}, \bibinfo {author} {\bibfnamefont
  {A.}~\bibnamefont {{Kirk}}}, \bibinfo {author} {\bibfnamefont {I.~P.}\
  \bibnamefont {{Lehane}}}, \bibinfo {author} {\bibfnamefont {S.}~\bibnamefont
  {{Lisgo}}}, \bibinfo {author} {\bibfnamefont {B.}~\bibnamefont {{Lloyd}}},
  \bibinfo {author} {\bibfnamefont {F.}~\bibnamefont {{Lott}}}, \bibinfo
  {author} {\bibfnamefont {G.~P.}\ \bibnamefont {{Maddison}}}, \bibinfo
  {author} {\bibfnamefont {S.~J.}\ \bibnamefont {{Manhood}}}, \bibinfo {author}
  {\bibfnamefont {R.}~\bibnamefont {{Martin}}}, \bibinfo {author}
  {\bibfnamefont {G.~J.}\ \bibnamefont {{McArdle}}}, \bibinfo {author}
  {\bibfnamefont {K.~G.}\ \bibnamefont {{McClements}}}, \bibinfo {author}
  {\bibfnamefont {H.}~\bibnamefont {{Meyer}}}, \bibinfo {author} {\bibfnamefont
  {A.~W.}\ \bibnamefont {{Morris}}}, \bibinfo {author} {\bibfnamefont
  {M.}~\bibnamefont {{Nelson}}}, \bibinfo {author} {\bibfnamefont {M.~R.}\
  \bibnamefont {{O'Brien}}}, \bibinfo {author} {\bibfnamefont {A.}~\bibnamefont
  {{Patel}}}, \bibinfo {author} {\bibfnamefont {T.}~\bibnamefont {{Pinfold}}},
  \bibinfo {author} {\bibfnamefont {J.}~\bibnamefont {{Preinhaelter}}},
  \bibinfo {author} {\bibfnamefont {M.~N.}\ \bibnamefont {{Price}}}, \bibinfo
  {author} {\bibfnamefont {C.~M.}\ \bibnamefont {{Roach}}}, \bibinfo {author}
  {\bibfnamefont {V.}~\bibnamefont {{Rozhansky}}}, \bibinfo {author}
  {\bibfnamefont {S.}~\bibnamefont {{Saarelma}}}, \bibinfo {author}
  {\bibfnamefont {A.}~\bibnamefont {{Saveliev}}}, \bibinfo {author}
  {\bibfnamefont {R.}~\bibnamefont {{Scannell}}}, \bibinfo {author}
  {\bibfnamefont {S.}~\bibnamefont {{Sharapov}}}, \bibinfo {author}
  {\bibfnamefont {V.}~\bibnamefont {{Shevchenko}}}, \bibinfo {author}
  {\bibfnamefont {S.}~\bibnamefont {{Shibaev}}}, \bibinfo {author}
  {\bibfnamefont {K.}~\bibnamefont {{Stammers}}}, \bibinfo {author}
  {\bibfnamefont {J.}~\bibnamefont {{Storrs}}}, \bibinfo {author}
  {\bibfnamefont {A.}~\bibnamefont {{Sykes}}}, \bibinfo {author} {\bibfnamefont
  {A.}~\bibnamefont {{Tabasso}}}, \bibinfo {author} {\bibfnamefont
  {S.}~\bibnamefont {{Tallents}}}, \bibinfo {author} {\bibfnamefont
  {D.}~\bibnamefont {{Taylor}}}, \bibinfo {author} {\bibfnamefont {M.~R.}\
  \bibnamefont {{Tournianski}}}, \bibinfo {author} {\bibfnamefont
  {A.}~\bibnamefont {{Turner}}}, \bibinfo {author} {\bibfnamefont
  {G.}~\bibnamefont {{Turri}}}, \bibinfo {author} {\bibfnamefont
  {M.}~\bibnamefont {{Valovic}}}, \bibinfo {author} {\bibfnamefont
  {F.}~\bibnamefont {{Volpe}}}, \bibinfo {author} {\bibfnamefont
  {G.}~\bibnamefont {{Voss}}}, \bibinfo {author} {\bibfnamefont {M.~J.}\
  \bibnamefont {{Walsh}}}, \bibinfo {author} {\bibfnamefont {J.~R.}\
  \bibnamefont {{Watkins}}}, \bibinfo {author} {\bibfnamefont {H.~R.}\
  \bibnamefont {{Wilson}}}, \bibinfo {author} {\bibfnamefont {M.}~\bibnamefont
  {{Wisse}}}, \bibinfo {author} {\bibfnamefont {t.}~\bibnamefont {{MAST}}},
  \bibinfo {author} {\bibnamefont {{NBI}}}, \ and\ \bibinfo {author}
  {\bibnamefont {{ECRH Teams}}},\ }\bibfield  {title} {\enquote {\bibinfo
  {title} {{Overview of MAST results}},}\ }\href {\doibase
  10.1088/0029-5515/45/10/S13} {\bibfield  {journal} {\bibinfo  {journal}
  {Nuclear Fusion}\ }\textbf {\bibinfo {volume} {45}},\ \bibinfo {pages}
  {S157--S167} (\bibinfo {year} {2005})}\BibitemShut {NoStop}%
\bibitem [{\citenamefont {{Chapman}}\ \emph {et~al.}(2015)\citenamefont
  {{Chapman}}, \citenamefont {{Adamek}}, \citenamefont {{Akers}}, \citenamefont
  {{Allan}}, \citenamefont {{Appel}}, \citenamefont {{Asunta}}, \citenamefont
  {{Barnes}}, \citenamefont {{Ben Ayed}}, \citenamefont {{Bigelow}},
  \citenamefont {{Boeglin}}, \citenamefont {{Bradley}}, \citenamefont
  {{Br{\"u}nner}}, \citenamefont {{Cahyna}}, \citenamefont {{Carr}},
  \citenamefont {{Caughman}}, \citenamefont {{Cecconello}}, \citenamefont
  {{Challis}}, \citenamefont {{Chapman}}, \citenamefont {{Chorley}},
  \citenamefont {{Colyer}}, \citenamefont {{Conway}}, \citenamefont {{Cooper}},
  \citenamefont {{Cox}}, \citenamefont {{Crocker}}, \citenamefont {{Crowley}},
  \citenamefont {{Cunningham}}, \citenamefont {{Danilov}}, \citenamefont
  {{Darrow}}, \citenamefont {{Dendy}}, \citenamefont {{Diallo}}, \citenamefont
  {{Dickinson}}, \citenamefont {{Diem}}, \citenamefont {{Dorland}},
  \citenamefont {{Dudson}}, \citenamefont {{Dunai}}, \citenamefont {{Easy}},
  \citenamefont {{Elmore}}, \citenamefont {{Field}}, \citenamefont
  {{Fishpool}}, \citenamefont {{Fox}}, \citenamefont {{Fredrickson}},
  \citenamefont {{Freethy}}, \citenamefont {{Garzotti}}, \citenamefont
  {{Ghim}}, \citenamefont {{Gibson}}, \citenamefont {{Graves}}, \citenamefont
  {{Gurl}}, \citenamefont {{Guttenfelder}}, \citenamefont {{Ham}},
  \citenamefont {{Harrison}}, \citenamefont {{Harting}}, \citenamefont
  {{Havlickova}}, \citenamefont {{Hawke}}, \citenamefont {{Hawkes}},
  \citenamefont {{Hender}}, \citenamefont {{Henderson}}, \citenamefont
  {{Highcock}}, \citenamefont {{Hillesheim}}, \citenamefont {{Hnat}},
  \citenamefont {{Holgate}}, \citenamefont {{Horacek}}, \citenamefont
  {{Howard}}, \citenamefont {{Huang}}, \citenamefont {{Imada}}, \citenamefont
  {{Jones}}, \citenamefont {{Kaye}}, \citenamefont {{Keeling}}, \citenamefont
  {{Kirk}}, \citenamefont {{Klimek}}, \citenamefont {{Kocan}}, \citenamefont
  {{Leggate}}, \citenamefont {{Lilley}}, \citenamefont {{Lipschultz}},
  \citenamefont {{Lisgo}}, \citenamefont {{Liu}}, \citenamefont {{Lloyd}},
  \citenamefont {{Lomanowski}}, \citenamefont {{Lupelli}}, \citenamefont
  {{Maddison}}, \citenamefont {{Mailloux}}, \citenamefont {{Martin}},
  \citenamefont {{McArdle}}, \citenamefont {{McClements}}, \citenamefont
  {{McMillan}}, \citenamefont {{Meakins}}, \citenamefont {{Meyer}},
  \citenamefont {{Michael}}, \citenamefont {{Militello}}, \citenamefont
  {{Milnes}}, \citenamefont {{Morris}}, \citenamefont {{Motojima}},
  \citenamefont {{Muir}}, \citenamefont {{Nardon}}, \citenamefont {{Naulin}},
  \citenamefont {{Naylor}}, \citenamefont {{Nielsen}}, \citenamefont
  {{O'Brien}}, \citenamefont {{O'Gorman}}, \citenamefont {{Ono}}, \citenamefont
  {{Oliver}}, \citenamefont {{Pamela}}, \citenamefont {{Pangione}},
  \citenamefont {{Parra}}, \citenamefont {{Patel}}, \citenamefont {{Peebles}},
  \citenamefont {{Peng}}, \citenamefont {{Perez}}, \citenamefont {{Pinches}},
  \citenamefont {{Piron}}, \citenamefont {{Podesta}}, \citenamefont {{Price}},
  \citenamefont {{Reinke}}, \citenamefont {{Ren}}, \citenamefont {{Roach}},
  \citenamefont {{Robinson}}, \citenamefont {{Romanelli}}, \citenamefont
  {{Rozhansky}}, \citenamefont {{Saarelma}}, \citenamefont {{Sangaroon}},
  \citenamefont {{Saveliev}}, \citenamefont {{Scannell}}, \citenamefont
  {{Schekochihin}}, \citenamefont {{Sharapov}}, \citenamefont {{Sharples}},
  \citenamefont {{Shevchenko}}, \citenamefont {{Silburn}}, \citenamefont
  {{Simpson}}, \citenamefont {{Storrs}}, \citenamefont {{Takase}},
  \citenamefont {{Tanabe}}, \citenamefont {{Tanaka}}, \citenamefont {{Taylor}},
  \citenamefont {{Taylor}}, \citenamefont {{Thomas}}, \citenamefont
  {{Thomas-Davies}}, \citenamefont {{Thornton}}, \citenamefont {{Turnyanskiy}},
  \citenamefont {{Valovic}}, \citenamefont {{Vann}}, \citenamefont {{Walkden}},
  \citenamefont {{Wilson}}, \citenamefont {{van Wyk}}, \citenamefont
  {{Yamada}}, \citenamefont {{Zoletnik}}, \citenamefont {{MAST}},\ and\
  \citenamefont {{MAST Upgrade Teams}}}]{MAST2015}%
  \BibitemOpen
  \bibfield  {author} {\bibinfo {author} {\bibfnamefont {I.~T.}\ \bibnamefont
  {{Chapman}}}, \bibinfo {author} {\bibfnamefont {J.}~\bibnamefont {{Adamek}}},
  \bibinfo {author} {\bibfnamefont {R.~J.}\ \bibnamefont {{Akers}}}, \bibinfo
  {author} {\bibfnamefont {S.}~\bibnamefont {{Allan}}}, \bibinfo {author}
  {\bibfnamefont {L.}~\bibnamefont {{Appel}}}, \bibinfo {author} {\bibfnamefont
  {O.}~\bibnamefont {{Asunta}}}, \bibinfo {author} {\bibfnamefont
  {M.}~\bibnamefont {{Barnes}}}, \bibinfo {author} {\bibfnamefont
  {N.}~\bibnamefont {{Ben Ayed}}}, \bibinfo {author} {\bibfnamefont
  {T.}~\bibnamefont {{Bigelow}}}, \bibinfo {author} {\bibfnamefont
  {W.}~\bibnamefont {{Boeglin}}}, \bibinfo {author} {\bibfnamefont
  {J.}~\bibnamefont {{Bradley}}}, \bibinfo {author} {\bibfnamefont
  {J.}~\bibnamefont {{Br{\"u}nner}}}, \bibinfo {author} {\bibfnamefont
  {P.}~\bibnamefont {{Cahyna}}}, \bibinfo {author} {\bibfnamefont
  {M.}~\bibnamefont {{Carr}}}, \bibinfo {author} {\bibfnamefont
  {J.}~\bibnamefont {{Caughman}}}, \bibinfo {author} {\bibfnamefont
  {M.}~\bibnamefont {{Cecconello}}}, \bibinfo {author} {\bibfnamefont
  {C.}~\bibnamefont {{Challis}}}, \bibinfo {author} {\bibfnamefont
  {S.}~\bibnamefont {{Chapman}}}, \bibinfo {author} {\bibfnamefont
  {J.}~\bibnamefont {{Chorley}}}, \bibinfo {author} {\bibfnamefont
  {G.}~\bibnamefont {{Colyer}}}, \bibinfo {author} {\bibfnamefont
  {N.}~\bibnamefont {{Conway}}}, \bibinfo {author} {\bibfnamefont {W.~A.}\
  \bibnamefont {{Cooper}}}, \bibinfo {author} {\bibfnamefont {M.}~\bibnamefont
  {{Cox}}}, \bibinfo {author} {\bibfnamefont {N.}~\bibnamefont {{Crocker}}},
  \bibinfo {author} {\bibfnamefont {B.}~\bibnamefont {{Crowley}}}, \bibinfo
  {author} {\bibfnamefont {G.}~\bibnamefont {{Cunningham}}}, \bibinfo {author}
  {\bibfnamefont {A.}~\bibnamefont {{Danilov}}}, \bibinfo {author}
  {\bibfnamefont {D.}~\bibnamefont {{Darrow}}}, \bibinfo {author}
  {\bibfnamefont {R.}~\bibnamefont {{Dendy}}}, \bibinfo {author} {\bibfnamefont
  {A.}~\bibnamefont {{Diallo}}}, \bibinfo {author} {\bibfnamefont
  {D.}~\bibnamefont {{Dickinson}}}, \bibinfo {author} {\bibfnamefont
  {S.}~\bibnamefont {{Diem}}}, \bibinfo {author} {\bibfnamefont
  {W.}~\bibnamefont {{Dorland}}}, \bibinfo {author} {\bibfnamefont
  {B.}~\bibnamefont {{Dudson}}}, \bibinfo {author} {\bibfnamefont
  {D.}~\bibnamefont {{Dunai}}}, \bibinfo {author} {\bibfnamefont
  {L.}~\bibnamefont {{Easy}}}, \bibinfo {author} {\bibfnamefont
  {S.}~\bibnamefont {{Elmore}}}, \bibinfo {author} {\bibfnamefont
  {A.}~\bibnamefont {{Field}}}, \bibinfo {author} {\bibfnamefont
  {G.}~\bibnamefont {{Fishpool}}}, \bibinfo {author} {\bibfnamefont
  {M.}~\bibnamefont {{Fox}}}, \bibinfo {author} {\bibfnamefont
  {E.}~\bibnamefont {{Fredrickson}}}, \bibinfo {author} {\bibfnamefont
  {S.}~\bibnamefont {{Freethy}}}, \bibinfo {author} {\bibfnamefont
  {L.}~\bibnamefont {{Garzotti}}}, \bibinfo {author} {\bibfnamefont {Y.~C.}\
  \bibnamefont {{Ghim}}}, \bibinfo {author} {\bibfnamefont {K.}~\bibnamefont
  {{Gibson}}}, \bibinfo {author} {\bibfnamefont {J.}~\bibnamefont {{Graves}}},
  \bibinfo {author} {\bibfnamefont {C.}~\bibnamefont {{Gurl}}}, \bibinfo
  {author} {\bibfnamefont {W.}~\bibnamefont {{Guttenfelder}}}, \bibinfo
  {author} {\bibfnamefont {C.}~\bibnamefont {{Ham}}}, \bibinfo {author}
  {\bibfnamefont {J.}~\bibnamefont {{Harrison}}}, \bibinfo {author}
  {\bibfnamefont {D.}~\bibnamefont {{Harting}}}, \bibinfo {author}
  {\bibfnamefont {E.}~\bibnamefont {{Havlickova}}}, \bibinfo {author}
  {\bibfnamefont {J.}~\bibnamefont {{Hawke}}}, \bibinfo {author} {\bibfnamefont
  {N.}~\bibnamefont {{Hawkes}}}, \bibinfo {author} {\bibfnamefont
  {T.}~\bibnamefont {{Hender}}}, \bibinfo {author} {\bibfnamefont
  {S.}~\bibnamefont {{Henderson}}}, \bibinfo {author} {\bibfnamefont
  {E.}~\bibnamefont {{Highcock}}}, \bibinfo {author} {\bibfnamefont
  {J.}~\bibnamefont {{Hillesheim}}}, \bibinfo {author} {\bibfnamefont
  {B.}~\bibnamefont {{Hnat}}}, \bibinfo {author} {\bibfnamefont
  {J.}~\bibnamefont {{Holgate}}}, \bibinfo {author} {\bibfnamefont
  {J.}~\bibnamefont {{Horacek}}}, \bibinfo {author} {\bibfnamefont
  {J.}~\bibnamefont {{Howard}}}, \bibinfo {author} {\bibfnamefont
  {B.}~\bibnamefont {{Huang}}}, \bibinfo {author} {\bibfnamefont
  {K.}~\bibnamefont {{Imada}}}, \bibinfo {author} {\bibfnamefont
  {O.}~\bibnamefont {{Jones}}}, \bibinfo {author} {\bibfnamefont
  {S.}~\bibnamefont {{Kaye}}}, \bibinfo {author} {\bibfnamefont
  {D.}~\bibnamefont {{Keeling}}}, \bibinfo {author} {\bibfnamefont
  {A.}~\bibnamefont {{Kirk}}}, \bibinfo {author} {\bibfnamefont
  {I.}~\bibnamefont {{Klimek}}}, \bibinfo {author} {\bibfnamefont
  {M.}~\bibnamefont {{Kocan}}}, \bibinfo {author} {\bibfnamefont
  {H.}~\bibnamefont {{Leggate}}}, \bibinfo {author} {\bibfnamefont
  {M.}~\bibnamefont {{Lilley}}}, \bibinfo {author} {\bibfnamefont
  {B.}~\bibnamefont {{Lipschultz}}}, \bibinfo {author} {\bibfnamefont
  {S.}~\bibnamefont {{Lisgo}}}, \bibinfo {author} {\bibfnamefont {Y.~Q.}\
  \bibnamefont {{Liu}}}, \bibinfo {author} {\bibfnamefont {B.}~\bibnamefont
  {{Lloyd}}}, \bibinfo {author} {\bibfnamefont {B.}~\bibnamefont
  {{Lomanowski}}}, \bibinfo {author} {\bibfnamefont {I.}~\bibnamefont
  {{Lupelli}}}, \bibinfo {author} {\bibfnamefont {G.}~\bibnamefont
  {{Maddison}}}, \bibinfo {author} {\bibfnamefont {J.}~\bibnamefont
  {{Mailloux}}}, \bibinfo {author} {\bibfnamefont {R.}~\bibnamefont
  {{Martin}}}, \bibinfo {author} {\bibfnamefont {G.}~\bibnamefont {{McArdle}}},
  \bibinfo {author} {\bibfnamefont {K.}~\bibnamefont {{McClements}}}, \bibinfo
  {author} {\bibfnamefont {B.}~\bibnamefont {{McMillan}}}, \bibinfo {author}
  {\bibfnamefont {A.}~\bibnamefont {{Meakins}}}, \bibinfo {author}
  {\bibfnamefont {H.}~\bibnamefont {{Meyer}}}, \bibinfo {author} {\bibfnamefont
  {C.}~\bibnamefont {{Michael}}}, \bibinfo {author} {\bibfnamefont
  {F.}~\bibnamefont {{Militello}}}, \bibinfo {author} {\bibfnamefont
  {J.}~\bibnamefont {{Milnes}}}, \bibinfo {author} {\bibfnamefont {A.~W.}\
  \bibnamefont {{Morris}}}, \bibinfo {author} {\bibfnamefont {G.}~\bibnamefont
  {{Motojima}}}, \bibinfo {author} {\bibfnamefont {D.}~\bibnamefont {{Muir}}},
  \bibinfo {author} {\bibfnamefont {E.}~\bibnamefont {{Nardon}}}, \bibinfo
  {author} {\bibfnamefont {V.}~\bibnamefont {{Naulin}}}, \bibinfo {author}
  {\bibfnamefont {G.}~\bibnamefont {{Naylor}}}, \bibinfo {author}
  {\bibfnamefont {A.}~\bibnamefont {{Nielsen}}}, \bibinfo {author}
  {\bibfnamefont {M.}~\bibnamefont {{O'Brien}}}, \bibinfo {author}
  {\bibfnamefont {T.}~\bibnamefont {{O'Gorman}}}, \bibinfo {author}
  {\bibfnamefont {Y.}~\bibnamefont {{Ono}}}, \bibinfo {author} {\bibfnamefont
  {H.}~\bibnamefont {{Oliver}}}, \bibinfo {author} {\bibfnamefont
  {S.}~\bibnamefont {{Pamela}}}, \bibinfo {author} {\bibfnamefont
  {L.}~\bibnamefont {{Pangione}}}, \bibinfo {author} {\bibfnamefont
  {F.}~\bibnamefont {{Parra}}}, \bibinfo {author} {\bibfnamefont
  {A.}~\bibnamefont {{Patel}}}, \bibinfo {author} {\bibfnamefont
  {W.}~\bibnamefont {{Peebles}}}, \bibinfo {author} {\bibfnamefont
  {M.}~\bibnamefont {{Peng}}}, \bibinfo {author} {\bibfnamefont
  {R.}~\bibnamefont {{Perez}}}, \bibinfo {author} {\bibfnamefont
  {S.}~\bibnamefont {{Pinches}}}, \bibinfo {author} {\bibfnamefont
  {L.}~\bibnamefont {{Piron}}}, \bibinfo {author} {\bibfnamefont
  {M.}~\bibnamefont {{Podesta}}}, \bibinfo {author} {\bibfnamefont
  {M.}~\bibnamefont {{Price}}}, \bibinfo {author} {\bibfnamefont
  {M.}~\bibnamefont {{Reinke}}}, \bibinfo {author} {\bibfnamefont
  {Y.}~\bibnamefont {{Ren}}}, \bibinfo {author} {\bibfnamefont
  {C.}~\bibnamefont {{Roach}}}, \bibinfo {author} {\bibfnamefont
  {J.}~\bibnamefont {{Robinson}}}, \bibinfo {author} {\bibfnamefont
  {M.}~\bibnamefont {{Romanelli}}}, \bibinfo {author} {\bibfnamefont
  {V.}~\bibnamefont {{Rozhansky}}}, \bibinfo {author} {\bibfnamefont
  {S.}~\bibnamefont {{Saarelma}}}, \bibinfo {author} {\bibfnamefont
  {S.}~\bibnamefont {{Sangaroon}}}, \bibinfo {author} {\bibfnamefont
  {A.}~\bibnamefont {{Saveliev}}}, \bibinfo {author} {\bibfnamefont
  {R.}~\bibnamefont {{Scannell}}}, \bibinfo {author} {\bibfnamefont
  {A.}~\bibnamefont {{Schekochihin}}}, \bibinfo {author} {\bibfnamefont
  {S.}~\bibnamefont {{Sharapov}}}, \bibinfo {author} {\bibfnamefont
  {R.}~\bibnamefont {{Sharples}}}, \bibinfo {author} {\bibfnamefont
  {V.}~\bibnamefont {{Shevchenko}}}, \bibinfo {author} {\bibfnamefont
  {S.}~\bibnamefont {{Silburn}}}, \bibinfo {author} {\bibfnamefont
  {J.}~\bibnamefont {{Simpson}}}, \bibinfo {author} {\bibfnamefont
  {J.}~\bibnamefont {{Storrs}}}, \bibinfo {author} {\bibfnamefont
  {Y.}~\bibnamefont {{Takase}}}, \bibinfo {author} {\bibfnamefont
  {H.}~\bibnamefont {{Tanabe}}}, \bibinfo {author} {\bibfnamefont
  {H.}~\bibnamefont {{Tanaka}}}, \bibinfo {author} {\bibfnamefont
  {D.}~\bibnamefont {{Taylor}}}, \bibinfo {author} {\bibfnamefont
  {G.}~\bibnamefont {{Taylor}}}, \bibinfo {author} {\bibfnamefont
  {D.}~\bibnamefont {{Thomas}}}, \bibinfo {author} {\bibfnamefont
  {N.}~\bibnamefont {{Thomas-Davies}}}, \bibinfo {author} {\bibfnamefont
  {A.}~\bibnamefont {{Thornton}}}, \bibinfo {author} {\bibfnamefont
  {M.}~\bibnamefont {{Turnyanskiy}}}, \bibinfo {author} {\bibfnamefont
  {M.}~\bibnamefont {{Valovic}}}, \bibinfo {author} {\bibfnamefont
  {R.}~\bibnamefont {{Vann}}}, \bibinfo {author} {\bibfnamefont
  {N.}~\bibnamefont {{Walkden}}}, \bibinfo {author} {\bibfnamefont
  {H.}~\bibnamefont {{Wilson}}}, \bibinfo {author} {\bibfnamefont
  {F.}~\bibnamefont {{van Wyk}}}, \bibinfo {author} {\bibfnamefont
  {T.}~\bibnamefont {{Yamada}}}, \bibinfo {author} {\bibfnamefont
  {S.}~\bibnamefont {{Zoletnik}}}, \bibinfo {author} {\bibnamefont {{MAST}}}, \
  and\ \bibinfo {author} {\bibnamefont {{MAST Upgrade Teams}}},\ }\bibfield
  {title} {\enquote {\bibinfo {title} {{Overview of MAST results}},}\ }\href
  {\doibase 10.1088/0029-5515/55/10/104008} {\bibfield  {journal} {\bibinfo
  {journal} {Nuclear Fusion}\ }\textbf {\bibinfo {volume} {55}},\ \bibinfo
  {eid} {104008} (\bibinfo {year} {2015})}\BibitemShut {NoStop}%
\bibitem [{\citenamefont {{Fishpool}}\ \emph {et~al.}(2013)\citenamefont
  {{Fishpool}}, \citenamefont {{Canik}}, \citenamefont {{Cunningham}},
  \citenamefont {{Harrison}}, \citenamefont {{Katramados}}, \citenamefont
  {{Kirk}}, \citenamefont {{Kovari}}, \citenamefont {{Meyer}}, \citenamefont
  {{Scannell}},\ and\ \citenamefont {{MAST-upgrade Team}}}]{Fishpool2013}%
  \BibitemOpen
  \bibfield  {author} {\bibinfo {author} {\bibfnamefont {G.}~\bibnamefont
  {{Fishpool}}}, \bibinfo {author} {\bibfnamefont {J.}~\bibnamefont {{Canik}}},
  \bibinfo {author} {\bibfnamefont {G.}~\bibnamefont {{Cunningham}}}, \bibinfo
  {author} {\bibfnamefont {J.}~\bibnamefont {{Harrison}}}, \bibinfo {author}
  {\bibfnamefont {I.}~\bibnamefont {{Katramados}}}, \bibinfo {author}
  {\bibfnamefont {A.}~\bibnamefont {{Kirk}}}, \bibinfo {author} {\bibfnamefont
  {M.}~\bibnamefont {{Kovari}}}, \bibinfo {author} {\bibfnamefont
  {H.}~\bibnamefont {{Meyer}}}, \bibinfo {author} {\bibfnamefont
  {R.}~\bibnamefont {{Scannell}}}, \ and\ \bibinfo {author} {\bibnamefont
  {{MAST-upgrade Team}}},\ }\bibfield  {title} {\enquote {\bibinfo {title}
  {{MAST-upgrade divertor facility and assessing performance of long-legged
  divertors}},}\ }\href {\doibase 10.1016/j.jnucmat.2013.01.067} {\bibfield
  {journal} {\bibinfo  {journal} {Journal of Nuclear Materials}\ }\textbf
  {\bibinfo {volume} {438}},\ \bibinfo {pages} {S356--S359} (\bibinfo {year}
  {2013})},\ \Eprint {http://arxiv.org/abs/1306.6774} {arXiv:1306.6774
  [physics.plasm-ph]} \BibitemShut {NoStop}%
\bibitem [{\citenamefont {{Strait}}\ \emph {et~al.}(1995)\citenamefont
  {{Strait}}, \citenamefont {{Lao}}, \citenamefont {{Mauel}}, \citenamefont
  {{Rice}}, \citenamefont {{Taylor}}, \citenamefont {{Burrell}}, \citenamefont
  {{Chu}}, \citenamefont {{Lazarus}}, \citenamefont {{Osborne}}, \citenamefont
  {{Thompson}},\ and\ \citenamefont {{Turnbull}}}]{DiiiD1995}%
  \BibitemOpen
  \bibfield  {author} {\bibinfo {author} {\bibfnamefont {E.~J.}\ \bibnamefont
  {{Strait}}}, \bibinfo {author} {\bibfnamefont {L.~L.}\ \bibnamefont {{Lao}}},
  \bibinfo {author} {\bibfnamefont {M.~E.}\ \bibnamefont {{Mauel}}}, \bibinfo
  {author} {\bibfnamefont {B.~W.}\ \bibnamefont {{Rice}}}, \bibinfo {author}
  {\bibfnamefont {T.~S.}\ \bibnamefont {{Taylor}}}, \bibinfo {author}
  {\bibfnamefont {K.~H.}\ \bibnamefont {{Burrell}}}, \bibinfo {author}
  {\bibfnamefont {M.~S.}\ \bibnamefont {{Chu}}}, \bibinfo {author}
  {\bibfnamefont {E.~A.}\ \bibnamefont {{Lazarus}}}, \bibinfo {author}
  {\bibfnamefont {T.~H.}\ \bibnamefont {{Osborne}}}, \bibinfo {author}
  {\bibfnamefont {S.~J.}\ \bibnamefont {{Thompson}}}, \ and\ \bibinfo {author}
  {\bibfnamefont {A.~D.}\ \bibnamefont {{Turnbull}}},\ }\bibfield  {title}
  {\enquote {\bibinfo {title} {{Enhanced confinement and stability in DIII-D
  discharges with reversed magnetic shear}},}\ }\href {\doibase
  10.1103/PhysRevLett.75.4421} {\bibfield  {journal} {\bibinfo  {journal}
  {Phys.Rev.L}\ }\textbf {\bibinfo {volume} {75}},\ \bibinfo {pages}
  {4421--4424} (\bibinfo {year} {1995})}\BibitemShut {NoStop}%
\bibitem [{\citenamefont {{Reimerdes}}\ \emph {et~al.}(2022)\citenamefont
  {{Reimerdes}}, \citenamefont {{Agostini}}, \citenamefont {{Alessi}},
  \citenamefont {{Alberti}}, \citenamefont {{Andrebe}}, \citenamefont
  {{Arnichand}}, \citenamefont {{Balbin}}, \citenamefont {{Bagnato}},
  \citenamefont {{Baquero-Ruiz}}, \citenamefont {{Bernert}}, \citenamefont
  {{Bin}}, \citenamefont {{Blanchard}}, \citenamefont {{Blanken}},
  \citenamefont {{Boedo}}, \citenamefont {{Brida}}, \citenamefont {{Brunner}},
  \citenamefont {{Bogar}}, \citenamefont {{Bogar}}, \citenamefont
  {{Bolzonella}}, \citenamefont {{Bombarda}}, \citenamefont {{Bouquey}},
  \citenamefont {{Bowman}}, \citenamefont {{Brunetti}}, \citenamefont
  {{Buermans}}, \citenamefont {{Bufferand}}, \citenamefont {{Calacci}},
  \citenamefont {{Camenen}}, \citenamefont {{Carli}}, \citenamefont
  {{Carnevale}}, \citenamefont {{Carpanese}}, \citenamefont {{Causa}},
  \citenamefont {{Cavalier}}, \citenamefont {{Cavedon}}, \citenamefont
  {{Cazabonne}}, \citenamefont {{Cerovsky}}, \citenamefont {{Chandra}},
  \citenamefont {{Chandrarajan Jayalekshmi}}, \citenamefont {{Chella{\"\i}}},
  \citenamefont {{Chmielewski}}, \citenamefont {{Choi}}, \citenamefont
  {{Ciraolo}}, \citenamefont {{Classen}}, \citenamefont {{Coda}}, \citenamefont
  {{Colandrea}}, \citenamefont {{Dal Molin}}, \citenamefont {{David}},
  \citenamefont {{de Baar}}, \citenamefont {{Decker}}, \citenamefont
  {{Dekeyser}}, \citenamefont {{de Oliveira}}, \citenamefont {{Douai}},
  \citenamefont {{Dreval}}, \citenamefont {{Dunne}}, \citenamefont {{Duval}},
  \citenamefont {{Elmore}}, \citenamefont {{Embreus}}, \citenamefont
  {{Eriksson}}, \citenamefont {{Faitsch}}, \citenamefont {{Falchetto}},
  \citenamefont {{Farnik}}, \citenamefont {{Fasoli}}, \citenamefont
  {{Fedorczak}}, \citenamefont {{Felici}}, \citenamefont {{F{\'e}vrier}},
  \citenamefont {{Ficker}}, \citenamefont {{Fil}}, \citenamefont {{Fontana}},
  \citenamefont {{Fransson}}, \citenamefont {{Frassinetti}}, \citenamefont
  {{Furno}}, \citenamefont {{Gahle}}, \citenamefont {{Galassi}}, \citenamefont
  {{Galazka}}, \citenamefont {{Galperti}}, \citenamefont {{Garavaglia}},
  \citenamefont {{Garcia-Munoz}}, \citenamefont {{Geiger}}, \citenamefont
  {{Giacomin}}, \citenamefont {{Giruzzi}}, \citenamefont {{Gobbin}},
  \citenamefont {{Golfinopoulos}}, \citenamefont {{Goodman}}, \citenamefont
  {{Gorno}}, \citenamefont {{Granucci}}, \citenamefont {{Graves}},
  \citenamefont {{Griener}}, \citenamefont {{Gruca}}, \citenamefont
  {{Gyergyek}}, \citenamefont {{Haelterman}}, \citenamefont {{Hakola}},
  \citenamefont {{Han}}, \citenamefont {{Happel}}, \citenamefont {{Harrer}},
  \citenamefont {{Harrison}}, \citenamefont {{Henderson}}, \citenamefont
  {{Hogeweij}}, \citenamefont {{Hogge}}, \citenamefont {{Hoppe}}, \citenamefont
  {{Horacek}}, \citenamefont {{Huang}}, \citenamefont {{Iantchenko}},
  \citenamefont {{Innocente}}, \citenamefont {{Insulander Bj{\"o}rk}},
  \citenamefont {{Ionita-Schrittweiser}}, \citenamefont {{Isliker}},
  \citenamefont {{Jardin}}, \citenamefont {{Jaspers}}, \citenamefont
  {{Karimov}}, \citenamefont {{Karpushov}}, \citenamefont {{Kazakov}},
  \citenamefont {{Komm}}, \citenamefont {{Kong}}, \citenamefont {{Kovacic}},
  \citenamefont {{Krutkin}}, \citenamefont {{Kudlacek}}, \citenamefont
  {{Kumar}}, \citenamefont {{Kwiatkowski}}, \citenamefont {{Labit}},
  \citenamefont {{Laguardia}}, \citenamefont {{Lammers}}, \citenamefont
  {{Laribi}}, \citenamefont {{Laszynska}}, \citenamefont {{Lazaros}},
  \citenamefont {{Linder}}, \citenamefont {{Linehan}}, \citenamefont
  {{Lipschultz}}, \citenamefont {{Llobet}}, \citenamefont {{Loizu}},
  \citenamefont {{Lunt}}, \citenamefont {{Macusova}}, \citenamefont
  {{Marandet}}, \citenamefont {{Maraschek}}, \citenamefont {{Marceca}},
  \citenamefont {{Marchetto}}, \citenamefont {{Marchioni}}, \citenamefont
  {{Marmar}}, \citenamefont {{Martin}}, \citenamefont {{Martinelli}},
  \citenamefont {{Matos}}, \citenamefont {{Maurizio}}, \citenamefont
  {{Mayoral}}, \citenamefont {{Mazon}}, \citenamefont {{Menkovski}},
  \citenamefont {{Merle}}, \citenamefont {{Merlo}}, \citenamefont {{Meyer}},
  \citenamefont {{Mikszuta-Michalik}}, \citenamefont {{Molina Cabrera}},
  \citenamefont {{Morales}}, \citenamefont {{Moret}}, \citenamefont {{Moro}},
  \citenamefont {{Moulton}}, \citenamefont {{Muhammed}}, \citenamefont
  {{Myatra}}, \citenamefont {{Mykytchuk}}, \citenamefont {{Napoli}},
  \citenamefont {{Nem}}, \citenamefont {{Nielsen}}, \citenamefont {{Nocente}},
  \citenamefont {{Nowak}}, \citenamefont {{Offeddu}}, \citenamefont {{Olsen}},
  \citenamefont {{Orsitto}}, \citenamefont {{Pan}}, \citenamefont {{Papp}},
  \citenamefont {{Pau}}, \citenamefont {{Perek}}, \citenamefont {{Pesamosca}},
  \citenamefont {{Peysson}}, \citenamefont {{Pigatto}}, \citenamefont
  {{Piron}}, \citenamefont {{Poradzinski}}, \citenamefont {{Porte}},
  \citenamefont {{P{\"u}tterich}}, \citenamefont {{Rabinski}}, \citenamefont
  {{Raj}}, \citenamefont {{Rasmussen}}, \citenamefont {{Ratt{\'a}}},
  \citenamefont {{Ravensbergen}}, \citenamefont {{Ricci}}, \citenamefont
  {{Ricci}}, \citenamefont {{Rispoli}}, \citenamefont {{Riva}}, \citenamefont
  {{Rivero-Rodriguez}}, \citenamefont {{Salewski}}, \citenamefont {{Sauter}},
  \citenamefont {{Schmidt}}, \citenamefont {{Schrittweiser}}, \citenamefont
  {{Sharapov}}, \citenamefont {{Sheikh}}, \citenamefont {{Sieglin}},
  \citenamefont {{Silva}}, \citenamefont {{Smolders}}, \citenamefont
  {{Snicker}}, \citenamefont {{Sozzi}}, \citenamefont {{Spolaore}},
  \citenamefont {{Stagni}}, \citenamefont {{Stipani}}, \citenamefont {{Sun}},
  \citenamefont {{Tala}}, \citenamefont {{Tamain}}, \citenamefont {{Tanaka}},
  \citenamefont {{Tema Biwole}}, \citenamefont {{Terranova}}, \citenamefont
  {{Terry}}, \citenamefont {{Testa}}, \citenamefont {{Theiler}}, \citenamefont
  {{Thornton}}, \citenamefont {{Thrys{\o}e}}, \citenamefont {{Torreblanca}},
  \citenamefont {{Tsui}}, \citenamefont {{Vaccaro}}, \citenamefont {{Vallar}},
  \citenamefont {{van Berkel}}, \citenamefont {{Van Eester}}, \citenamefont
  {{van Kampen}}, \citenamefont {{Van Mulders}}, \citenamefont {{Verhaegh}},
  \citenamefont {{Verhaeghe}}, \citenamefont {{Vianello}}, \citenamefont
  {{Villone}}, \citenamefont {{Viezzer}}, \citenamefont {{Vincent}},
  \citenamefont {{Voitsekhovitch}}, \citenamefont {{Vu}}, \citenamefont
  {{Walkden}}, \citenamefont {{Wauters}}, \citenamefont {{Weisen}},
  \citenamefont {{Wendler}}, \citenamefont {{Wensing}}, \citenamefont
  {{Widmer}}, \citenamefont {{Wiesen}}, \citenamefont {{Wischmeier}},
  \citenamefont {{Wijkamp}}, \citenamefont {{W{\"u}nderlich}}, \citenamefont
  {{W{\"u}thrich}}, \citenamefont {{Yanovskiy}}, \citenamefont {{Zebrowski}},\
  and\ \citenamefont {{EUROfusion MST1 Team}}}]{2022NucFu..62d2018R}%
  \BibitemOpen
  \bibfield  {author} {\bibinfo {author} {\bibfnamefont {H.}~\bibnamefont
  {{Reimerdes}}}, \bibinfo {author} {\bibfnamefont {M.}~\bibnamefont
  {{Agostini}}}, \bibinfo {author} {\bibfnamefont {E.}~\bibnamefont
  {{Alessi}}}, \bibinfo {author} {\bibfnamefont {S.}~\bibnamefont {{Alberti}}},
  \bibinfo {author} {\bibfnamefont {Y.}~\bibnamefont {{Andrebe}}}, \bibinfo
  {author} {\bibfnamefont {H.}~\bibnamefont {{Arnichand}}}, \bibinfo {author}
  {\bibfnamefont {J.}~\bibnamefont {{Balbin}}}, \bibinfo {author}
  {\bibfnamefont {F.}~\bibnamefont {{Bagnato}}}, \bibinfo {author}
  {\bibfnamefont {M.}~\bibnamefont {{Baquero-Ruiz}}}, \bibinfo {author}
  {\bibfnamefont {M.}~\bibnamefont {{Bernert}}}, \bibinfo {author}
  {\bibfnamefont {W.}~\bibnamefont {{Bin}}}, \bibinfo {author} {\bibfnamefont
  {P.}~\bibnamefont {{Blanchard}}}, \bibinfo {author} {\bibfnamefont {T.~C.}\
  \bibnamefont {{Blanken}}}, \bibinfo {author} {\bibfnamefont {J.~A.}\
  \bibnamefont {{Boedo}}}, \bibinfo {author} {\bibfnamefont {D.}~\bibnamefont
  {{Brida}}}, \bibinfo {author} {\bibfnamefont {S.}~\bibnamefont {{Brunner}}},
  \bibinfo {author} {\bibfnamefont {C.}~\bibnamefont {{Bogar}}}, \bibinfo
  {author} {\bibfnamefont {O.}~\bibnamefont {{Bogar}}}, \bibinfo {author}
  {\bibfnamefont {T.}~\bibnamefont {{Bolzonella}}}, \bibinfo {author}
  {\bibfnamefont {F.}~\bibnamefont {{Bombarda}}}, \bibinfo {author}
  {\bibfnamefont {F.}~\bibnamefont {{Bouquey}}}, \bibinfo {author}
  {\bibfnamefont {C.}~\bibnamefont {{Bowman}}}, \bibinfo {author}
  {\bibfnamefont {D.}~\bibnamefont {{Brunetti}}}, \bibinfo {author}
  {\bibfnamefont {J.}~\bibnamefont {{Buermans}}}, \bibinfo {author}
  {\bibfnamefont {H.}~\bibnamefont {{Bufferand}}}, \bibinfo {author}
  {\bibfnamefont {L.}~\bibnamefont {{Calacci}}}, \bibinfo {author}
  {\bibfnamefont {Y.}~\bibnamefont {{Camenen}}}, \bibinfo {author}
  {\bibfnamefont {S.}~\bibnamefont {{Carli}}}, \bibinfo {author} {\bibfnamefont
  {D.}~\bibnamefont {{Carnevale}}}, \bibinfo {author} {\bibfnamefont
  {F.}~\bibnamefont {{Carpanese}}}, \bibinfo {author} {\bibfnamefont
  {F.}~\bibnamefont {{Causa}}}, \bibinfo {author} {\bibfnamefont
  {J.}~\bibnamefont {{Cavalier}}}, \bibinfo {author} {\bibfnamefont
  {M.}~\bibnamefont {{Cavedon}}}, \bibinfo {author} {\bibfnamefont {J.~A.}\
  \bibnamefont {{Cazabonne}}}, \bibinfo {author} {\bibfnamefont
  {J.}~\bibnamefont {{Cerovsky}}}, \bibinfo {author} {\bibfnamefont
  {R.}~\bibnamefont {{Chandra}}}, \bibinfo {author} {\bibfnamefont
  {A.}~\bibnamefont {{Chandrarajan Jayalekshmi}}}, \bibinfo {author}
  {\bibfnamefont {O.}~\bibnamefont {{Chella{\"\i}}}}, \bibinfo {author}
  {\bibfnamefont {P.}~\bibnamefont {{Chmielewski}}}, \bibinfo {author}
  {\bibfnamefont {D.}~\bibnamefont {{Choi}}}, \bibinfo {author} {\bibfnamefont
  {G.}~\bibnamefont {{Ciraolo}}}, \bibinfo {author} {\bibfnamefont {I.~G.~J.}\
  \bibnamefont {{Classen}}}, \bibinfo {author} {\bibfnamefont {S.}~\bibnamefont
  {{Coda}}}, \bibinfo {author} {\bibfnamefont {C.}~\bibnamefont {{Colandrea}}},
  \bibinfo {author} {\bibfnamefont {A.}~\bibnamefont {{Dal Molin}}}, \bibinfo
  {author} {\bibfnamefont {P.}~\bibnamefont {{David}}}, \bibinfo {author}
  {\bibfnamefont {M.~R.}\ \bibnamefont {{de Baar}}}, \bibinfo {author}
  {\bibfnamefont {J.}~\bibnamefont {{Decker}}}, \bibinfo {author}
  {\bibfnamefont {W.}~\bibnamefont {{Dekeyser}}}, \bibinfo {author}
  {\bibfnamefont {H.}~\bibnamefont {{de Oliveira}}}, \bibinfo {author}
  {\bibfnamefont {D.}~\bibnamefont {{Douai}}}, \bibinfo {author} {\bibfnamefont
  {M.}~\bibnamefont {{Dreval}}}, \bibinfo {author} {\bibfnamefont {M.~G.}\
  \bibnamefont {{Dunne}}}, \bibinfo {author} {\bibfnamefont {B.~P.}\
  \bibnamefont {{Duval}}}, \bibinfo {author} {\bibfnamefont {S.}~\bibnamefont
  {{Elmore}}}, \bibinfo {author} {\bibfnamefont {O.}~\bibnamefont {{Embreus}}},
  \bibinfo {author} {\bibfnamefont {F.}~\bibnamefont {{Eriksson}}}, \bibinfo
  {author} {\bibfnamefont {M.}~\bibnamefont {{Faitsch}}}, \bibinfo {author}
  {\bibfnamefont {G.}~\bibnamefont {{Falchetto}}}, \bibinfo {author}
  {\bibfnamefont {M.}~\bibnamefont {{Farnik}}}, \bibinfo {author}
  {\bibfnamefont {A.}~\bibnamefont {{Fasoli}}}, \bibinfo {author}
  {\bibfnamefont {N.}~\bibnamefont {{Fedorczak}}}, \bibinfo {author}
  {\bibfnamefont {F.}~\bibnamefont {{Felici}}}, \bibinfo {author}
  {\bibfnamefont {O.}~\bibnamefont {{F{\'e}vrier}}}, \bibinfo {author}
  {\bibfnamefont {O.}~\bibnamefont {{Ficker}}}, \bibinfo {author}
  {\bibfnamefont {A.}~\bibnamefont {{Fil}}}, \bibinfo {author} {\bibfnamefont
  {M.}~\bibnamefont {{Fontana}}}, \bibinfo {author} {\bibfnamefont
  {E.}~\bibnamefont {{Fransson}}}, \bibinfo {author} {\bibfnamefont
  {L.}~\bibnamefont {{Frassinetti}}}, \bibinfo {author} {\bibfnamefont
  {I.}~\bibnamefont {{Furno}}}, \bibinfo {author} {\bibfnamefont {D.~S.}\
  \bibnamefont {{Gahle}}}, \bibinfo {author} {\bibfnamefont {D.}~\bibnamefont
  {{Galassi}}}, \bibinfo {author} {\bibfnamefont {K.}~\bibnamefont
  {{Galazka}}}, \bibinfo {author} {\bibfnamefont {C.}~\bibnamefont
  {{Galperti}}}, \bibinfo {author} {\bibfnamefont {S.}~\bibnamefont
  {{Garavaglia}}}, \bibinfo {author} {\bibfnamefont {M.}~\bibnamefont
  {{Garcia-Munoz}}}, \bibinfo {author} {\bibfnamefont {B.}~\bibnamefont
  {{Geiger}}}, \bibinfo {author} {\bibfnamefont {M.}~\bibnamefont
  {{Giacomin}}}, \bibinfo {author} {\bibfnamefont {G.}~\bibnamefont
  {{Giruzzi}}}, \bibinfo {author} {\bibfnamefont {M.}~\bibnamefont {{Gobbin}}},
  \bibinfo {author} {\bibfnamefont {T.}~\bibnamefont {{Golfinopoulos}}},
  \bibinfo {author} {\bibfnamefont {T.}~\bibnamefont {{Goodman}}}, \bibinfo
  {author} {\bibfnamefont {S.}~\bibnamefont {{Gorno}}}, \bibinfo {author}
  {\bibfnamefont {G.}~\bibnamefont {{Granucci}}}, \bibinfo {author}
  {\bibfnamefont {J.~P.}\ \bibnamefont {{Graves}}}, \bibinfo {author}
  {\bibfnamefont {M.}~\bibnamefont {{Griener}}}, \bibinfo {author}
  {\bibfnamefont {M.}~\bibnamefont {{Gruca}}}, \bibinfo {author} {\bibfnamefont
  {T.}~\bibnamefont {{Gyergyek}}}, \bibinfo {author} {\bibfnamefont
  {R.}~\bibnamefont {{Haelterman}}}, \bibinfo {author} {\bibfnamefont
  {A.}~\bibnamefont {{Hakola}}}, \bibinfo {author} {\bibfnamefont
  {W.}~\bibnamefont {{Han}}}, \bibinfo {author} {\bibfnamefont
  {T.}~\bibnamefont {{Happel}}}, \bibinfo {author} {\bibfnamefont
  {G.}~\bibnamefont {{Harrer}}}, \bibinfo {author} {\bibfnamefont {J.~R.}\
  \bibnamefont {{Harrison}}}, \bibinfo {author} {\bibfnamefont
  {S.}~\bibnamefont {{Henderson}}}, \bibinfo {author} {\bibfnamefont
  {G.~M.~D.}\ \bibnamefont {{Hogeweij}}}, \bibinfo {author} {\bibfnamefont
  {J.~P.}\ \bibnamefont {{Hogge}}}, \bibinfo {author} {\bibfnamefont
  {M.}~\bibnamefont {{Hoppe}}}, \bibinfo {author} {\bibfnamefont
  {J.}~\bibnamefont {{Horacek}}}, \bibinfo {author} {\bibfnamefont
  {Z.}~\bibnamefont {{Huang}}}, \bibinfo {author} {\bibfnamefont
  {A.}~\bibnamefont {{Iantchenko}}}, \bibinfo {author} {\bibfnamefont
  {P.}~\bibnamefont {{Innocente}}}, \bibinfo {author} {\bibfnamefont
  {K.}~\bibnamefont {{Insulander Bj{\"o}rk}}}, \bibinfo {author} {\bibfnamefont
  {C.}~\bibnamefont {{Ionita-Schrittweiser}}}, \bibinfo {author} {\bibfnamefont
  {H.}~\bibnamefont {{Isliker}}}, \bibinfo {author} {\bibfnamefont
  {A.}~\bibnamefont {{Jardin}}}, \bibinfo {author} {\bibfnamefont {R.~J.~E.}\
  \bibnamefont {{Jaspers}}}, \bibinfo {author} {\bibfnamefont {R.}~\bibnamefont
  {{Karimov}}}, \bibinfo {author} {\bibfnamefont {A.~N.}\ \bibnamefont
  {{Karpushov}}}, \bibinfo {author} {\bibfnamefont {Y.}~\bibnamefont
  {{Kazakov}}}, \bibinfo {author} {\bibfnamefont {M.}~\bibnamefont {{Komm}}},
  \bibinfo {author} {\bibfnamefont {M.}~\bibnamefont {{Kong}}}, \bibinfo
  {author} {\bibfnamefont {J.}~\bibnamefont {{Kovacic}}}, \bibinfo {author}
  {\bibfnamefont {O.}~\bibnamefont {{Krutkin}}}, \bibinfo {author}
  {\bibfnamefont {O.}~\bibnamefont {{Kudlacek}}}, \bibinfo {author}
  {\bibfnamefont {U.}~\bibnamefont {{Kumar}}}, \bibinfo {author} {\bibfnamefont
  {R.}~\bibnamefont {{Kwiatkowski}}}, \bibinfo {author} {\bibfnamefont
  {B.}~\bibnamefont {{Labit}}}, \bibinfo {author} {\bibfnamefont
  {L.}~\bibnamefont {{Laguardia}}}, \bibinfo {author} {\bibfnamefont {J.~T.}\
  \bibnamefont {{Lammers}}}, \bibinfo {author} {\bibfnamefont {E.}~\bibnamefont
  {{Laribi}}}, \bibinfo {author} {\bibfnamefont {E.}~\bibnamefont
  {{Laszynska}}}, \bibinfo {author} {\bibfnamefont {A.}~\bibnamefont
  {{Lazaros}}}, \bibinfo {author} {\bibfnamefont {O.}~\bibnamefont {{Linder}}},
  \bibinfo {author} {\bibfnamefont {B.}~\bibnamefont {{Linehan}}}, \bibinfo
  {author} {\bibfnamefont {B.}~\bibnamefont {{Lipschultz}}}, \bibinfo {author}
  {\bibfnamefont {X.}~\bibnamefont {{Llobet}}}, \bibinfo {author}
  {\bibfnamefont {J.}~\bibnamefont {{Loizu}}}, \bibinfo {author} {\bibfnamefont
  {T.}~\bibnamefont {{Lunt}}}, \bibinfo {author} {\bibfnamefont
  {E.}~\bibnamefont {{Macusova}}}, \bibinfo {author} {\bibfnamefont
  {Y.}~\bibnamefont {{Marandet}}}, \bibinfo {author} {\bibfnamefont
  {M.}~\bibnamefont {{Maraschek}}}, \bibinfo {author} {\bibfnamefont
  {G.}~\bibnamefont {{Marceca}}}, \bibinfo {author} {\bibfnamefont
  {C.}~\bibnamefont {{Marchetto}}}, \bibinfo {author} {\bibfnamefont
  {S.}~\bibnamefont {{Marchioni}}}, \bibinfo {author} {\bibfnamefont {E.~S.}\
  \bibnamefont {{Marmar}}}, \bibinfo {author} {\bibfnamefont {Y.}~\bibnamefont
  {{Martin}}}, \bibinfo {author} {\bibfnamefont {L.}~\bibnamefont
  {{Martinelli}}}, \bibinfo {author} {\bibfnamefont {F.}~\bibnamefont
  {{Matos}}}, \bibinfo {author} {\bibfnamefont {R.}~\bibnamefont {{Maurizio}}},
  \bibinfo {author} {\bibfnamefont {M.~L.}\ \bibnamefont {{Mayoral}}}, \bibinfo
  {author} {\bibfnamefont {D.}~\bibnamefont {{Mazon}}}, \bibinfo {author}
  {\bibfnamefont {V.}~\bibnamefont {{Menkovski}}}, \bibinfo {author}
  {\bibfnamefont {A.}~\bibnamefont {{Merle}}}, \bibinfo {author} {\bibfnamefont
  {G.}~\bibnamefont {{Merlo}}}, \bibinfo {author} {\bibfnamefont
  {H.}~\bibnamefont {{Meyer}}}, \bibinfo {author} {\bibfnamefont
  {K.}~\bibnamefont {{Mikszuta-Michalik}}}, \bibinfo {author} {\bibfnamefont
  {P.~A.}\ \bibnamefont {{Molina Cabrera}}}, \bibinfo {author} {\bibfnamefont
  {J.}~\bibnamefont {{Morales}}}, \bibinfo {author} {\bibfnamefont {J.~M.}\
  \bibnamefont {{Moret}}}, \bibinfo {author} {\bibfnamefont {A.}~\bibnamefont
  {{Moro}}}, \bibinfo {author} {\bibfnamefont {D.}~\bibnamefont {{Moulton}}},
  \bibinfo {author} {\bibfnamefont {H.}~\bibnamefont {{Muhammed}}}, \bibinfo
  {author} {\bibfnamefont {O.}~\bibnamefont {{Myatra}}}, \bibinfo {author}
  {\bibfnamefont {D.}~\bibnamefont {{Mykytchuk}}}, \bibinfo {author}
  {\bibfnamefont {F.}~\bibnamefont {{Napoli}}}, \bibinfo {author}
  {\bibfnamefont {R.~D.}\ \bibnamefont {{Nem}}}, \bibinfo {author}
  {\bibfnamefont {A.~H.}\ \bibnamefont {{Nielsen}}}, \bibinfo {author}
  {\bibfnamefont {M.}~\bibnamefont {{Nocente}}}, \bibinfo {author}
  {\bibfnamefont {S.}~\bibnamefont {{Nowak}}}, \bibinfo {author} {\bibfnamefont
  {N.}~\bibnamefont {{Offeddu}}}, \bibinfo {author} {\bibfnamefont
  {J.}~\bibnamefont {{Olsen}}}, \bibinfo {author} {\bibfnamefont {F.~P.}\
  \bibnamefont {{Orsitto}}}, \bibinfo {author} {\bibfnamefont {O.}~\bibnamefont
  {{Pan}}}, \bibinfo {author} {\bibfnamefont {G.}~\bibnamefont {{Papp}}},
  \bibinfo {author} {\bibfnamefont {A.}~\bibnamefont {{Pau}}}, \bibinfo
  {author} {\bibfnamefont {A.}~\bibnamefont {{Perek}}}, \bibinfo {author}
  {\bibfnamefont {F.}~\bibnamefont {{Pesamosca}}}, \bibinfo {author}
  {\bibfnamefont {Y.}~\bibnamefont {{Peysson}}}, \bibinfo {author}
  {\bibfnamefont {L.}~\bibnamefont {{Pigatto}}}, \bibinfo {author}
  {\bibfnamefont {C.}~\bibnamefont {{Piron}}}, \bibinfo {author} {\bibfnamefont
  {M.}~\bibnamefont {{Poradzinski}}}, \bibinfo {author} {\bibfnamefont
  {L.}~\bibnamefont {{Porte}}}, \bibinfo {author} {\bibfnamefont
  {T.}~\bibnamefont {{P{\"u}tterich}}}, \bibinfo {author} {\bibfnamefont
  {M.}~\bibnamefont {{Rabinski}}}, \bibinfo {author} {\bibfnamefont
  {H.}~\bibnamefont {{Raj}}}, \bibinfo {author} {\bibfnamefont {J.~J.}\
  \bibnamefont {{Rasmussen}}}, \bibinfo {author} {\bibfnamefont {G.~A.}\
  \bibnamefont {{Ratt{\'a}}}}, \bibinfo {author} {\bibfnamefont
  {T.}~\bibnamefont {{Ravensbergen}}}, \bibinfo {author} {\bibfnamefont
  {D.}~\bibnamefont {{Ricci}}}, \bibinfo {author} {\bibfnamefont
  {P.}~\bibnamefont {{Ricci}}}, \bibinfo {author} {\bibfnamefont
  {N.}~\bibnamefont {{Rispoli}}}, \bibinfo {author} {\bibfnamefont
  {F.}~\bibnamefont {{Riva}}}, \bibinfo {author} {\bibfnamefont {J.~F.}\
  \bibnamefont {{Rivero-Rodriguez}}}, \bibinfo {author} {\bibfnamefont
  {M.}~\bibnamefont {{Salewski}}}, \bibinfo {author} {\bibfnamefont
  {O.}~\bibnamefont {{Sauter}}}, \bibinfo {author} {\bibfnamefont {B.~S.}\
  \bibnamefont {{Schmidt}}}, \bibinfo {author} {\bibfnamefont {R.}~\bibnamefont
  {{Schrittweiser}}}, \bibinfo {author} {\bibfnamefont {S.}~\bibnamefont
  {{Sharapov}}}, \bibinfo {author} {\bibfnamefont {U.~A.}\ \bibnamefont
  {{Sheikh}}}, \bibinfo {author} {\bibfnamefont {B.}~\bibnamefont {{Sieglin}}},
  \bibinfo {author} {\bibfnamefont {M.}~\bibnamefont {{Silva}}}, \bibinfo
  {author} {\bibfnamefont {A.}~\bibnamefont {{Smolders}}}, \bibinfo {author}
  {\bibfnamefont {A.}~\bibnamefont {{Snicker}}}, \bibinfo {author}
  {\bibfnamefont {C.}~\bibnamefont {{Sozzi}}}, \bibinfo {author} {\bibfnamefont
  {M.}~\bibnamefont {{Spolaore}}}, \bibinfo {author} {\bibfnamefont
  {A.}~\bibnamefont {{Stagni}}}, \bibinfo {author} {\bibfnamefont
  {L.}~\bibnamefont {{Stipani}}}, \bibinfo {author} {\bibfnamefont
  {G.}~\bibnamefont {{Sun}}}, \bibinfo {author} {\bibfnamefont
  {T.}~\bibnamefont {{Tala}}}, \bibinfo {author} {\bibfnamefont
  {P.}~\bibnamefont {{Tamain}}}, \bibinfo {author} {\bibfnamefont
  {K.}~\bibnamefont {{Tanaka}}}, \bibinfo {author} {\bibfnamefont
  {A.}~\bibnamefont {{Tema Biwole}}}, \bibinfo {author} {\bibfnamefont
  {D.}~\bibnamefont {{Terranova}}}, \bibinfo {author} {\bibfnamefont {J.~L.}\
  \bibnamefont {{Terry}}}, \bibinfo {author} {\bibfnamefont {D.}~\bibnamefont
  {{Testa}}}, \bibinfo {author} {\bibfnamefont {C.}~\bibnamefont {{Theiler}}},
  \bibinfo {author} {\bibfnamefont {A.}~\bibnamefont {{Thornton}}}, \bibinfo
  {author} {\bibfnamefont {A.}~\bibnamefont {{Thrys{\o}e}}}, \bibinfo {author}
  {\bibfnamefont {H.}~\bibnamefont {{Torreblanca}}}, \bibinfo {author}
  {\bibfnamefont {C.~K.}\ \bibnamefont {{Tsui}}}, \bibinfo {author}
  {\bibfnamefont {D.}~\bibnamefont {{Vaccaro}}}, \bibinfo {author}
  {\bibfnamefont {M.}~\bibnamefont {{Vallar}}}, \bibinfo {author}
  {\bibfnamefont {M.}~\bibnamefont {{van Berkel}}}, \bibinfo {author}
  {\bibfnamefont {D.}~\bibnamefont {{Van Eester}}}, \bibinfo {author}
  {\bibfnamefont {R.~J.~R.}\ \bibnamefont {{van Kampen}}}, \bibinfo {author}
  {\bibfnamefont {S.}~\bibnamefont {{Van Mulders}}}, \bibinfo {author}
  {\bibfnamefont {K.}~\bibnamefont {{Verhaegh}}}, \bibinfo {author}
  {\bibfnamefont {T.}~\bibnamefont {{Verhaeghe}}}, \bibinfo {author}
  {\bibfnamefont {N.}~\bibnamefont {{Vianello}}}, \bibinfo {author}
  {\bibfnamefont {F.}~\bibnamefont {{Villone}}}, \bibinfo {author}
  {\bibfnamefont {E.}~\bibnamefont {{Viezzer}}}, \bibinfo {author}
  {\bibfnamefont {B.}~\bibnamefont {{Vincent}}}, \bibinfo {author}
  {\bibfnamefont {I.}~\bibnamefont {{Voitsekhovitch}}}, \bibinfo {author}
  {\bibfnamefont {N.~M.~T.}\ \bibnamefont {{Vu}}}, \bibinfo {author}
  {\bibfnamefont {N.}~\bibnamefont {{Walkden}}}, \bibinfo {author}
  {\bibfnamefont {T.}~\bibnamefont {{Wauters}}}, \bibinfo {author}
  {\bibfnamefont {H.}~\bibnamefont {{Weisen}}}, \bibinfo {author}
  {\bibfnamefont {N.}~\bibnamefont {{Wendler}}}, \bibinfo {author}
  {\bibfnamefont {M.}~\bibnamefont {{Wensing}}}, \bibinfo {author}
  {\bibfnamefont {F.}~\bibnamefont {{Widmer}}}, \bibinfo {author}
  {\bibfnamefont {S.}~\bibnamefont {{Wiesen}}}, \bibinfo {author}
  {\bibfnamefont {M.}~\bibnamefont {{Wischmeier}}}, \bibinfo {author}
  {\bibfnamefont {T.~A.}\ \bibnamefont {{Wijkamp}}}, \bibinfo {author}
  {\bibfnamefont {D.}~\bibnamefont {{W{\"u}nderlich}}}, \bibinfo {author}
  {\bibfnamefont {C.}~\bibnamefont {{W{\"u}thrich}}}, \bibinfo {author}
  {\bibfnamefont {V.}~\bibnamefont {{Yanovskiy}}}, \bibinfo {author}
  {\bibfnamefont {J.}~\bibnamefont {{Zebrowski}}}, \ and\ \bibinfo {author}
  {\bibnamefont {{EUROfusion MST1 Team}}},\ }\bibfield  {title} {\enquote
  {\bibinfo {title} {{Overview of the TCV tokamak experimental programme}},}\
  }\href {\doibase 10.1088/1741-4326/ac369b} {\bibfield  {journal} {\bibinfo
  {journal} {Nuclear Fusion}\ }\textbf {\bibinfo {volume} {62}},\ \bibinfo
  {eid} {042018} (\bibinfo {year} {2022})}\BibitemShut {NoStop}%
\bibitem [{\citenamefont {{Gusev}}\ \emph {et~al.}(2009)\citenamefont
  {{Gusev}}, \citenamefont {{Aleksandrov}}, \citenamefont {{Alimov}},
  \citenamefont {{Arkhipov}}, \citenamefont {{Ayushin}}, \citenamefont
  {{Barsukov}}, \citenamefont {{Ber}}, \citenamefont {{Chernyshev}},
  \citenamefont {{Chugunov}}, \citenamefont {{Dech}}, \citenamefont {{Golant}},
  \citenamefont {{Gorodetsky}}, \citenamefont {{Dyachenko}}, \citenamefont
  {{Kochergin}}, \citenamefont {{Kurskiev}}, \citenamefont {{Khitrov}},
  \citenamefont {{Khromov}}, \citenamefont {{Lebedev}}, \citenamefont
  {{Leonov}}, \citenamefont {{Litunovsky}}, \citenamefont {{Mazul}},
  \citenamefont {{Minaev}}, \citenamefont {{Mineev}}, \citenamefont
  {{Mironov}}, \citenamefont {{Miroshnikov}}, \citenamefont {{Mukhin}},
  \citenamefont {{Nikolaev}}, \citenamefont {{Novokhatsky}}, \citenamefont
  {{Panasenkov}}, \citenamefont {{Patrov}}, \citenamefont {{Petrov}},
  \citenamefont {{Petrov}}, \citenamefont {{Podushnikova}}, \citenamefont
  {{Rozhansky}}, \citenamefont {{Rozhdestvensky}}, \citenamefont {{Sakharov}},
  \citenamefont {{Shcherbinin}}, \citenamefont {{Senichenkov}}, \citenamefont
  {{Shevelev}}, \citenamefont {{Suhov}}, \citenamefont {{Trapesnikova}},
  \citenamefont {{Terukov}}, \citenamefont {{Tilinin}}, \citenamefont
  {{Tolstyakov}}, \citenamefont {{Varfolomeev}}, \citenamefont {{Voronin}},
  \citenamefont {{Zakharov}}, \citenamefont {{Zalavutdinov}}, \citenamefont
  {{Yagnov}}, \citenamefont {{Kuznetsov}},\ and\ \citenamefont
  {{Zhilin}}}]{GlobusM_2009}%
  \BibitemOpen
  \bibfield  {author} {\bibinfo {author} {\bibfnamefont {V.~K.}\ \bibnamefont
  {{Gusev}}}, \bibinfo {author} {\bibfnamefont {S.~E.}\ \bibnamefont
  {{Aleksandrov}}}, \bibinfo {author} {\bibfnamefont {V.~K.}\ \bibnamefont
  {{Alimov}}}, \bibinfo {author} {\bibfnamefont {I.~I.}\ \bibnamefont
  {{Arkhipov}}}, \bibinfo {author} {\bibfnamefont {B.~B.}\ \bibnamefont
  {{Ayushin}}}, \bibinfo {author} {\bibfnamefont {A.~G.}\ \bibnamefont
  {{Barsukov}}}, \bibinfo {author} {\bibfnamefont {B.~Y.}\ \bibnamefont
  {{Ber}}}, \bibinfo {author} {\bibfnamefont {F.~V.}\ \bibnamefont
  {{Chernyshev}}}, \bibinfo {author} {\bibfnamefont {I.~N.}\ \bibnamefont
  {{Chugunov}}}, \bibinfo {author} {\bibfnamefont {A.~V.}\ \bibnamefont
  {{Dech}}}, \bibinfo {author} {\bibfnamefont {V.~E.}\ \bibnamefont
  {{Golant}}}, \bibinfo {author} {\bibfnamefont {A.~E.}\ \bibnamefont
  {{Gorodetsky}}}, \bibinfo {author} {\bibfnamefont {V.~V.}\ \bibnamefont
  {{Dyachenko}}}, \bibinfo {author} {\bibfnamefont {M.~M.}\ \bibnamefont
  {{Kochergin}}}, \bibinfo {author} {\bibfnamefont {G.~S.}\ \bibnamefont
  {{Kurskiev}}}, \bibinfo {author} {\bibfnamefont {S.~A.}\ \bibnamefont
  {{Khitrov}}}, \bibinfo {author} {\bibfnamefont {N.~A.}\ \bibnamefont
  {{Khromov}}}, \bibinfo {author} {\bibfnamefont {V.~M.}\ \bibnamefont
  {{Lebedev}}}, \bibinfo {author} {\bibfnamefont {V.~M.}\ \bibnamefont
  {{Leonov}}}, \bibinfo {author} {\bibfnamefont {N.~V.}\ \bibnamefont
  {{Litunovsky}}}, \bibinfo {author} {\bibfnamefont {I.~V.}\ \bibnamefont
  {{Mazul}}}, \bibinfo {author} {\bibfnamefont {V.~B.}\ \bibnamefont
  {{Minaev}}}, \bibinfo {author} {\bibfnamefont {A.~B.}\ \bibnamefont
  {{Mineev}}}, \bibinfo {author} {\bibfnamefont {M.~I.}\ \bibnamefont
  {{Mironov}}}, \bibinfo {author} {\bibfnamefont {I.~V.}\ \bibnamefont
  {{Miroshnikov}}}, \bibinfo {author} {\bibfnamefont {E.~E.}\ \bibnamefont
  {{Mukhin}}}, \bibinfo {author} {\bibfnamefont {Y.~A.}\ \bibnamefont
  {{Nikolaev}}}, \bibinfo {author} {\bibfnamefont {A.~N.}\ \bibnamefont
  {{Novokhatsky}}}, \bibinfo {author} {\bibfnamefont {A.~A.}\ \bibnamefont
  {{Panasenkov}}}, \bibinfo {author} {\bibfnamefont {M.~I.}\ \bibnamefont
  {{Patrov}}}, \bibinfo {author} {\bibfnamefont {M.~P.}\ \bibnamefont
  {{Petrov}}}, \bibinfo {author} {\bibfnamefont {Y.~V.}\ \bibnamefont
  {{Petrov}}}, \bibinfo {author} {\bibfnamefont {K.~A.}\ \bibnamefont
  {{Podushnikova}}}, \bibinfo {author} {\bibfnamefont {V.~A.}\ \bibnamefont
  {{Rozhansky}}}, \bibinfo {author} {\bibfnamefont {V.~V.}\ \bibnamefont
  {{Rozhdestvensky}}}, \bibinfo {author} {\bibfnamefont {N.~V.}\ \bibnamefont
  {{Sakharov}}}, \bibinfo {author} {\bibfnamefont {O.~N.}\ \bibnamefont
  {{Shcherbinin}}}, \bibinfo {author} {\bibfnamefont {I.~Y.}\ \bibnamefont
  {{Senichenkov}}}, \bibinfo {author} {\bibfnamefont {A.~E.}\ \bibnamefont
  {{Shevelev}}}, \bibinfo {author} {\bibfnamefont {E.~V.}\ \bibnamefont
  {{Suhov}}}, \bibinfo {author} {\bibfnamefont {I.~N.}\ \bibnamefont
  {{Trapesnikova}}}, \bibinfo {author} {\bibfnamefont {E.~I.}\ \bibnamefont
  {{Terukov}}}, \bibinfo {author} {\bibfnamefont {G.~N.}\ \bibnamefont
  {{Tilinin}}}, \bibinfo {author} {\bibfnamefont {S.~Y.}\ \bibnamefont
  {{Tolstyakov}}}, \bibinfo {author} {\bibfnamefont {V.~I.}\ \bibnamefont
  {{Varfolomeev}}}, \bibinfo {author} {\bibfnamefont {A.~V.}\ \bibnamefont
  {{Voronin}}}, \bibinfo {author} {\bibfnamefont {A.~P.}\ \bibnamefont
  {{Zakharov}}}, \bibinfo {author} {\bibfnamefont {R.~K.}\ \bibnamefont
  {{Zalavutdinov}}}, \bibinfo {author} {\bibfnamefont {V.~A.}\ \bibnamefont
  {{Yagnov}}}, \bibinfo {author} {\bibfnamefont {E.~A.}\ \bibnamefont
  {{Kuznetsov}}}, \ and\ \bibinfo {author} {\bibfnamefont {E.~G.}\ \bibnamefont
  {{Zhilin}}},\ }\bibfield  {title} {\enquote {\bibinfo {title} {{Overview of
  results obtained at the Globus-M spherical tokamak}},}\ }\href {\doibase
  10.1088/0029-5515/49/10/104021} {\bibfield  {journal} {\bibinfo  {journal}
  {Nuclear Fusion}\ }\textbf {\bibinfo {volume} {49}},\ \bibinfo {eid} {104021}
  (\bibinfo {year} {2009})}\BibitemShut {NoStop}%
\bibitem [{\citenamefont {{Kurskiev}}\ \emph {et~al.}(2019)\citenamefont
  {{Kurskiev}}, \citenamefont {{Bakharev}}, \citenamefont {{Bulanin}},
  \citenamefont {{Chernyshev}}, \citenamefont {{Gusev}}, \citenamefont
  {{Khromov}}, \citenamefont {{Kiselev}}, \citenamefont {{Minaev}},
  \citenamefont {{Miroshnikov}}, \citenamefont {{Mukhin}}, \citenamefont
  {{Patrov}}, \citenamefont {{Petrov}}, \citenamefont {{Petrov}}, \citenamefont
  {{Sakharov}}, \citenamefont {{Shchegolev}}, \citenamefont {{Sladkomedova}},
  \citenamefont {{Solokha}}, \citenamefont {{Telnova}}, \citenamefont
  {{Tolstyakov}}, \citenamefont {{Tokarev}},\ and\ \citenamefont
  {{Yashin}}}]{GlobusM_2019}%
  \BibitemOpen
  \bibfield  {author} {\bibinfo {author} {\bibfnamefont {G.~S.}\ \bibnamefont
  {{Kurskiev}}}, \bibinfo {author} {\bibfnamefont {N.~N.}\ \bibnamefont
  {{Bakharev}}}, \bibinfo {author} {\bibfnamefont {V.~V.}\ \bibnamefont
  {{Bulanin}}}, \bibinfo {author} {\bibfnamefont {F.~V.}\ \bibnamefont
  {{Chernyshev}}}, \bibinfo {author} {\bibfnamefont {V.~K.}\ \bibnamefont
  {{Gusev}}}, \bibinfo {author} {\bibfnamefont {N.~A.}\ \bibnamefont
  {{Khromov}}}, \bibinfo {author} {\bibfnamefont {E.~O.}\ \bibnamefont
  {{Kiselev}}}, \bibinfo {author} {\bibfnamefont {V.~B.}\ \bibnamefont
  {{Minaev}}}, \bibinfo {author} {\bibfnamefont {I.~V.}\ \bibnamefont
  {{Miroshnikov}}}, \bibinfo {author} {\bibfnamefont {E.~E.}\ \bibnamefont
  {{Mukhin}}}, \bibinfo {author} {\bibfnamefont {M.~I.}\ \bibnamefont
  {{Patrov}}}, \bibinfo {author} {\bibfnamefont {A.~V.}\ \bibnamefont
  {{Petrov}}}, \bibinfo {author} {\bibfnamefont {Y.~V.}\ \bibnamefont
  {{Petrov}}}, \bibinfo {author} {\bibfnamefont {N.~V.}\ \bibnamefont
  {{Sakharov}}}, \bibinfo {author} {\bibfnamefont {P.~B.}\ \bibnamefont
  {{Shchegolev}}}, \bibinfo {author} {\bibfnamefont {A.~D.}\ \bibnamefont
  {{Sladkomedova}}}, \bibinfo {author} {\bibfnamefont {V.~V.}\ \bibnamefont
  {{Solokha}}}, \bibinfo {author} {\bibfnamefont {A.~Y.}\ \bibnamefont
  {{Telnova}}}, \bibinfo {author} {\bibfnamefont {S.~Y.}\ \bibnamefont
  {{Tolstyakov}}}, \bibinfo {author} {\bibfnamefont {V.~A.}\ \bibnamefont
  {{Tokarev}}}, \ and\ \bibinfo {author} {\bibfnamefont {A.~Y.}\ \bibnamefont
  {{Yashin}}},\ }\bibfield  {title} {\enquote {\bibinfo {title} {{Thermal
  energy confinement at the Globus-M spherical tokamak}},}\ }\href {\doibase
  10.1088/1741-4326/ab15c5} {\bibfield  {journal} {\bibinfo  {journal} {Nuclear
  Fusion}\ }\textbf {\bibinfo {volume} {59}},\ \bibinfo {eid} {066032}
  (\bibinfo {year} {2019})}\BibitemShut {NoStop}%
\bibitem [{\citenamefont {{Dnestrovskij}}, \citenamefont {{Connor}},\ and\
  \citenamefont {{Gryaznevich}}(2019)}]{ST40_2019}%
  \BibitemOpen
  \bibfield  {author} {\bibinfo {author} {\bibfnamefont {A.~Y.}\ \bibnamefont
  {{Dnestrovskij}}}, \bibinfo {author} {\bibfnamefont {J.~W.}\ \bibnamefont
  {{Connor}}}, \ and\ \bibinfo {author} {\bibfnamefont {M.~P.}\ \bibnamefont
  {{Gryaznevich}}},\ }\bibfield  {title} {\enquote {\bibinfo {title} {{On the
  confinement modeling of a high field spherical tokamak ST40}},}\ }\href
  {\doibase 10.1088/1361-6587/ab0bf8} {\bibfield  {journal} {\bibinfo
  {journal} {Plasma Physics and Controlled Fusion}\ }\textbf {\bibinfo {volume}
  {61}},\ \bibinfo {eid} {055009} (\bibinfo {year} {2019})}\BibitemShut
  {NoStop}%
\bibitem [{\citenamefont {{McNamara}}\ \emph {et~al.}(2023)\citenamefont
  {{McNamara}}, \citenamefont {{Asunta}}, \citenamefont {{Bland}},
  \citenamefont {{Buxton}}, \citenamefont {{Colgan}}, \citenamefont
  {{Dnestrovskii}}, \citenamefont {{Gemmell}}, \citenamefont {{Gryaznevich}},
  \citenamefont {{Hoffman}}, \citenamefont {{Janky}}, \citenamefont {{Lister}},
  \citenamefont {{Lowe}}, \citenamefont {{Mirfayzi}}, \citenamefont {{Naylor}},
  \citenamefont {{Nemytov}}, \citenamefont {{Njau}}, \citenamefont
  {{Pyragius}}, \citenamefont {{Rengle}}, \citenamefont {{Romanelli}},
  \citenamefont {{Romero}}, \citenamefont {{Sertoli}}, \citenamefont
  {{Shevchenko}}, \citenamefont {{Sinha}}, \citenamefont {{Sladkomedova}},
  \citenamefont {{Sridhar}}, \citenamefont {{Takase}}, \citenamefont
  {{Thomas}}, \citenamefont {{Varje}}, \citenamefont {{Vincent}}, \citenamefont
  {{Willett}}, \citenamefont {{Wood}}, \citenamefont {{Zakhar}}, \citenamefont
  {{Battaglia}}, \citenamefont {{Kaye}}, \citenamefont {{Delgado-Aparicio}},
  \citenamefont {{Maingi}}, \citenamefont {{Mueller}}, \citenamefont
  {{Podesta}}, \citenamefont {{Delabie}}, \citenamefont {{Lomanowski}},
  \citenamefont {{Marchuk}},\ and\ \citenamefont {{ST40 Team}}}]{ST40_2023}%
  \BibitemOpen
  \bibfield  {author} {\bibinfo {author} {\bibfnamefont {S.~A.~M.}\
  \bibnamefont {{McNamara}}}, \bibinfo {author} {\bibfnamefont
  {O.}~\bibnamefont {{Asunta}}}, \bibinfo {author} {\bibfnamefont
  {J.}~\bibnamefont {{Bland}}}, \bibinfo {author} {\bibfnamefont {P.~F.}\
  \bibnamefont {{Buxton}}}, \bibinfo {author} {\bibfnamefont {C.}~\bibnamefont
  {{Colgan}}}, \bibinfo {author} {\bibfnamefont {A.}~\bibnamefont
  {{Dnestrovskii}}}, \bibinfo {author} {\bibfnamefont {M.}~\bibnamefont
  {{Gemmell}}}, \bibinfo {author} {\bibfnamefont {M.}~\bibnamefont
  {{Gryaznevich}}}, \bibinfo {author} {\bibfnamefont {D.}~\bibnamefont
  {{Hoffman}}}, \bibinfo {author} {\bibfnamefont {F.}~\bibnamefont {{Janky}}},
  \bibinfo {author} {\bibfnamefont {J.~B.}\ \bibnamefont {{Lister}}}, \bibinfo
  {author} {\bibfnamefont {H.~F.}\ \bibnamefont {{Lowe}}}, \bibinfo {author}
  {\bibfnamefont {R.~S.}\ \bibnamefont {{Mirfayzi}}}, \bibinfo {author}
  {\bibfnamefont {G.}~\bibnamefont {{Naylor}}}, \bibinfo {author}
  {\bibfnamefont {V.}~\bibnamefont {{Nemytov}}}, \bibinfo {author}
  {\bibfnamefont {J.}~\bibnamefont {{Njau}}}, \bibinfo {author} {\bibfnamefont
  {T.}~\bibnamefont {{Pyragius}}}, \bibinfo {author} {\bibfnamefont
  {A.}~\bibnamefont {{Rengle}}}, \bibinfo {author} {\bibfnamefont
  {M.}~\bibnamefont {{Romanelli}}}, \bibinfo {author} {\bibfnamefont
  {C.}~\bibnamefont {{Romero}}}, \bibinfo {author} {\bibfnamefont
  {M.}~\bibnamefont {{Sertoli}}}, \bibinfo {author} {\bibfnamefont
  {V.}~\bibnamefont {{Shevchenko}}}, \bibinfo {author} {\bibfnamefont
  {J.}~\bibnamefont {{Sinha}}}, \bibinfo {author} {\bibfnamefont
  {A.}~\bibnamefont {{Sladkomedova}}}, \bibinfo {author} {\bibfnamefont
  {S.}~\bibnamefont {{Sridhar}}}, \bibinfo {author} {\bibfnamefont
  {Y.}~\bibnamefont {{Takase}}}, \bibinfo {author} {\bibfnamefont
  {P.}~\bibnamefont {{Thomas}}}, \bibinfo {author} {\bibfnamefont
  {J.}~\bibnamefont {{Varje}}}, \bibinfo {author} {\bibfnamefont
  {B.}~\bibnamefont {{Vincent}}}, \bibinfo {author} {\bibfnamefont {H.~V.}\
  \bibnamefont {{Willett}}}, \bibinfo {author} {\bibfnamefont {J.}~\bibnamefont
  {{Wood}}}, \bibinfo {author} {\bibfnamefont {D.}~\bibnamefont {{Zakhar}}},
  \bibinfo {author} {\bibfnamefont {D.~J.}\ \bibnamefont {{Battaglia}}},
  \bibinfo {author} {\bibfnamefont {S.~M.}\ \bibnamefont {{Kaye}}}, \bibinfo
  {author} {\bibfnamefont {L.~F.}\ \bibnamefont {{Delgado-Aparicio}}}, \bibinfo
  {author} {\bibfnamefont {R.}~\bibnamefont {{Maingi}}}, \bibinfo {author}
  {\bibfnamefont {D.}~\bibnamefont {{Mueller}}}, \bibinfo {author}
  {\bibfnamefont {M.}~\bibnamefont {{Podesta}}}, \bibinfo {author}
  {\bibfnamefont {E.}~\bibnamefont {{Delabie}}}, \bibinfo {author}
  {\bibfnamefont {B.}~\bibnamefont {{Lomanowski}}}, \bibinfo {author}
  {\bibfnamefont {O.}~\bibnamefont {{Marchuk}}}, \ and\ \bibinfo {author}
  {\bibnamefont {{ST40 Team}}},\ }\bibfield  {title} {\enquote {\bibinfo
  {title} {{Achievement of ion temperatures in excess of 100 million degrees
  Kelvin in the compact high-field spherical tokamak ST40}},}\ }\href {\doibase
  10.1088/1741-4326/acbec8} {\bibfield  {journal} {\bibinfo  {journal} {Nuclear
  Fusion}\ }\textbf {\bibinfo {volume} {63}},\ \bibinfo {eid} {054002}
  (\bibinfo {year} {2023})}\BibitemShut {NoStop}%
\bibitem [{\citenamefont {{Ariola}}\ \emph {et~al.}(2008)\citenamefont
  {{Ariola}}, \citenamefont {{De Tommasi}}, \citenamefont {{Mazon}},
  \citenamefont {{Moreau}}, \citenamefont {{Piccolo}}, \citenamefont
  {{Pironti}}, \citenamefont {{Sartori}},\ and\ \citenamefont
  {{Zabeo}}}]{Ariola2008}%
  \BibitemOpen
  \bibfield  {author} {\bibinfo {author} {\bibfnamefont {M.}~\bibnamefont
  {{Ariola}}}, \bibinfo {author} {\bibfnamefont {G.}~\bibnamefont {{De
  Tommasi}}}, \bibinfo {author} {\bibfnamefont {D.}~\bibnamefont {{Mazon}}},
  \bibinfo {author} {\bibfnamefont {D.}~\bibnamefont {{Moreau}}}, \bibinfo
  {author} {\bibfnamefont {F.}~\bibnamefont {{Piccolo}}}, \bibinfo {author}
  {\bibfnamefont {A.}~\bibnamefont {{Pironti}}}, \bibinfo {author}
  {\bibfnamefont {F.}~\bibnamefont {{Sartori}}}, \ and\ \bibinfo {author}
  {\bibfnamefont {L.}~\bibnamefont {{Zabeo}}},\ }\bibfield  {title} {\enquote
  {\bibinfo {title} {{Integrated Plasma Shape and Boundary Flux Control on JET
  Tokamak}},}\ }\href {\doibase 10.13182/FST08-A1735} {\bibfield  {journal}
  {\bibinfo  {journal} {Fusion Science and Technology}\ }\textbf {\bibinfo
  {volume} {53}},\ \bibinfo {pages} {789--805} (\bibinfo {year}
  {2008})}\BibitemShut {NoStop}%
\bibitem [{\citenamefont {Seo}\ \emph {et~al.}(2021)\citenamefont {Seo},
  \citenamefont {Na}, \citenamefont {Kim}, \citenamefont {Lee}, \citenamefont
  {Park}, \citenamefont {Park},\ and\ \citenamefont {Lee}}]{Seo_2021}%
  \BibitemOpen
  \bibfield  {author} {\bibinfo {author} {\bibfnamefont {J.}~\bibnamefont
  {Seo}}, \bibinfo {author} {\bibfnamefont {Y.-S.}\ \bibnamefont {Na}},
  \bibinfo {author} {\bibfnamefont {B.}~\bibnamefont {Kim}}, \bibinfo {author}
  {\bibfnamefont {C.}~\bibnamefont {Lee}}, \bibinfo {author} {\bibfnamefont
  {M.}~\bibnamefont {Park}}, \bibinfo {author} {\bibfnamefont {S.}~\bibnamefont
  {Park}}, \ and\ \bibinfo {author} {\bibfnamefont {Y.}~\bibnamefont {Lee}},\
  }\bibfield  {title} {\enquote {\bibinfo {title} {Feedforward beta control in
  the kstar tokamak by deep reinforcement learning},}\ }\href {\doibase
  10.1088/1741-4326/ac121b} {\bibfield  {journal} {\bibinfo  {journal} {Nuclear
  Fusion}\ }\textbf {\bibinfo {volume} {61}},\ \bibinfo {pages} {106010}
  (\bibinfo {year} {2021})}\BibitemShut {NoStop}%
\bibitem [{\citenamefont {{Wai}}, \citenamefont {{Boyer}},\ and\ \citenamefont
  {{Kolemen}}(2022)}]{Wai2022}%
  \BibitemOpen
  \bibfield  {author} {\bibinfo {author} {\bibfnamefont {J.~T.}\ \bibnamefont
  {{Wai}}}, \bibinfo {author} {\bibfnamefont {M.~D.}\ \bibnamefont {{Boyer}}},
  \ and\ \bibinfo {author} {\bibfnamefont {E.}~\bibnamefont {{Kolemen}}},\
  }\bibfield  {title} {\enquote {\bibinfo {title} {{Neural net modeling of
  equilibria in NSTX-U}},}\ }\href {\doibase 10.1088/1741-4326/ac77e6}
  {\bibfield  {journal} {\bibinfo  {journal} {Nuclear Fusion}\ }\textbf
  {\bibinfo {volume} {62}},\ \bibinfo {eid} {086042} (\bibinfo {year}
  {2022})},\ \Eprint {http://arxiv.org/abs/2202.13915} {arXiv:2202.13915
  [physics.plasm-ph]} \BibitemShut {NoStop}%
\bibitem [{\citenamefont {{Wan}}\ \emph {et~al.}(2023)\citenamefont {{Wan}},
  \citenamefont {{Yu}}, \citenamefont {{Pau}}, \citenamefont {{Sauter}},
  \citenamefont {{Liu}}, \citenamefont {{Yuan}},\ and\ \citenamefont
  {{Li}}}]{Wan2023}%
  \BibitemOpen
  \bibfield  {author} {\bibinfo {author} {\bibfnamefont {C.}~\bibnamefont
  {{Wan}}}, \bibinfo {author} {\bibfnamefont {Z.}~\bibnamefont {{Yu}}},
  \bibinfo {author} {\bibfnamefont {A.}~\bibnamefont {{Pau}}}, \bibinfo
  {author} {\bibfnamefont {O.}~\bibnamefont {{Sauter}}}, \bibinfo {author}
  {\bibfnamefont {X.}~\bibnamefont {{Liu}}}, \bibinfo {author} {\bibfnamefont
  {Q.}~\bibnamefont {{Yuan}}}, \ and\ \bibinfo {author} {\bibfnamefont
  {J.}~\bibnamefont {{Li}}},\ }\bibfield  {title} {\enquote {\bibinfo {title}
  {{A machine-learning-based tool for last closed-flux surface reconstruction
  on tokamaks}},}\ }\href {\doibase 10.1088/1741-4326/acbfcc} {\bibfield
  {journal} {\bibinfo  {journal} {Nuclear Fusion}\ }\textbf {\bibinfo {volume}
  {63}},\ \bibinfo {eid} {056019} (\bibinfo {year} {2023})},\ \Eprint
  {http://arxiv.org/abs/2207.05695} {arXiv:2207.05695 [physics.plasm-ph]}
  \BibitemShut {NoStop}%
\bibitem [{\citenamefont {Wei}\ \emph {et~al.}(2023)\citenamefont {Wei},
  \citenamefont {Sun}, \citenamefont {Tang}, \citenamefont {Lin}, \citenamefont
  {Du},\ and\ \citenamefont {Dong}}]{Wei2023}%
  \BibitemOpen
  \bibfield  {author} {\bibinfo {author} {\bibfnamefont {X.}~\bibnamefont
  {Wei}}, \bibinfo {author} {\bibfnamefont {S.}~\bibnamefont {Sun}}, \bibinfo
  {author} {\bibfnamefont {W.}~\bibnamefont {Tang}}, \bibinfo {author}
  {\bibfnamefont {Z.}~\bibnamefont {Lin}}, \bibinfo {author} {\bibfnamefont
  {H.}~\bibnamefont {Du}}, \ and\ \bibinfo {author} {\bibfnamefont
  {G.}~\bibnamefont {Dong}},\ }\bibfield  {title} {\enquote {\bibinfo {title}
  {Reconstruction of tokamak plasma safety factor profile using deep
  learning},}\ }\href {\doibase 10.1088/1741-4326/acdf00} {\bibfield  {journal}
  {\bibinfo  {journal} {Nuclear Fusion}\ }\textbf {\bibinfo {volume} {63}},\
  \bibinfo {pages} {086020} (\bibinfo {year} {2023})}\BibitemShut {NoStop}%
\bibitem [{\citenamefont {{Lister}}\ \emph {et~al.}(2002)\citenamefont
  {{Lister}}, \citenamefont {{Sharma}}, \citenamefont {{Limebeer}},
  \citenamefont {{Nakamura}}, \citenamefont {{Wainwright}},\ and\ \citenamefont
  {{Yoshino}}}]{Lister_2002}%
  \BibitemOpen
  \bibfield  {author} {\bibinfo {author} {\bibfnamefont {J.~B.}\ \bibnamefont
  {{Lister}}}, \bibinfo {author} {\bibfnamefont {A.}~\bibnamefont {{Sharma}}},
  \bibinfo {author} {\bibfnamefont {D.~J.~N.}\ \bibnamefont {{Limebeer}}},
  \bibinfo {author} {\bibfnamefont {Y.}~\bibnamefont {{Nakamura}}}, \bibinfo
  {author} {\bibfnamefont {J.~P.}\ \bibnamefont {{Wainwright}}}, \ and\
  \bibinfo {author} {\bibfnamefont {R.}~\bibnamefont {{Yoshino}}},\ }\bibfield
  {title} {\enquote {\bibinfo {title} {{Plasma equilibrium response modelling
  and validation on JT-60U}},}\ }\href {\doibase 10.1088/0029-5515/42/6/309}
  {\bibfield  {journal} {\bibinfo  {journal} {Nuclear Fusion}\ }\textbf
  {\bibinfo {volume} {42}},\ \bibinfo {pages} {708--724} (\bibinfo {year}
  {2002})}\BibitemShut {NoStop}%
\bibitem [{\citenamefont {Albanese}(2015)}]{albanese2015create}%
  \BibitemOpen
  \bibfield  {author} {\bibinfo {author} {\bibfnamefont {R.}~\bibnamefont
  {Albanese}},\ }\bibfield  {title} {\enquote {\bibinfo {title} {Create-nl+: A
  robust control-oriented free boundary dynamic plasma equilibrium solver},}\
  }\href@noop {} {\bibfield  {journal} {\bibinfo  {journal} {Fusion Engineering
  and Design}\ }\textbf {\bibinfo {volume} {96}},\ \bibinfo {pages} {664--667}
  (\bibinfo {year} {2015})}\BibitemShut {NoStop}%
\bibitem [{\citenamefont {{Humphreys}}\ \emph {et~al.}(2007)\citenamefont
  {{Humphreys}}, \citenamefont {{Ferron}}, \citenamefont {{Bakhtiari}},
  \citenamefont {{Blair}}, \citenamefont {{In}}, \citenamefont {{Jackson}},
  \citenamefont {{Jhang}}, \citenamefont {{Johnson}}, \citenamefont {{Kim}},
  \citenamefont {{La Haye}}, \citenamefont {{Leuer}}, \citenamefont
  {{Penaflor}}, \citenamefont {{Schuster}}, \citenamefont {{Walker}},
  \citenamefont {{Wang}}, \citenamefont {{Welander}},\ and\ \citenamefont
  {{Whyte}}}]{Humphreys2007}%
  \BibitemOpen
  \bibfield  {author} {\bibinfo {author} {\bibfnamefont {D.~A.}\ \bibnamefont
  {{Humphreys}}}, \bibinfo {author} {\bibfnamefont {J.~R.}\ \bibnamefont
  {{Ferron}}}, \bibinfo {author} {\bibfnamefont {M.}~\bibnamefont
  {{Bakhtiari}}}, \bibinfo {author} {\bibfnamefont {J.~A.}\ \bibnamefont
  {{Blair}}}, \bibinfo {author} {\bibfnamefont {Y.}~\bibnamefont {{In}}},
  \bibinfo {author} {\bibfnamefont {G.~L.}\ \bibnamefont {{Jackson}}}, \bibinfo
  {author} {\bibfnamefont {H.}~\bibnamefont {{Jhang}}}, \bibinfo {author}
  {\bibfnamefont {R.~D.}\ \bibnamefont {{Johnson}}}, \bibinfo {author}
  {\bibfnamefont {J.~S.}\ \bibnamefont {{Kim}}}, \bibinfo {author}
  {\bibfnamefont {R.~J.}\ \bibnamefont {{La Haye}}}, \bibinfo {author}
  {\bibfnamefont {J.~A.}\ \bibnamefont {{Leuer}}}, \bibinfo {author}
  {\bibfnamefont {B.~G.}\ \bibnamefont {{Penaflor}}}, \bibinfo {author}
  {\bibfnamefont {E.}~\bibnamefont {{Schuster}}}, \bibinfo {author}
  {\bibfnamefont {M.~L.}\ \bibnamefont {{Walker}}}, \bibinfo {author}
  {\bibfnamefont {H.}~\bibnamefont {{Wang}}}, \bibinfo {author} {\bibfnamefont
  {A.~S.}\ \bibnamefont {{Welander}}}, \ and\ \bibinfo {author} {\bibfnamefont
  {D.~G.}\ \bibnamefont {{Whyte}}},\ }\bibfield  {title} {\enquote {\bibinfo
  {title} {{Development of ITER-relevant plasma control solutions at
  DIII-D}},}\ }\href {\doibase 10.1088/0029-5515/47/8/028} {\bibfield
  {journal} {\bibinfo  {journal} {Nuclear Fusion}\ }\textbf {\bibinfo {volume}
  {47}},\ \bibinfo {pages} {943--951} (\bibinfo {year} {2007})}\BibitemShut
  {NoStop}%
\bibitem [{\citenamefont {Eldon}\ \emph {et~al.}(2020)\citenamefont {Eldon},
  \citenamefont {Hyatt}, \citenamefont {Covele}, \citenamefont {Eidietis},
  \citenamefont {Guo}, \citenamefont {Humphreys}, \citenamefont {Moser},
  \citenamefont {Sammuli},\ and\ \citenamefont {Walker}}]{ELDON2020111797}%
  \BibitemOpen
  \bibfield  {author} {\bibinfo {author} {\bibfnamefont {D.}~\bibnamefont
  {Eldon}}, \bibinfo {author} {\bibfnamefont {A.}~\bibnamefont {Hyatt}},
  \bibinfo {author} {\bibfnamefont {B.}~\bibnamefont {Covele}}, \bibinfo
  {author} {\bibfnamefont {N.}~\bibnamefont {Eidietis}}, \bibinfo {author}
  {\bibfnamefont {H.}~\bibnamefont {Guo}}, \bibinfo {author} {\bibfnamefont
  {D.}~\bibnamefont {Humphreys}}, \bibinfo {author} {\bibfnamefont
  {A.}~\bibnamefont {Moser}}, \bibinfo {author} {\bibfnamefont
  {B.}~\bibnamefont {Sammuli}}, \ and\ \bibinfo {author} {\bibfnamefont
  {M.}~\bibnamefont {Walker}},\ }\bibfield  {title} {\enquote {\bibinfo {title}
  {High precision strike point control to support experiments in the diii-d
  small angle slot divertor},}\ }\href {\doibase
  https://doi.org/10.1016/j.fusengdes.2020.111797} {\bibfield  {journal}
  {\bibinfo  {journal} {Fusion Engineering and Design}\ }\textbf {\bibinfo
  {volume} {160}},\ \bibinfo {pages} {111797} (\bibinfo {year}
  {2020})}\BibitemShut {NoStop}%
\bibitem [{\citenamefont {McArdle}, \citenamefont {Pangione},\ and\
  \citenamefont {Kochan}(2020)}]{MCARDLE2020111764}%
  \BibitemOpen
  \bibfield  {author} {\bibinfo {author} {\bibfnamefont {G.}~\bibnamefont
  {McArdle}}, \bibinfo {author} {\bibfnamefont {L.}~\bibnamefont {Pangione}}, \
  and\ \bibinfo {author} {\bibfnamefont {M.}~\bibnamefont {Kochan}},\
  }\bibfield  {title} {\enquote {\bibinfo {title} {The mast upgrade plasma
  control system},}\ }\href {\doibase
  https://doi.org/10.1016/j.fusengdes.2020.111764} {\bibfield  {journal}
  {\bibinfo  {journal} {Fusion Engineering and Design}\ }\textbf {\bibinfo
  {volume} {159}},\ \bibinfo {pages} {111764} (\bibinfo {year}
  {2020})}\BibitemShut {NoStop}%
\bibitem [{\citenamefont {Anand}\ \emph {et~al.}(2023)\citenamefont {Anand},
  \citenamefont {Bardsley}, \citenamefont {Humphreys}, \citenamefont
  {Lennholm}, \citenamefont {Welander}, \citenamefont {Xing}, \citenamefont
  {Barr}, \citenamefont {Walker}, \citenamefont {Mitchell},\ and\ \citenamefont
  {Meyer}}]{ANAND2023113724}%
  \BibitemOpen
  \bibfield  {author} {\bibinfo {author} {\bibfnamefont {H.}~\bibnamefont
  {Anand}}, \bibinfo {author} {\bibfnamefont {O.}~\bibnamefont {Bardsley}},
  \bibinfo {author} {\bibfnamefont {D.}~\bibnamefont {Humphreys}}, \bibinfo
  {author} {\bibfnamefont {M.}~\bibnamefont {Lennholm}}, \bibinfo {author}
  {\bibfnamefont {A.}~\bibnamefont {Welander}}, \bibinfo {author}
  {\bibfnamefont {Z.}~\bibnamefont {Xing}}, \bibinfo {author} {\bibfnamefont
  {J.}~\bibnamefont {Barr}}, \bibinfo {author} {\bibfnamefont {M.}~\bibnamefont
  {Walker}}, \bibinfo {author} {\bibfnamefont {J.}~\bibnamefont {Mitchell}}, \
  and\ \bibinfo {author} {\bibfnamefont {H.}~\bibnamefont {Meyer}},\ }\bibfield
   {title} {\enquote {\bibinfo {title} {Modelling, design and simulation of
  plasma magnetic control for the spherical tokamak for energy production
  (step)},}\ }\href {\doibase https://doi.org/10.1016/j.fusengdes.2023.113724}
  {\bibfield  {journal} {\bibinfo  {journal} {Fusion Engineering and Design}\
  }\textbf {\bibinfo {volume} {194}},\ \bibinfo {pages} {113724} (\bibinfo
  {year} {2023})}\BibitemShut {NoStop}%
\bibitem [{\citenamefont {{Pitcher}}\ and\ \citenamefont
  {{Stangeby}}(1997)}]{Pitcher1997}%
  \BibitemOpen
  \bibfield  {author} {\bibinfo {author} {\bibfnamefont {C.~S.}\ \bibnamefont
  {{Pitcher}}}\ and\ \bibinfo {author} {\bibfnamefont {P.~C.}\ \bibnamefont
  {{Stangeby}}},\ }\bibfield  {title} {\enquote {\bibinfo {title} {{REVIEW
  ARTICLE: Experimental divertor physics}},}\ }\href {\doibase
  10.1088/0741-3335/39/6/001} {\bibfield  {journal} {\bibinfo  {journal}
  {Plasma Physics and Controlled Fusion}\ }\textbf {\bibinfo {volume} {39}},\
  \bibinfo {pages} {779--930} (\bibinfo {year} {1997})}\BibitemShut {NoStop}%
\bibitem [{\citenamefont {{Ambrosino}}\ \emph {et~al.}(2008)\citenamefont
  {{Ambrosino}}, \citenamefont {{Ariola}}, \citenamefont {{de Tommasi}},
  \citenamefont {{Pironti}}, \citenamefont {{Sartori}}, \citenamefont
  {{Joffrin}},\ and\ \citenamefont {{Villone}}}]{Ambrosino2008}%
  \BibitemOpen
  \bibfield  {author} {\bibinfo {author} {\bibfnamefont {G.}~\bibnamefont
  {{Ambrosino}}}, \bibinfo {author} {\bibfnamefont {M.}~\bibnamefont
  {{Ariola}}}, \bibinfo {author} {\bibfnamefont {G.}~\bibnamefont {{de
  Tommasi}}}, \bibinfo {author} {\bibfnamefont {A.}~\bibnamefont {{Pironti}}},
  \bibinfo {author} {\bibfnamefont {F.}~\bibnamefont {{Sartori}}}, \bibinfo
  {author} {\bibfnamefont {E.}~\bibnamefont {{Joffrin}}}, \ and\ \bibinfo
  {author} {\bibfnamefont {F.}~\bibnamefont {{Villone}}},\ }\bibfield  {title}
  {\enquote {\bibinfo {title} {{Plasma Strike-Point Sweeping on JET Tokamak
  With the eXtreme Shape Controller}},}\ }\href {\doibase
  10.1109/TPS.2008.922920} {\bibfield  {journal} {\bibinfo  {journal} {IEEE
  Transactions on Plasma Science}\ }\textbf {\bibinfo {volume} {36}},\ \bibinfo
  {pages} {834--840} (\bibinfo {year} {2008})}\BibitemShut {NoStop}%
\bibitem [{\citenamefont {{Goldston}}(2010)}]{Goldston2010}%
  \BibitemOpen
  \bibfield  {author} {\bibinfo {author} {\bibfnamefont {R.~J.}\ \bibnamefont
  {{Goldston}}},\ }\bibfield  {title} {\enquote {\bibinfo {title} {{Downstream
  heat flux profile versus midplane T profile in tokamaks}},}\ }\href {\doibase
  10.1063/1.3280011} {\bibfield  {journal} {\bibinfo  {journal} {Physics of
  Plasmas}\ }\textbf {\bibinfo {volume} {17}},\ \bibinfo {eid} {012503}
  (\bibinfo {year} {2010})}\BibitemShut {NoStop}%
\bibitem [{\citenamefont {{Petrie}}\ \emph {et~al.}(2013)\citenamefont
  {{Petrie}}, \citenamefont {{Canik}}, \citenamefont {{Lasnier}}, \citenamefont
  {{Leonard}}, \citenamefont {{Mahdavi}}, \citenamefont {{Watkins}},
  \citenamefont {{Fenstermacher}}, \citenamefont {{Ferron}}, \citenamefont
  {{Groebner}}, \citenamefont {{Hill}}, \citenamefont {{Hyatt}}, \citenamefont
  {{Holcomb}}, \citenamefont {{Luce}}, \citenamefont {{Moyer}},\ and\
  \citenamefont {{Stangeby}}}]{Petrie2013}%
  \BibitemOpen
  \bibfield  {author} {\bibinfo {author} {\bibfnamefont {T.~W.}\ \bibnamefont
  {{Petrie}}}, \bibinfo {author} {\bibfnamefont {J.~M.}\ \bibnamefont
  {{Canik}}}, \bibinfo {author} {\bibfnamefont {C.~J.}\ \bibnamefont
  {{Lasnier}}}, \bibinfo {author} {\bibfnamefont {A.~W.}\ \bibnamefont
  {{Leonard}}}, \bibinfo {author} {\bibfnamefont {M.~A.}\ \bibnamefont
  {{Mahdavi}}}, \bibinfo {author} {\bibfnamefont {J.~G.}\ \bibnamefont
  {{Watkins}}}, \bibinfo {author} {\bibfnamefont {M.~E.}\ \bibnamefont
  {{Fenstermacher}}}, \bibinfo {author} {\bibfnamefont {J.~R.}\ \bibnamefont
  {{Ferron}}}, \bibinfo {author} {\bibfnamefont {R.~J.}\ \bibnamefont
  {{Groebner}}}, \bibinfo {author} {\bibfnamefont {D.~N.}\ \bibnamefont
  {{Hill}}}, \bibinfo {author} {\bibfnamefont {A.~W.}\ \bibnamefont {{Hyatt}}},
  \bibinfo {author} {\bibfnamefont {C.~T.}\ \bibnamefont {{Holcomb}}}, \bibinfo
  {author} {\bibfnamefont {T.~C.}\ \bibnamefont {{Luce}}}, \bibinfo {author}
  {\bibfnamefont {R.~A.}\ \bibnamefont {{Moyer}}}, \ and\ \bibinfo {author}
  {\bibfnamefont {P.~C.}\ \bibnamefont {{Stangeby}}},\ }\bibfield  {title}
  {\enquote {\bibinfo {title} {{Effect of separatrix magnetic geometry on
  divertor behavior in DIII-D}},}\ }\href {\doibase
  10.1016/j.jnucmat.2013.01.051} {\bibfield  {journal} {\bibinfo  {journal}
  {Journal of Nuclear Materials}\ }\textbf {\bibinfo {volume} {438}},\ \bibinfo
  {pages} {S166--S169} (\bibinfo {year} {2013})}\BibitemShut {NoStop}%
\bibitem [{\citenamefont {{Kotov}}\ and\ \citenamefont
  {{Reiter}}(2009)}]{Kotov2009}%
  \BibitemOpen
  \bibfield  {author} {\bibinfo {author} {\bibfnamefont {V.}~\bibnamefont
  {{Kotov}}}\ and\ \bibinfo {author} {\bibfnamefont {D.}~\bibnamefont
  {{Reiter}}},\ }\bibfield  {title} {\enquote {\bibinfo {title} {{Two-point
  analysis of the numerical modelling of detached divertor plasmas}},}\ }\href
  {\doibase 10.1088/0741-3335/51/11/115002} {\bibfield  {journal} {\bibinfo
  {journal} {Plasma Physics and Controlled Fusion}\ }\textbf {\bibinfo {volume}
  {51}},\ \bibinfo {eid} {115002} (\bibinfo {year} {2009})}\BibitemShut
  {NoStop}%
\bibitem [{\citenamefont {{Stangeby}}(2018)}]{Stangeby2018}%
  \BibitemOpen
  \bibfield  {author} {\bibinfo {author} {\bibfnamefont {P.~C.}\ \bibnamefont
  {{Stangeby}}},\ }\bibfield  {title} {\enquote {\bibinfo {title} {{Basic
  physical processes and reduced models for plasma detachment}},}\ }\href
  {\doibase 10.1088/1361-6587/aaacf6} {\bibfield  {journal} {\bibinfo
  {journal} {Plasma Physics and Controlled Fusion}\ }\textbf {\bibinfo {volume}
  {60}},\ \bibinfo {eid} {044022} (\bibinfo {year} {2018})}\BibitemShut
  {NoStop}%
\bibitem [{\citenamefont {{Chankin}}\ \emph {et~al.}(2006)\citenamefont
  {{Chankin}}, \citenamefont {{Coster}}, \citenamefont {{Dux}}, \citenamefont
  {{Fuchs}}, \citenamefont {{Haas}}, \citenamefont {{Herrmann}}, \citenamefont
  {{Horton}}, \citenamefont {{Kallenbach}}, \citenamefont {{Kaufmann}},
  \citenamefont {{Konz}}, \citenamefont {{Lackner}}, \citenamefont {{Maggi}},
  \citenamefont {{M{\"u}ller}}, \citenamefont {{Neuhauser}}, \citenamefont
  {{Pugno}}, \citenamefont {{Reich}},\ and\ \citenamefont
  {{Schneider}}}]{chankin2006}%
  \BibitemOpen
  \bibfield  {author} {\bibinfo {author} {\bibfnamefont {A.~V.}\ \bibnamefont
  {{Chankin}}}, \bibinfo {author} {\bibfnamefont {D.~P.}\ \bibnamefont
  {{Coster}}}, \bibinfo {author} {\bibfnamefont {R.}~\bibnamefont {{Dux}}},
  \bibinfo {author} {\bibfnamefont {C.}~\bibnamefont {{Fuchs}}}, \bibinfo
  {author} {\bibfnamefont {G.}~\bibnamefont {{Haas}}}, \bibinfo {author}
  {\bibfnamefont {A.}~\bibnamefont {{Herrmann}}}, \bibinfo {author}
  {\bibfnamefont {L.~D.}\ \bibnamefont {{Horton}}}, \bibinfo {author}
  {\bibfnamefont {A.}~\bibnamefont {{Kallenbach}}}, \bibinfo {author}
  {\bibfnamefont {M.}~\bibnamefont {{Kaufmann}}}, \bibinfo {author}
  {\bibfnamefont {C.}~\bibnamefont {{Konz}}}, \bibinfo {author} {\bibfnamefont
  {K.}~\bibnamefont {{Lackner}}}, \bibinfo {author} {\bibfnamefont
  {C.}~\bibnamefont {{Maggi}}}, \bibinfo {author} {\bibfnamefont {H.~W.}\
  \bibnamefont {{M{\"u}ller}}}, \bibinfo {author} {\bibfnamefont
  {J.}~\bibnamefont {{Neuhauser}}}, \bibinfo {author} {\bibfnamefont
  {R.}~\bibnamefont {{Pugno}}}, \bibinfo {author} {\bibfnamefont
  {M.}~\bibnamefont {{Reich}}}, \ and\ \bibinfo {author} {\bibfnamefont
  {W.}~\bibnamefont {{Schneider}}},\ }\bibfield  {title} {\enquote {\bibinfo
  {title} {{SOLPS modelling of ASDEX upgrade H-mode plasma}},}\ }\href
  {\doibase 10.1088/0741-3335/48/6/010} {\bibfield  {journal} {\bibinfo
  {journal} {Plasma Physics and Controlled Fusion}\ }\textbf {\bibinfo {volume}
  {48}},\ \bibinfo {pages} {839--868} (\bibinfo {year} {2006})}\BibitemShut
  {NoStop}%
\bibitem [{\citenamefont {{Havl{\'\i}{\v{c}}kov{\'a}}}\ \emph
  {et~al.}(2015)\citenamefont {{Havl{\'\i}{\v{c}}kov{\'a}}}, \citenamefont
  {{Harrison}}, \citenamefont {{Lipschultz}}, \citenamefont {{Fishpool}},
  \citenamefont {{Kirk}}, \citenamefont {{Thornton}}, \citenamefont
  {{Wischmeier}}, \citenamefont {{Elmore}},\ and\ \citenamefont
  {{Allan}}}]{havlichkova2015}%
  \BibitemOpen
  \bibfield  {author} {\bibinfo {author} {\bibfnamefont {E.}~\bibnamefont
  {{Havl{\'\i}{\v{c}}kov{\'a}}}}, \bibinfo {author} {\bibfnamefont
  {J.}~\bibnamefont {{Harrison}}}, \bibinfo {author} {\bibfnamefont
  {B.}~\bibnamefont {{Lipschultz}}}, \bibinfo {author} {\bibfnamefont
  {G.}~\bibnamefont {{Fishpool}}}, \bibinfo {author} {\bibfnamefont
  {A.}~\bibnamefont {{Kirk}}}, \bibinfo {author} {\bibfnamefont
  {A.}~\bibnamefont {{Thornton}}}, \bibinfo {author} {\bibfnamefont
  {M.}~\bibnamefont {{Wischmeier}}}, \bibinfo {author} {\bibfnamefont
  {S.}~\bibnamefont {{Elmore}}}, \ and\ \bibinfo {author} {\bibfnamefont
  {S.}~\bibnamefont {{Allan}}},\ }\bibfield  {title} {\enquote {\bibinfo
  {title} {{SOLPS analysis of the MAST-U divertor with the effect of heating
  power and pumping on the access to detachment in the Super-x
  configuration}},}\ }\href {\doibase 10.1088/0741-3335/57/11/115001}
  {\bibfield  {journal} {\bibinfo  {journal} {Plasma Physics and Controlled
  Fusion}\ }\textbf {\bibinfo {volume} {57}},\ \bibinfo {eid} {115001}
  (\bibinfo {year} {2015})}\BibitemShut {NoStop}%
\bibitem [{\citenamefont {{Dudson}}\ \emph {et~al.}(2023)\citenamefont
  {{Dudson}}, \citenamefont {{Kryjak}}, \citenamefont {{Muhammed}},
  \citenamefont {{Hill}},\ and\ \citenamefont {{Omotani}}}]{Dudson2023}%
  \BibitemOpen
  \bibfield  {author} {\bibinfo {author} {\bibfnamefont {B.}~\bibnamefont
  {{Dudson}}}, \bibinfo {author} {\bibfnamefont {M.}~\bibnamefont {{Kryjak}}},
  \bibinfo {author} {\bibfnamefont {H.}~\bibnamefont {{Muhammed}}}, \bibinfo
  {author} {\bibfnamefont {P.}~\bibnamefont {{Hill}}}, \ and\ \bibinfo {author}
  {\bibfnamefont {J.}~\bibnamefont {{Omotani}}},\ }\bibfield  {title} {\enquote
  {\bibinfo {title} {{Hermes-3: Multi-component plasma simulations with
  BOUT++}},}\ }\href {\doibase 10.48550/arXiv.2303.12131} {\bibfield  {journal}
  {\bibinfo  {journal} {arXiv e-prints}\ ,\ \bibinfo {eid} {arXiv:2303.12131}}
  (\bibinfo {year} {2023})},\ \Eprint {http://arxiv.org/abs/2303.12131}
  {arXiv:2303.12131 [physics.plasm-ph]} \BibitemShut {NoStop}%
\bibitem [{\citenamefont {{Myatra}}\ \emph {et~al.}(2023)\citenamefont
  {{Myatra}}, \citenamefont {{Lipschultz}}, \citenamefont {{Moulton}},
  \citenamefont {{Verhaegh}}, \citenamefont {{Dudson}}, \citenamefont
  {{Orchard}}, \citenamefont {{Fil}},\ and\ \citenamefont
  {{Cowley}}}]{Myatra2023}%
  \BibitemOpen
  \bibfield  {author} {\bibinfo {author} {\bibfnamefont {O.}~\bibnamefont
  {{Myatra}}}, \bibinfo {author} {\bibfnamefont {B.}~\bibnamefont
  {{Lipschultz}}}, \bibinfo {author} {\bibfnamefont {D.}~\bibnamefont
  {{Moulton}}}, \bibinfo {author} {\bibfnamefont {K.}~\bibnamefont
  {{Verhaegh}}}, \bibinfo {author} {\bibfnamefont {B.}~\bibnamefont
  {{Dudson}}}, \bibinfo {author} {\bibfnamefont {S.}~\bibnamefont {{Orchard}}},
  \bibinfo {author} {\bibfnamefont {A.}~\bibnamefont {{Fil}}}, \ and\ \bibinfo
  {author} {\bibfnamefont {C.}~\bibnamefont {{Cowley}}},\ }\bibfield  {title}
  {\enquote {\bibinfo {title} {{Predictive SOLPS-ITER simulations to study the
  role of divertor magnetic geometry in detachment control in the MAST-U
  Super-X configuration}},}\ }\href {\doibase 10.1088/1741-4326/acea33}
  {\bibfield  {journal} {\bibinfo  {journal} {Nuclear Fusion}\ }\textbf
  {\bibinfo {volume} {63}},\ \bibinfo {eid} {096018} (\bibinfo {year}
  {2023})}\BibitemShut {NoStop}%
\bibitem [{\citenamefont {{Moulton}}\ \emph {et~al.}(2017)\citenamefont
  {{Moulton}}, \citenamefont {{Harrison}}, \citenamefont {{Lipschultz}},\ and\
  \citenamefont {{Coster}}}]{Moulton2017}%
  \BibitemOpen
  \bibfield  {author} {\bibinfo {author} {\bibfnamefont {D.}~\bibnamefont
  {{Moulton}}}, \bibinfo {author} {\bibfnamefont {J.}~\bibnamefont
  {{Harrison}}}, \bibinfo {author} {\bibfnamefont {B.}~\bibnamefont
  {{Lipschultz}}}, \ and\ \bibinfo {author} {\bibfnamefont {D.}~\bibnamefont
  {{Coster}}},\ }\bibfield  {title} {\enquote {\bibinfo {title} {{Using SOLPS
  to confirm the importance of total flux expansion in Super-X divertors}},}\
  }\href {\doibase 10.1088/1361-6587/aa6b13} {\bibfield  {journal} {\bibinfo
  {journal} {Plasma Physics and Controlled Fusion}\ }\textbf {\bibinfo {volume}
  {59}},\ \bibinfo {eid} {065011} (\bibinfo {year} {2017})}\BibitemShut
  {NoStop}%
\bibitem [{Note1()}]{Note1}%
  \BibitemOpen
  \bibinfo {note} {{https://github.com/freegs-plasma/freegs}}\BibitemShut
  {NoStop}%
\bibitem [{Note2()}]{Note2}%
  \BibitemOpen
  \bibinfo {note} {{https://github.com/freegs-plasma/freegs}}\BibitemShut
  {NoStop}%
\bibitem [{\citenamefont {{Amorisco}}\ \emph {et~al.}(2023)\citenamefont
  {{Amorisco}}, \citenamefont {{Agnello}}, \citenamefont {{Holt}},
  \citenamefont {{Mars}}, \citenamefont {{Buchanan}},\ and\ \citenamefont
  {{Pamela}}}]{Amorisco2023}%
  \BibitemOpen
  \bibfield  {author} {\bibinfo {author} {\bibfnamefont {N.}~\bibnamefont
  {{Amorisco}}}, \bibinfo {author} {\bibfnamefont {A.}~\bibnamefont
  {{Agnello}}}, \bibinfo {author} {\bibfnamefont {G.}~\bibnamefont {{Holt}}},
  \bibinfo {author} {\bibfnamefont {M.}~\bibnamefont {{Mars}}}, \bibinfo
  {author} {\bibfnamefont {J.}~\bibnamefont {{Buchanan}}}, \ and\ \bibinfo
  {author} {\bibfnamefont {S.}~\bibnamefont {{Pamela}}},\ }\bibfield  {title}
  {\enquote {\bibinfo {title} {Freegsnke: a python-based dynamic free-boundary
  toroidal plasma equilibrium solver},}\ }\href@noop {} {\bibfield  {journal}
  {\bibinfo  {journal} {Physics of Plasmas, ICDDPS-4 special issue}\ }\textbf
  {\bibinfo {volume} {0}},\ \bibinfo {pages} {(subm.)} (\bibinfo {year}
  {2023})}\BibitemShut {NoStop}%
\bibitem [{\citenamefont {{Lackner}}(1976)}]{Lackner1976}%
  \BibitemOpen
  \bibfield  {author} {\bibinfo {author} {\bibfnamefont {K.}~\bibnamefont
  {{Lackner}}},\ }\bibfield  {title} {\enquote {\bibinfo {title} {{Computation
  of ideal MHD equilibria}},}\ }\href {\doibase 10.1016/0010-4655(76)90008-4}
  {\bibfield  {journal} {\bibinfo  {journal} {Computer Physics Communications}\
  }\textbf {\bibinfo {volume} {12}},\ \bibinfo {pages} {33} (\bibinfo {year}
  {1976})}\BibitemShut {NoStop}%
\bibitem [{\citenamefont {{Jeon}}(2015)}]{Jeon2015}%
  \BibitemOpen
  \bibfield  {author} {\bibinfo {author} {\bibfnamefont {Y.~M.}\ \bibnamefont
  {{Jeon}}},\ }\bibfield  {title} {\enquote {\bibinfo {title} {{Development of
  a free-boundary tokamak equilibrium solver for advanced study of tokamak
  equilibria}},}\ }\href {\doibase 10.3938/jkps.67.843} {\bibfield  {journal}
  {\bibinfo  {journal} {Journal of Korean Physical Society}\ }\textbf {\bibinfo
  {volume} {67}},\ \bibinfo {pages} {843--853} (\bibinfo {year} {2015})},\
  \Eprint {http://arxiv.org/abs/1503.03135} {arXiv:1503.03135
  [physics.plasm-ph]} \BibitemShut {NoStop}%
\bibitem [{\citenamefont {{Goodman}}\ and\ \citenamefont
  {{Weare}}(2010)}]{GW2010}%
  \BibitemOpen
  \bibfield  {author} {\bibinfo {author} {\bibfnamefont {J.}~\bibnamefont
  {{Goodman}}}\ and\ \bibinfo {author} {\bibfnamefont {J.}~\bibnamefont
  {{Weare}}},\ }\bibfield  {title} {\enquote {\bibinfo {title} {{Ensemble
  samplers with affine invariance}},}\ }\href {\doibase
  10.2140/camcos.2010.5.65} {\bibfield  {journal} {\bibinfo  {journal}
  {Communications in Applied Mathematics and Computational Science}\ }\textbf
  {\bibinfo {volume} {5}},\ \bibinfo {pages} {65--80} (\bibinfo {year}
  {2010})}\BibitemShut {NoStop}%
\bibitem [{\citenamefont {{Jaderberg}}\ \emph {et~al.}(2017)\citenamefont
  {{Jaderberg}}, \citenamefont {{Dalibard}}, \citenamefont {{Osindero}},
  \citenamefont {{Czarnecki}}, \citenamefont {{Donahue}}, \citenamefont
  {{Razavi}}, \citenamefont {{Vinyals}}, \citenamefont {{Green}}, \citenamefont
  {{Dunning}}, \citenamefont {{Simonyan}}, \citenamefont {{Fernando}},\ and\
  \citenamefont {{Kavukcuoglu}}}]{PBT}%
  \BibitemOpen
  \bibfield  {author} {\bibinfo {author} {\bibfnamefont {M.}~\bibnamefont
  {{Jaderberg}}}, \bibinfo {author} {\bibfnamefont {V.}~\bibnamefont
  {{Dalibard}}}, \bibinfo {author} {\bibfnamefont {S.}~\bibnamefont
  {{Osindero}}}, \bibinfo {author} {\bibfnamefont {W.~M.}\ \bibnamefont
  {{Czarnecki}}}, \bibinfo {author} {\bibfnamefont {J.}~\bibnamefont
  {{Donahue}}}, \bibinfo {author} {\bibfnamefont {A.}~\bibnamefont {{Razavi}}},
  \bibinfo {author} {\bibfnamefont {O.}~\bibnamefont {{Vinyals}}}, \bibinfo
  {author} {\bibfnamefont {T.}~\bibnamefont {{Green}}}, \bibinfo {author}
  {\bibfnamefont {I.}~\bibnamefont {{Dunning}}}, \bibinfo {author}
  {\bibfnamefont {K.}~\bibnamefont {{Simonyan}}}, \bibinfo {author}
  {\bibfnamefont {C.}~\bibnamefont {{Fernando}}}, \ and\ \bibinfo {author}
  {\bibfnamefont {K.}~\bibnamefont {{Kavukcuoglu}}},\ }\bibfield  {title}
  {\enquote {\bibinfo {title} {{Population Based Training of Neural
  Networks}},}\ }\href {\doibase 10.48550/arXiv.1711.09846} {\bibfield
  {journal} {\bibinfo  {journal} {arXiv e-prints}\ ,\ \bibinfo {eid}
  {arXiv:1711.09846}} (\bibinfo {year} {2017})},\ \Eprint
  {http://arxiv.org/abs/1711.09846} {arXiv:1711.09846 [cs.LG]} \BibitemShut
  {NoStop}%
\bibitem [{\citenamefont {{Li}}\ \emph {et~al.}(2016)\citenamefont {{Li}},
  \citenamefont {{Jamieson}}, \citenamefont {{DeSalvo}}, \citenamefont
  {{Rostamizadeh}},\ and\ \citenamefont {{Talwalkar}}}]{hyperband}%
  \BibitemOpen
  \bibfield  {author} {\bibinfo {author} {\bibfnamefont {L.}~\bibnamefont
  {{Li}}}, \bibinfo {author} {\bibfnamefont {K.}~\bibnamefont {{Jamieson}}},
  \bibinfo {author} {\bibfnamefont {G.}~\bibnamefont {{DeSalvo}}}, \bibinfo
  {author} {\bibfnamefont {A.}~\bibnamefont {{Rostamizadeh}}}, \ and\ \bibinfo
  {author} {\bibfnamefont {A.}~\bibnamefont {{Talwalkar}}},\ }\bibfield
  {title} {\enquote {\bibinfo {title} {{Hyperband: A Novel Bandit-Based
  Approach to Hyperparameter Optimization}},}\ }\href {\doibase
  10.48550/arXiv.1603.06560} {\bibfield  {journal} {\bibinfo  {journal} {arXiv
  e-prints}\ ,\ \bibinfo {eid} {arXiv:1603.06560}} (\bibinfo {year} {2016})},\
  \Eprint {http://arxiv.org/abs/1603.06560} {arXiv:1603.06560 [cs.LG]}
  \BibitemShut {NoStop}%
\bibitem [{\citenamefont {{Akiba}}\ \emph {et~al.}(2019)\citenamefont
  {{Akiba}}, \citenamefont {{Sano}}, \citenamefont {{Yanase}}, \citenamefont
  {{Ohta}},\ and\ \citenamefont {{Koyama}}}]{optuna}%
  \BibitemOpen
  \bibfield  {author} {\bibinfo {author} {\bibfnamefont {T.}~\bibnamefont
  {{Akiba}}}, \bibinfo {author} {\bibfnamefont {S.}~\bibnamefont {{Sano}}},
  \bibinfo {author} {\bibfnamefont {T.}~\bibnamefont {{Yanase}}}, \bibinfo
  {author} {\bibfnamefont {T.}~\bibnamefont {{Ohta}}}, \ and\ \bibinfo {author}
  {\bibfnamefont {M.}~\bibnamefont {{Koyama}}},\ }\bibfield  {title} {\enquote
  {\bibinfo {title} {{Optuna: A Next-generation Hyperparameter Optimization
  Framework}},}\ }\href {\doibase 10.48550/arXiv.1907.10902} {\bibfield
  {journal} {\bibinfo  {journal} {arXiv e-prints}\ ,\ \bibinfo {eid}
  {arXiv:1907.10902}} (\bibinfo {year} {2019})},\ \Eprint
  {http://arxiv.org/abs/1907.10902} {arXiv:1907.10902 [cs.LG]} \BibitemShut
  {NoStop}%
\bibitem [{\citenamefont {{Mc Carthy}}\ \emph {et~al.}(2018)\citenamefont {{Mc
  Carthy}}, \citenamefont {{Riedel}}, \citenamefont {{Kardaun}}, \citenamefont
  {{Murmann}},\ and\ \citenamefont {{Lackner}}}]{McCarthy2018}%
  \BibitemOpen
  \bibfield  {author} {\bibinfo {author} {\bibfnamefont {P.~J.}\ \bibnamefont
  {{Mc Carthy}}}, \bibinfo {author} {\bibfnamefont {K.~S.}\ \bibnamefont
  {{Riedel}}}, \bibinfo {author} {\bibfnamefont {O.~J.~W.~F.}\ \bibnamefont
  {{Kardaun}}}, \bibinfo {author} {\bibfnamefont {H.}~\bibnamefont
  {{Murmann}}}, \ and\ \bibinfo {author} {\bibfnamefont {K.}~\bibnamefont
  {{Lackner}}},\ }\bibfield  {title} {\enquote {\bibinfo {title} {{Scalings and
  Plasma Profile Parameterisation of ASDEX High Density Ohmic Discharges}},}\
  }\href {\doibase 10.48550/arXiv.1803.10398} {\bibfield  {journal} {\bibinfo
  {journal} {arXiv e-prints}\ ,\ \bibinfo {eid} {arXiv:1803.10398}} (\bibinfo
  {year} {2018})},\ \Eprint {http://arxiv.org/abs/1803.10398} {arXiv:1803.10398
  [physics.plasm-ph]} \BibitemShut {NoStop}%
\bibitem [{\citenamefont {{Verdoolaege}}\ \emph {et~al.}(2021)\citenamefont
  {{Verdoolaege}}, \citenamefont {{Kaye}}, \citenamefont {{Angioni}},
  \citenamefont {{Kardaun}}, \citenamefont {{Maslov}}, \citenamefont
  {{Romanelli}}, \citenamefont {{Ryter}}, \citenamefont {{Thomsen}},
  \citenamefont {{ASDEX Upgrade Team}}, \citenamefont {{Eurofusion Mst1
  Team}},\ and\ \citenamefont {{Contributors}}}]{Verdoolaege2021}%
  \BibitemOpen
  \bibfield  {author} {\bibinfo {author} {\bibfnamefont {G.}~\bibnamefont
  {{Verdoolaege}}}, \bibinfo {author} {\bibfnamefont {S.~M.}\ \bibnamefont
  {{Kaye}}}, \bibinfo {author} {\bibfnamefont {C.}~\bibnamefont {{Angioni}}},
  \bibinfo {author} {\bibfnamefont {O.~J.~W.~F.}\ \bibnamefont {{Kardaun}}},
  \bibinfo {author} {\bibfnamefont {M.}~\bibnamefont {{Maslov}}}, \bibinfo
  {author} {\bibfnamefont {M.}~\bibnamefont {{Romanelli}}}, \bibinfo {author}
  {\bibfnamefont {F.}~\bibnamefont {{Ryter}}}, \bibinfo {author} {\bibfnamefont
  {K.}~\bibnamefont {{Thomsen}}}, \bibinfo {author} {\bibfnamefont
  {T.}~\bibnamefont {{ASDEX Upgrade Team}}}, \bibinfo {author} {\bibfnamefont
  {T.}~\bibnamefont {{Eurofusion Mst1 Team}}}, \ and\ \bibinfo {author}
  {\bibfnamefont {J.}~\bibnamefont {{Contributors}}},\ }\bibfield  {title}
  {\enquote {\bibinfo {title} {{The updated ITPA global H-mode confinement
  database: description and analysis}},}\ }\href {\doibase
  10.1088/1741-4326/abdb91} {\bibfield  {journal} {\bibinfo  {journal} {Nuclear
  Fusion}\ }\textbf {\bibinfo {volume} {61}},\ \bibinfo {eid} {076006}
  (\bibinfo {year} {2021})}\BibitemShut {NoStop}%
\bibitem [{\citenamefont {Graves}(2011)}]{Graves2011}%
  \BibitemOpen
  \bibfield  {author} {\bibinfo {author} {\bibfnamefont {A.}~\bibnamefont
  {Graves}},\ }\bibfield  {title} {\enquote {\bibinfo {title} {Practical
  variational inference for neural networks},}\ }in\ \href
  {https://proceedings.neurips.cc/paper_files/paper/2011/file/7eb3c8be3d411e8ebfab08eba5f49632-Paper.pdf}
  {\emph {\bibinfo {booktitle} {Advances in Neural Information Processing
  Systems}}},\ Vol.~\bibinfo {volume} {24},\ \bibinfo {editor} {edited by\
  \bibinfo {editor} {\bibfnamefont {J.}~\bibnamefont {Shawe-Taylor}}, \bibinfo
  {editor} {\bibfnamefont {R.}~\bibnamefont {Zemel}}, \bibinfo {editor}
  {\bibfnamefont {P.}~\bibnamefont {Bartlett}}, \bibinfo {editor}
  {\bibfnamefont {F.}~\bibnamefont {Pereira}}, \ and\ \bibinfo {editor}
  {\bibfnamefont {K.}~\bibnamefont {Weinberger}}}\ (\bibinfo  {publisher}
  {Curran Associates, Inc.},\ \bibinfo {year} {2011})\BibitemShut {NoStop}%
\bibitem [{\citenamefont {Kameoka}\ \emph {et~al.}(2019)\citenamefont
  {Kameoka}, \citenamefont {Li}, \citenamefont {Inoue},\ and\ \citenamefont
  {Makino}}]{Kameoka2019}%
  \BibitemOpen
  \bibfield  {author} {\bibinfo {author} {\bibfnamefont {H.}~\bibnamefont
  {Kameoka}}, \bibinfo {author} {\bibfnamefont {L.}~\bibnamefont {Li}},
  \bibinfo {author} {\bibfnamefont {S.}~\bibnamefont {Inoue}}, \ and\ \bibinfo
  {author} {\bibfnamefont {S.}~\bibnamefont {Makino}},\ }\bibfield  {title}
  {\enquote {\bibinfo {title} {{Supervised Determined Source Separation with
  Multichannel Variational Autoencoder}},}\ }\href {\doibase
  10.1162/neco_a_01217} {\bibfield  {journal} {\bibinfo  {journal} {Neural
  Computation}\ }\textbf {\bibinfo {volume} {31}},\ \bibinfo {pages}
  {1891--1914} (\bibinfo {year} {2019})},\ \Eprint
  {http://arxiv.org/abs/https://direct.mit.edu/neco/article-pdf/31/9/1891/1053426/neco\_a\_01217.pdf}
  {https://direct.mit.edu/neco/article-pdf/31/9/1891/1053426/neco\_a\_01217.pdf}
  \BibitemShut {NoStop}%
\bibitem [{\citenamefont {{de Tommasi}}\ \emph {et~al.}(2007)\citenamefont {{de
  Tommasi}}, \citenamefont {{Albanese}}, \citenamefont {{Ambrosino}},
  \citenamefont {{Ariola}}, \citenamefont {{Mattei}}, \citenamefont
  {{Pironti}}, \citenamefont {{Sartori}},\ and\ \citenamefont
  {{Villone}}}]{deTommasi2007}%
  \BibitemOpen
  \bibfield  {author} {\bibinfo {author} {\bibfnamefont {G.}~\bibnamefont {{de
  Tommasi}}}, \bibinfo {author} {\bibfnamefont {R.}~\bibnamefont {{Albanese}}},
  \bibinfo {author} {\bibfnamefont {G.}~\bibnamefont {{Ambrosino}}}, \bibinfo
  {author} {\bibfnamefont {M.}~\bibnamefont {{Ariola}}}, \bibinfo {author}
  {\bibfnamefont {M.}~\bibnamefont {{Mattei}}}, \bibinfo {author}
  {\bibfnamefont {A.}~\bibnamefont {{Pironti}}}, \bibinfo {author}
  {\bibfnamefont {F.}~\bibnamefont {{Sartori}}}, \ and\ \bibinfo {author}
  {\bibfnamefont {F.}~\bibnamefont {{Villone}}},\ }\bibfield  {title} {\enquote
  {\bibinfo {title} {{XSC Tools: A Software Suite for Tokamak Plasma Shape
  Control Design and Validation}},}\ }\href {\doibase 10.1109/TPS.2007.896989}
  {\bibfield  {journal} {\bibinfo  {journal} {IEEE Transactions on Plasma
  Science}\ }\textbf {\bibinfo {volume} {35}},\ \bibinfo {pages} {709--723}
  (\bibinfo {year} {2007})}\BibitemShut {NoStop}%
\bibitem [{\citenamefont {{Felici}}\ \emph {et~al.}(2010)\citenamefont
  {{Felici}}, \citenamefont {{Sauter}}, \citenamefont {{Goodman}},\ and\
  \citenamefont {{Paley}}}]{Felici_2010}%
  \BibitemOpen
  \bibfield  {author} {\bibinfo {author} {\bibfnamefont {F.}~\bibnamefont
  {{Felici}}}, \bibinfo {author} {\bibfnamefont {O.}~\bibnamefont {{Sauter}}},
  \bibinfo {author} {\bibfnamefont {T.}~\bibnamefont {{Goodman}}}, \ and\
  \bibinfo {author} {\bibfnamefont {J.}~\bibnamefont {{Paley}}},\ }\bibfield
  {title} {\enquote {\bibinfo {title} {{RAPTOR: Optimization, real-time
  simulation and control of the tokamak q profile evolution using a simplified
  transport model}},}\ }in\ \href@noop {} {\emph {\bibinfo {booktitle} {APS
  Division of Plasma Physics Meeting Abstracts}}},\ \bibinfo {series} {APS
  Meeting Abstracts}, Vol.~\bibinfo {volume} {52}\ (\bibinfo {year} {2010})\
  p.\ \bibinfo {pages} {BP9.090}\BibitemShut {NoStop}%
\bibitem [{\citenamefont {{Carpanese}}\ \emph {et~al.}(2020)\citenamefont
  {{Carpanese}}, \citenamefont {{Felici}}, \citenamefont {{Galperti}},
  \citenamefont {{Merle}}, \citenamefont {{Moret}}, \citenamefont {{Sauter}},\
  and\ \citenamefont {{the TCV Team}}}]{Carpanese_2020}%
  \BibitemOpen
  \bibfield  {author} {\bibinfo {author} {\bibfnamefont {F.}~\bibnamefont
  {{Carpanese}}}, \bibinfo {author} {\bibfnamefont {F.}~\bibnamefont
  {{Felici}}}, \bibinfo {author} {\bibfnamefont {C.}~\bibnamefont
  {{Galperti}}}, \bibinfo {author} {\bibfnamefont {A.}~\bibnamefont {{Merle}}},
  \bibinfo {author} {\bibfnamefont {J.~M.}\ \bibnamefont {{Moret}}}, \bibinfo
  {author} {\bibfnamefont {O.}~\bibnamefont {{Sauter}}}, \ and\ \bibinfo
  {author} {\bibnamefont {{the TCV Team}}},\ }\bibfield  {title} {\enquote
  {\bibinfo {title} {{First demonstration of real-time kinetic equilibrium
  reconstruction on TCV by coupling LIUQE and RAPTOR}},}\ }\href {\doibase
  10.1088/1741-4326/ab81ac} {\bibfield  {journal} {\bibinfo  {journal} {Nuclear
  Fusion}\ }\textbf {\bibinfo {volume} {60}},\ \bibinfo {eid} {066020}
  (\bibinfo {year} {2020})}\BibitemShut {NoStop}%
\bibitem [{\citenamefont {{Degrave}}\ \emph {et~al.}(2022)\citenamefont
  {{Degrave}}, \citenamefont {{Felici}}, \citenamefont {{Buchli}},
  \citenamefont {{Neunert}}, \citenamefont {{Tracey}}, \citenamefont
  {{Carpanese}}, \citenamefont {{Ewalds}}, \citenamefont {{Hafner}},
  \citenamefont {{Abdolmaleki}}, \citenamefont {{de las Casas}}, \citenamefont
  {{Donner}}, \citenamefont {{Fritz}}, \citenamefont {{Galperti}},
  \citenamefont {{Huber}}, \citenamefont {{Keeling}}, \citenamefont
  {{Tsimpoukelli}}, \citenamefont {{Kay}}, \citenamefont {{Merle}},
  \citenamefont {{Moret}}, \citenamefont {{Noury}}, \citenamefont
  {{Pesamosca}}, \citenamefont {{Pfau}}, \citenamefont {{Sauter}},
  \citenamefont {{Sommariva}}, \citenamefont {{Coda}}, \citenamefont {{Duval}},
  \citenamefont {{Fasoli}}, \citenamefont {{Kohli}}, \citenamefont
  {{Kavukcuoglu}}, \citenamefont {{Hassabis}},\ and\ \citenamefont
  {{Riedmiller}}}]{Degrave_2022}%
  \BibitemOpen
  \bibfield  {author} {\bibinfo {author} {\bibfnamefont {J.}~\bibnamefont
  {{Degrave}}}, \bibinfo {author} {\bibfnamefont {F.}~\bibnamefont {{Felici}}},
  \bibinfo {author} {\bibfnamefont {J.}~\bibnamefont {{Buchli}}}, \bibinfo
  {author} {\bibfnamefont {M.}~\bibnamefont {{Neunert}}}, \bibinfo {author}
  {\bibfnamefont {B.}~\bibnamefont {{Tracey}}}, \bibinfo {author}
  {\bibfnamefont {F.}~\bibnamefont {{Carpanese}}}, \bibinfo {author}
  {\bibfnamefont {T.}~\bibnamefont {{Ewalds}}}, \bibinfo {author}
  {\bibfnamefont {R.}~\bibnamefont {{Hafner}}}, \bibinfo {author}
  {\bibfnamefont {A.}~\bibnamefont {{Abdolmaleki}}}, \bibinfo {author}
  {\bibfnamefont {D.}~\bibnamefont {{de las Casas}}}, \bibinfo {author}
  {\bibfnamefont {C.}~\bibnamefont {{Donner}}}, \bibinfo {author}
  {\bibfnamefont {L.}~\bibnamefont {{Fritz}}}, \bibinfo {author} {\bibfnamefont
  {C.}~\bibnamefont {{Galperti}}}, \bibinfo {author} {\bibfnamefont
  {A.}~\bibnamefont {{Huber}}}, \bibinfo {author} {\bibfnamefont
  {J.}~\bibnamefont {{Keeling}}}, \bibinfo {author} {\bibfnamefont
  {M.}~\bibnamefont {{Tsimpoukelli}}}, \bibinfo {author} {\bibfnamefont
  {J.}~\bibnamefont {{Kay}}}, \bibinfo {author} {\bibfnamefont
  {A.}~\bibnamefont {{Merle}}}, \bibinfo {author} {\bibfnamefont {J.-M.}\
  \bibnamefont {{Moret}}}, \bibinfo {author} {\bibfnamefont {S.}~\bibnamefont
  {{Noury}}}, \bibinfo {author} {\bibfnamefont {F.}~\bibnamefont
  {{Pesamosca}}}, \bibinfo {author} {\bibfnamefont {D.}~\bibnamefont {{Pfau}}},
  \bibinfo {author} {\bibfnamefont {O.}~\bibnamefont {{Sauter}}}, \bibinfo
  {author} {\bibfnamefont {C.}~\bibnamefont {{Sommariva}}}, \bibinfo {author}
  {\bibfnamefont {S.}~\bibnamefont {{Coda}}}, \bibinfo {author} {\bibfnamefont
  {B.}~\bibnamefont {{Duval}}}, \bibinfo {author} {\bibfnamefont
  {A.}~\bibnamefont {{Fasoli}}}, \bibinfo {author} {\bibfnamefont
  {P.}~\bibnamefont {{Kohli}}}, \bibinfo {author} {\bibfnamefont
  {K.}~\bibnamefont {{Kavukcuoglu}}}, \bibinfo {author} {\bibfnamefont
  {D.}~\bibnamefont {{Hassabis}}}, \ and\ \bibinfo {author} {\bibfnamefont
  {M.}~\bibnamefont {{Riedmiller}}},\ }\bibfield  {title} {\enquote {\bibinfo
  {title} {{Magnetic control of tokamak plasmas through deep reinforcement
  learning}},}\ }\href {\doibase 10.1038/s41586-021-04301-9} {\bibfield
  {journal} {\bibinfo  {journal} {Nature}\ }\textbf {\bibinfo {volume} {602}},\
  \bibinfo {pages} {414--419} (\bibinfo {year} {2022})}\BibitemShut {NoStop}%
\end{thebibliography}%

\appendix*
\section{}
\subsection{Choices of circuit equations for the plasma control design}
\label{sect:circuits}
In the massless-plasma MHD approximation, from Poynting's Theorem we can write the coupling of toroidal currents as
\begin{equation}
\frac{1}{2}\frac{\mathrm{d}}{\mathrm{d}t}(\underline{I}_{y}^{T}M_{yy}\underline{I}_{y})+\underline{I}_{y}^{T}R_{yy}\underline{I}_{y}+\underline{I}_{y}^{T}M_{my}\dot{\underline{I}}_{m}+\frac{\mathrm{d}}{\mathrm{d}t}W_{T}=0
\end{equation}
where $\underline{I}_{y}$ is the plasma current distribution over a discretised domain of cells "\textit{y}", $\underline{I}_{m}$ is the vector of currents in the metals (active coils and passive structures), $M_{yy}$ and $M_{my}$ are plasma-to-plasma and metal-to-plasma inductance matrices, $R_{yy}$ is a diagonal matrix of resistances for each plasma domain cell, and $W_{T}$ is the internal thermal energy. Upon defining
\begin{eqnarray}
M_{p}=\frac{1}{I_{p}^{2}}\underline{I}_{y}^{T}M_{yy}\underline{I}_{y}\ ,\ \underline{M}_{mp}^{T}=\frac{1}{I_{p}}\underline{I}_{y}^{T}M_{my}\ ,\\
R_{p}=\frac{1}{I_{p}^{2}}\underline{I}_{y}^{T}R_{yy}\underline{I}_{y}
\end{eqnarray}
the above can be rewritten as
\begin{equation}M_{p}\dot{I}_{p}+\dot{M}_{p}I_{p}/2+R_{p}I_{p}+\underline{M}_{mp}^{T}\dot{\underline{I}}_{m}+\dot{W}_{T}/I_{p}=0
\end{equation}
If we adopt a \textit{stateful representation}, where the plasma current is described by total plasma current, metal currents and other parameters "$\underline{\alpha}$" (e.g. $p_a$ in this paper), then we have
\begin{eqnarray}
\label{eq:lumped} L_{p}\mathrm{d}{I}_{p} 
+ \underline{L}_{mp}^{T}\mathrm{d}{\underline{I}}_{m} & = & -R_{p}I_{p}\mathrm{d}t-\frac{1}{I_{p}}\frac{\partial W_{T}}{\partial \underline{\alpha}}\mathrm{d}{\underline{\alpha}} +\mathrm{d}\epsilon_{W} 
\end{eqnarray}
where we allow for external disturbances $\mathrm{d}\epsilon_{W}$ in the thermal evolution, and where
\begin{eqnarray}
L_{p}=\left( M_{p}+\frac{I_{p}}{2}\frac{\partial M_{p}}{\partial I_{p}} +\frac{1}{I_{p}}\frac{\partial W_{T}}{\partial I_{p}}\right)\\
\underline{L}_{mp}^{T}=\left(\underline{M}_{mp}+\frac{I_{p}}{2}\frac{\partial M_{p}}{\partial \underline{I}_{m}} +\frac{1}{I_{p}}\frac{\partial W_{T}}{\partial \underline{I}_{m}}\right)
\end{eqnarray}
The control design equation is then obtained by \textcolor{black}{decomposing the plasma and coil currents into a designed and perturbed currents. The zeroth-order $\underline{I}^{(0)}(t)$ set of designed currents is from a sweep of poloidal-field coils (including the Ohmic circuits) that are necessary to drive the plasma currents and mantain a desired shape. These cancel the right-hand side of equation~(\ref{eq:lumped}), while the left-hand side involves only variations in the perturbed currents $\underline{I}^{(1)}(t).$ These are the current changes (still coupled between plasma and coils) that are required in order to correct a deviation in the shape targets from the one predicted by the programmed zeroth-order coil sweeps.}. \textcolor{black}{Under a small change $\delta \underline{I}_{c}$ in shaping-coil currents,  any target that depends on plasma- and coil-currents changes as 
\begin{equation}
\delta y \approx \frac{\partial y}{\partial I_{p}}\delta I_{p}+\frac{\partial y}{\partial \underline{I}_{c}}\cdot\delta\underline{I}_{c}\ , \label{eq:linresp}
\end{equation}
and the perturbed plasma current changes as $\delta I_{p}\approx-\underline{L}_{cp}^{T}\delta\underline{I}_{c}/L_{p},$
which yields the control-design equation (1).}

The RZIp model of rigid displacement \cite{Lister_2002} uses a similar notation for the inductances, although it uses a radial and vertical displacement to parameterise the plasma instead of the coil currents, getting slightly different expressions for the $R$ and $L$ terms. Other lumped-parameter models \cite{Freidberg_2008,albanese2015create} neglect the $W_{T}$ derivatives, and either prescribe the behaviour of other plasma parameters \cite{Carpanese_2020, Degrave_2022} or derive it from a 0D transport model or from the current-diffusion equation \cite{Felici_2010}. Other derivations \cite{Ariola2008} instead replace $I_{y}^{T}$ with an integration over the plasma domain, writing Faraday's law while also neglecting changes in the plasma current distribution, and with the definitions
\begin{equation}
L_{p}=\mathbf{1}_{y}^{T}M_{yy}\underline{I}_{y}/I_{p}\ ;\ L_{mp}=\mathbf{1}_{y}^{T}M_{mp}\ ;\ 
R_{p}=\mathbf{1}_{y}^{T}R_{yy}\underline{I}_{y}/I_{p}\ .
\end{equation}

\textcolor{black}{ The linearised-response model (\ref{eq:linresp}) delegates to the designed coil sweep and the MIMO controller the task of minimising the shielding response of passive structures, and is the most commonly used on tokamak control design\cite{Ariola2008}. One may be tempted to account directly for part of the passive structures that are strongly coupled to the coils, e.g. the coil-cases or the MAST-U divertor "nose" close to the Dp coil. In fact, the solution of the Grad-Shafranov equation only needs knowledge about the flux-function $\psi_{tok}$ from the tokamak in a domain that includes the plasma, and if the domain $\mathcal{D}$ is chosen as a torus enclosing the plasma and sufficiently far inside the vessel, the contribution of Dp to $\psi_{tok}$ is almost indistinguishable from the combined contribution of Dp and divertor "nose". If we consider the tokamak flux-function in the domain $\mathcal{D}$ as an array, $\underline{\psi}_{tok}=\mathbf{A}_{tok}\underline{I}_{c}\ ,$ we can then use the SVD decomposition $\mathbf{A}_{tok}=\mathbf{U}{\Sigma}\mathbf{V}^{T},$ where $\mathbf{V}$ has dimension $n_{c}\times n_{c}$ and $\underline{\psi}_{tok}$ can be projected onto the first $n_{c}$ principal components, that are the first $n_{c}$ rows of $\mathbf{U}.$ In the coupling between plasma and metals, the vessel can be solved for, which yields slightly different definitions for the effective inductances above. A change in active coil currents then brings a change to $\underline{\psi}_{tok},$ and the sensitivity matrix has similar entries as before, but is suitably rotated in the new basis of currents. One reason why this alternative approach is not common in the literature is that, ultimately, the controller deals with requested voltages to the power supplies, which are subject to safety limits and output limits: aiming to directly cancel the shielding of tighly-coupled passive structures at all times is not necessarily feasible within those limits, and can also result in unstable open-loop control. The classical approach of designing the sensitivity matrix and, separately, a MIMO controller for the voltages provides a wealth of established techniques to minimise the asymptotic error, and examine the stability and safety of closed-loop control. We also note that the alternative approach still requires a calculation of the derivatives $\partial y/\partial \underline{I},$ and ultimately the emulation of the targets as a function of plasma and coil currents, which is the content of this paper. }

\subsection{ \textcolor{black}{Example form of the score function}}
Upon defining
\begin{equation}
\varsigma(x):=(x-1)-\log\left({1 +\exp(x-1)}\right)\ ,
\end{equation}
one example of log-score used in the MCMC is
\begin{eqnarray}
\nonumber \log(\mathcal{L})& = & \varsigma((R_{in}/\mathrm{m}-0.28)/0.01)\\
\nonumber &+&\varsigma((1.6-R_{out}/\mathrm{m})/0.01)+3\varsigma(q_{a}-1)\\
\nonumber & - & \frac{1}{2}(Z_{a}/0.15\mathrm{m})^{2}+\min(\psi_{n}/0.1|"nose") \\
\nonumber & -& \frac{\mathbf{1}_{Ip>0.95\mathrm{MA}} }{2}\left(\frac{I_{p}/\mathrm{MA}-0.95}{0.25}\right)^{2}\\
 &+& \frac{B_{X}/B_{S}-1.5}{0.2}+2\log\left( \frac{p_{a}}{10^{4}\mathrm{Pa}} \right) \ .
\end{eqnarray}
The last term encourages larger flux-expansion, evaluated between the X-point and the strike point, and the $\min(\psi_{n}/0.1|"nose")$ is evaluated on the divertor "nose and baffle" region with $0.75<|Z|<1.6$ and $0.75<R<1.6$.

\subsection{MCMC Parameter distributions}

\begin{figure*}[h!t]
\includegraphics[width=0.95\textwidth]{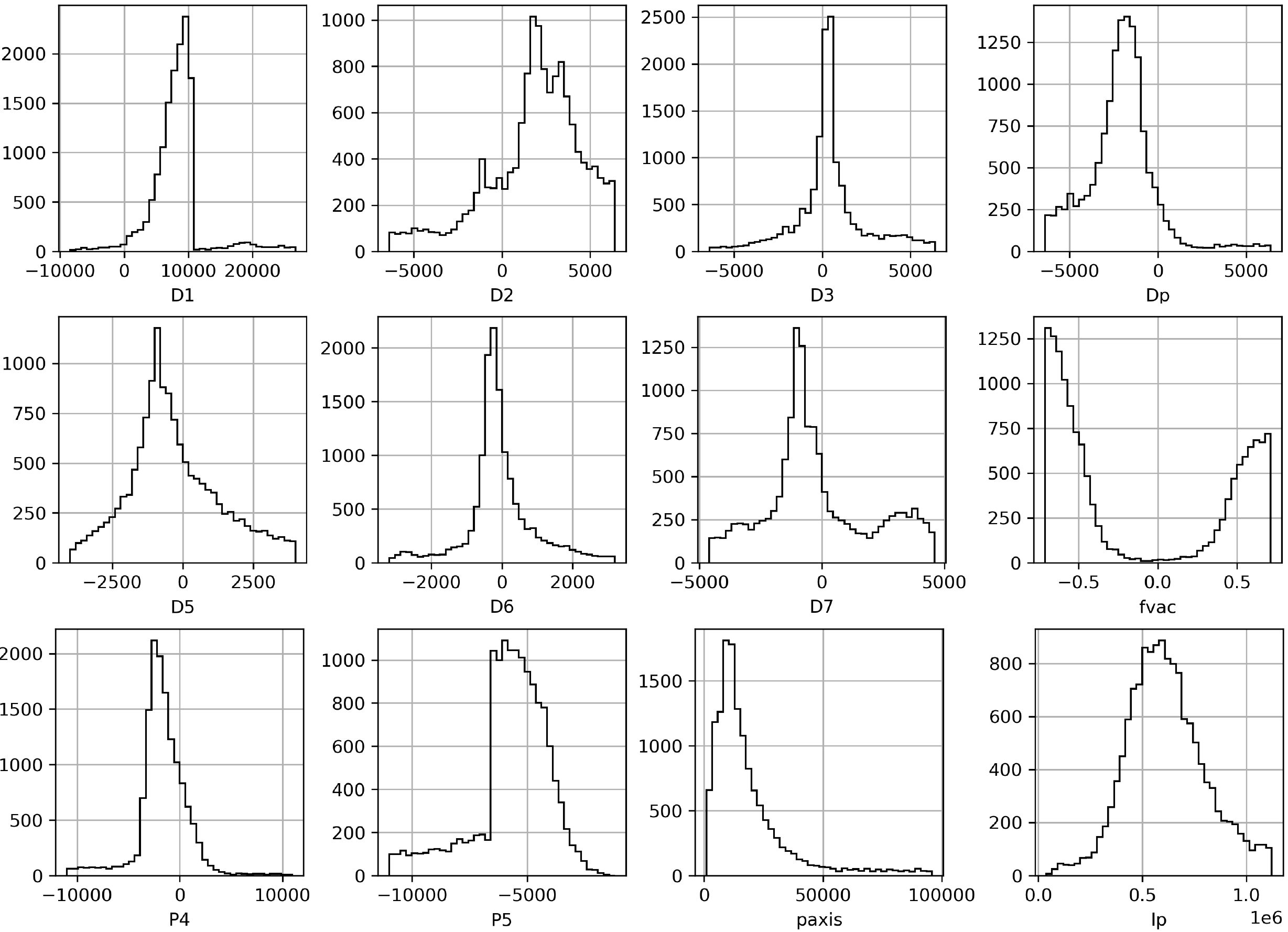}
\caption{\label{fig:mcmchist} Distribution of the plasma parameters, $f_{vac}$ and PF coil currents explored by the MCMC sampler. \textcolor{black}{Most chains are sampeld within the limits in Table~\textsc{ii}, except for one run with very high solenoid currents, whose high stray-field requires higher D1 currents}.}
\end{figure*}

The distributions of parameters from the sequential MCMC sampler are shown in Figure~\ref{fig:mcmchist}. Some of the PF coil currents (D1, P4, P5) and of plasma pressure and current are obliged by the need to keep the plasma separatrix confined, while other coil currents are more free to vary to produce finer changes in the resulting geometry of the separatrix and divertor legs. The distribution of $f_{vac}$ values is mostly reflective of the choice to privilege $q_a>1$ and longer connection in the score $\mathcal{L}(eq.)$.

\subsection{Power-law fits}
Given a tuple $\underline{\theta}^{\mathrm{T}}_{n}=(y_{1,n},...,y_{K,n})$ of $K$ targets values corresponding to the parameters $\underline{X}_{n}$ of equilibrium eq(n), a power-law model can be simply fit as an ordinary least-squares problem in log-space:
\begin{eqnarray}
\nonumber \mathrm{V}_{n,j}  & = & \log(X_{n,j}) - \frac{1}{N}\sum_{i=1}^{N} \log(X_{i,j})  \\
& & (n=1,...,N;\ j=1,...,dim_{in}) \\
\nonumber \mathrm{Y}_{n,k} & = & \log(\theta_{n,k})- \frac{1}{N}\sum_{i=1}^{N}\log(\theta_{i,j}) \\
& & (n=1,...,N; j=1,...,K) 
\end{eqnarray}
\begin{eqnarray}
{\beta}_{\mathrm{ols}}  & = &(\mathrm{V}^{\mathrm{T}}\mathrm{V})^{\dag}\mathrm{V}^{\mathrm{T}}\mathrm{Y}\\
\nonumber \underline{y}^{(PL)}_{(n)} & = & \frac{1}{N}\sum_{i=1}^{N} \log(\underline{\theta}_{i}) + \\
& & + \beta_{\mathrm{ols}} ^{\mathrm{T}} \left(\log(\underline{X_{n}}) -  \frac{1}{N}\sum_{i=1}^{N}\log(\underline{X}_{i}) \right)
\end{eqnarray}
Here, the dagger $^{\dag}$ indicates the Moore-Penrose pseudo-inverse. 
Therefore, in terms of the measurable targets $\theta,$ the entries in the $\beta_{\mathrm{ols}}$ matrix are the exponents of the scaling relations and the rms deviation from $y^{(PL)}$ is the relative uncertainty on the predicted target value given the parameters of the equilibrium configuration.


\end{document}